\documentclass[a4paper,11pt]{article}
\usepackage{jheppub} 
\usepackage{orcidlink}

\usepackage{mathtools}
\usepackage{amsmath}
\usepackage{amsfonts}
\usepackage{amssymb}
\usepackage{bbold}
\usepackage{bm}
\usepackage{esint}
\usepackage{orcidlink}

\usepackage{soul}
\usepackage{slashed}

\newcommand{\bB}{{\bf B}}
\newcommand{\bR}{{\bf R}}

\newcommand{\bE}{{\bf E}}
\newcommand{\bn}{{\bf n}}
\newcommand{\bk}{{\bf k}}

\newcommand{\bp}{{\bf p}}
\newcommand{\bq}{{\bf q}}
\newcommand{\ncm}{\ensuremath{n_{\mathrm{cm}^3}}}
\newcommand{\BG}{\ensuremath{B_{\mathrm{G}}}}
\newcommand{\BzG}{\ensuremath{B_{0,\mathrm{G}}}}
\newcommand{\Lcm}{\ensuremath{L_{\mathrm{cm}}}}
\newcommand{\mev}{\ensuremath{m_{\mathrm{eV}}}}

\long\def\exclude#1{}

\newcommand{\pF}{p_{\rm F}}
\newcommand{\wP}{\omega_{\rm P}}
\newcommand{\kS}{k_{\rm S}}

\newcommand{\Eav}{E_{\rm av}}

\title{Axion-photon conversion in transient compact stars:
Systematics, constraints, and opportunities}

\author[a]{Damiano F.\ G.\ Fiorillo \orcidlink{0000-0003-4927-9850},} 
\affiliation[a]{Deutsches Elektronen-Synchrotron DESY,
Platanenallee 6, 15738 Zeuthen, Germany}

\author[b,c]{\'Angel Gil Muyor
\orcidlink{0000-0003-0205-3010},}
\affiliation[b]{Dipartimento di Fisica e Astronomia, Universit\`a degli Studi di Padova,\\ Via Marzolo 8, 35131~Padova, Italy}
\affiliation[c]{Istituto Nazionale di Fisica Nucleare (INFN), Sezione di Padova,\\ Via Marzolo 8, 35131~Padova, Italy}

\author[d]{Hans-Thomas Janka
\orcidlink{0000-0002-0831-3330},}
\affiliation[d]{Max-Planck Institut f\"ur Astrophysik, Karl-Schwarzschild-Str.~1, 85748 Garching, Germany}

\author[e]{\\ Georg G.\ Raffelt
\orcidlink{0000-0002-0199-9560},}
\affiliation[e]{Max-Planck-Institut f\"ur Physik, Boltzmannstr.~8, 85748 Garching, Germany}

\author[b,c]{and Edoardo Vitagliano
\orcidlink{0000-0001-7847-1281}}

\abstract{We study magnetic conversion of ultra-relativistic axion-like particles (ALPs) into photons in compact-star environments, focusing on the hot, transient conditions of core-collapse supernova (SN) remnants and neutron-star mergers (NSMs). We address previously overlooked uncertainties, particularly the suppression caused by ejected matter near the stellar surface, a region crucial to the conversion process. We derive analytical expressions for the transition rate; they reveal the influence of key parameters and their uncertainties. We update constraints using historical gamma-ray data from SN~1987A and find $g_{a\gamma}<5\times10^{-12}~{\rm GeV}^{-1}$ for $m_a\lesssim10^{-9}$~eV. We also forecast sensitivities for a future Galactic SN and for NSMs, assuming observations with Fermi-LAT or similar gamma-ray instruments. We distinguish ALPs---defined as coupling only to photons and produced via Primakoff scattering---from axions, which also couple to nucleons and emerge through nuclear bremsstrahlung. We omit pionic axion production due to its large uncertainties and inconsistencies, though it could contribute comparably to bremsstrahlung under optimistic assumptions. For the compact sources, we adopt time-averaged one-zone models, guided by numerical simulations, to enable clear and reproducible parametric studies.
}

\begin{document}
\maketitle
\flushbottom

\section{Introduction}\label{sec:introduction}

One well-motivated class of new feebly interacting particles consists of axions and axion-like particles (ALPs) \cite{DiLuzio:2020wdo}. Depending on their fundamental origin, such pseudoscalar bosons couple with different strengths to different fermions and, unless there are special cancellations, generically to photons. This two-photon interaction, which is at the core of our investigation, is analogous to the one for neutral pions with the structure 
\begin{equation}\label{eq:interactionlagrangian}
    \mathcal{L}_{a\gamma}=-\frac{1}{4}g_{a\gamma}a F_{\mu\nu}\tilde{F}^{\mu\nu}=g_{a\gamma} a \bE\cdot \bB ,
\end{equation}
where $g_{a\gamma}$ is a coupling constant with the dimension of inverse mass. In astrophysical sources, this interaction engenders production through Primakoff scattering on charged particles, $\gamma+Ze\to Ze+a$, mediated by photon exchange \cite{Dicus:1978fp, Raffelt:1985nk}. The resulting energy loss, notably of globular cluster stars, would shorten the duration of advanced evolution stages, leading to a constraint of $g_{a\gamma}< 5\times10^{-11}~{\rm GeV}^{-1}$ \cite{Ayala:2014pea, Dolan:2022kul, Caputo:2024oqc,Carenza:2024ehj}. This bound provides a baseline for other arguments and searches, motivating us to use $g_{11}=g_{a\gamma}\times10^{11}~{\rm GeV}$ as a dimensionless parameter to express various scaling laws.

In axion or ALP searches, this extremely feeble interaction strength can be overcome by leveraging coherence of Primakoff conversion over the large volume of a macroscopic \hbox{$B$-field} rather than scattering on individual charged particles \cite{Sikivie:1983ip}. This seminal idea drives ongoing axion dark matter searches, which typically utilize microwave cavities placed in strong $B$-fields to detect power from axion-photon conversion. However, we here focus on another manifestation of the same principle, the conversion of relativistic ALPs as they propagate through a transverse $B$-field \cite{Sikivie:1983ip}. This effect can be conveniently framed as axion-photon ``flavor oscillations,''  where the two states mix by the external $B$-field \cite{Raffelt:1987im}. (This picture is equivalent to graviton-photon mixing in an external $B$-field known as the Gertsenshtein effect \cite{Gertsenshtein:1961}, which is based on the much weaker graviton-two photon coupling.) In particular, one can aim a long dipole magnet at the Sun to observe $X$-rays from the conversion of ALPs produced through Primakoff scattering of thermal photons in the solar core. The strongest limit from such a helioscope comes from the now-decommissioned CAST experiment, which provided $g_{11}<5.8$ at 95\% CL \cite{CAST:2024eil}, surprisingly similar to the globular-cluster bound. The upcoming IAXO helioscope \cite{IAXO:2019mpb} will strongly improve upon the CAST sensitivity.

It has long been recognized that a ``CAST in the sky'' approach can be more effective. One case to be reexamined here is axions emitted by the collapsed core of SN~1987A and their subsequent conversion into 100-MeV-range $\gamma$-rays within the $B$-field of the galaxy or of the progenitor star \cite{Brockway:1996yr, Grifols:1996id, Payez:2014xsa, Hoof:2022xbe, Manzari:2024jns}. Over the years, many studies have explored various astrophysical scenarios, ranging from the conversion of cosmic microwave photons all the way to 100~GeV $\gamma$-rays. The resulting constraints have been systematically compiled in what we call the ALPscape---a widely circulated plot that maps out the limits on $g_{a\gamma}$ as a function of the mass $m_a$ \cite{OHare}. This representation serves as a comprehensive reference, containing an extensive list of studies, too numerous to be detailed here. A portion of the ALPscape is shown in Fig.~\ref{fig:sn1987a}, where the constraints typically span the range $g_{11}=0.3$--3.

What are the key parameters for this general approach? Over a distance $L$ smaller than the axion-photon oscillation length, the transition probability is $P_{a\gamma}=(g_{a\gamma} B L/2)^2$.  Thus, one crucial scale is $BL$, which can be comparable across vastly different systems,  and numerically is $0.99~{\rm keV}\,\BG\Lcm$, where $\BG=B/{\rm Gauss}$ and $\Lcm=L/{\rm cm}$. For CAST, where $B\simeq 10~{\rm Tesla}=10^5~{\rm Gauss}$ and $L\simeq10~{\rm m}$, we obtain $(BL)_{\rm CAST}\simeq100~{\rm GeV}$. For a pulsar with $B=10^{12}~{\rm G}$ and a characteristic $L=10$~km, we find $(BL)_{\rm PSR}\simeq 10^{12}~{\rm GeV}$. In the Milky Way, a typical $B$ is of the order of $\mu$G with a coherence length of about 1~kpc, yielding $(BL)_{\rm MW}\simeq 3\times 10^{9}~{\rm GeV}$. 

The same $BL$ factor also appears in cosmic-ray acceleration because the maximum energy at which cosmic rays can be magnetically confined in a source can be conservatively estimated to be $Ze BL$, where $Ze$ is the charge of the particle. For protons to reach the highest observed cosmic-ray energies of $10^{20}~{\rm eV}$,  a value of $(BL)_{\rm CR}\simeq 3\times10^{11}~{\rm GeV}$ is needed, though such environments are rare. Nevertheless, many astrophysical systems with slightly lower but comparable $BL$ values exist on a diagonal band on the ``Hillas Plot,'' which is famous for compiling $B$ and $L$ for important astrophysical systems \cite{Hillas:1984ijl}. Therefore, in addition to setting a minimal limit on cosmic-ray confinement, the highest $BL$ values also constrain axion-photon conversion. Of course, the detailed sensitivities and constraints depend on more than just $BL$, as illustrated by CAST, which achieved strong constraints despite its modest~$BL$.

In conversion, the $a$--$\gamma$ momentum mismatch $\Delta k=\Delta m^2/2E$ is overcome by the limited distance $L$ between production and detection. Thus, the conversion reaches a maximum if $L\gtrsim 2E/\Delta m^2$, as the oscillation length becomes smaller than $L$. Therefore, larger astrophysical systems probe smaller $m_a$ values. This saturation effect is a major obstacle to searching for QCD axions by this method. For them, the mass is approximately given by the relation $m_a f_a\approx m_\pi f_\pi$, where $m_{\pi}=135$~MeV is the neutral pion mass and $f_\pi=92.4$~MeV the pion decay constant. Moreover, the photon coupling is $g_{a\gamma}=\alpha/(2\pi f_a)$, up to a model-dependent numerical factor, so that $g_{a\gamma}=(m_a/m_\pi)\,\alpha/(2\pi f_\pi)$, where $\alpha$ is the fine-structure constant. This linear relation between $g_{a\gamma}$ and $m_a$ is often called the {\em axion line\/} in the ALPscape, really a broad band after including model-dependent variations. For $\pi^0$ itself, $g_{\pi\gamma}$ and $m_{\pi}$ also lie on this band and sometimes it is useful to think of $\pi^0$ as a massive ALP. If the only coupling assumed is $g_{a\gamma}$, relativistic ALP-photon conversion in astrophysical systems never get close to the axion line.

While photons are massless, in a plasma and/or magnetic field, they acquire a nontrivial dispersion relation that can enhance or suppress the conversion depending on whether it makes the momentum mismatch smaller or larger. Even resonant conversion can occur if the photon plasma mass equals $m_a$. However, uncertainties in matter distributions within these systems introduce additional uncertainties, a key issue we will study in detail.

Our primary goal is to study such uncertainties and to examine the sensitivity region in the ALPscape coming from a traditional source, the historical SN~1987A, from the next galactic SN, and a new class, neutron star mergers (NSMs), where large $B$-fields are expected to be generated. The case of SN 1987A has recently received new attention by the intriguing idea of including the $B$-field of the progenitor star, which remains present for the few hours between collapse and explosion \cite{Manzari:2024jns}. The results depend on the assumed $B$-field as well as the matter distribution. Moreover, they depend on the production processes in the SN core, for which one could assume only the Primakoff process or, separately, interactions with the nuclear medium, notably bremsstrahlung and pion conversion. 

In addition, the Berkeley group \cite{Manzari:2024jns} has emphasized the detection opportunities provided by a future galactic SN or an extragalactic NSM, always assuming conversion in the $B$-field of these transient compact objects, and proposed a dedicated network GALAXIS of gamma-ray satellites, assumed to have similar capabilities as the existing Fermi-LAT instrument \cite{Fermi-LAT:2009ihh}. They suggested that, with optimistic assumptions, the high-mass end of the sensitivity forecast might actually cut into the QCD axion band, an exciting idea that warrants further examination. Moreover, following these ideas, NSMs were meanwhile studied in greater detail by the Bari group \cite{Lecce:2025dbz}, finding a somewhat pessimistic detection forecast for realistic QCD axions.

The Berkeley group has asserted that ALPs will never couple to photons alone because loop effects will always engender significant couplings to nucleons, implying that the traditional ALPscape, which is based on the photon coupling alone, is an overly pessimistic representation. We somewhat disagree with this perspective and prefer to study clearly separated cases, namely (i)~ALPs, defined to interact with photons alone and with unconstrained mass, (ii)~axions with relative couplings to nucleons and photons given by the usual axion relations, and (iii)~QCD axions, where in addition the mass is constrained to lie in the axion band, meaning that $m_a f_a\simeq f_\pi m_\pi$. Sensitivity to this region is the primary goal of any ALP search.

The main uncertainty of ALP-photon conversion near compact transient sources is dominated by uncertain $B$-field and matter distributions, and the primary goal of the first part of our study is to derive simple scaling laws that allow one to understand these effects. One concrete practical finding is that in NSMs, the ejected matter likely suppresses the conversion in the region close to the source, i.e., the critical region for the highest-mass ALPs. As a consequence, the perspective for QCD axion detection from NSMs is even more pessimistic than found by the Bari group \cite{Lecce:2025dbz}.

For the case of photon-only coupling, we find that the $B$-field of the SN~1987A progenitor does not exclude additional high-mass parameter space beyond the CAST and globular cluster limits, in contrast to the findings of the Berkeley group \cite{Manzari:2024jns}. Thus, the SN~1987A limit remains dominated by the Galactic $B$-field. We also reexamine constraints based on the latter, which slightly differ from previous ones.

On the other hand, the next galactic SN indeed provides an exciting QCD axion detection opportunity, although one would need to by lucky with regard to SN distance and $B$-field strength and orientation. In view of the general difficulties of reaching the axion band by any means, the proposed GALAXIS provides an exciting new perspective.

To substantiate these findings, we begin in Sec.~\ref{sec:axion-photon} with a general examination of ALP-photon conversion in stellar $B$-fields, assumed to have dipole structure. In Sec.~\ref{sec:production}, we describe the axion emission rates based on Primakoff scattering and bremsstrahlung, and briefly explain why we omit the pion process. We construct our compact-source models in Sec.~\ref{sec:sources}, where we adopt average properties guided by numerical simulations. We demonstrate that a more detailed description is not necessary. We turn to ALP-photon conversion in concrete astrophysical $B$-fields, including the Galactic one, in Sec.~\ref{sec:sources}, whereas we summarize the properties of the gamma-ray satellites SMM and Fermi-LAT in Sec.~\ref{sec:satellites}. Our practical results follow in Sec.~\ref{sec:results}, namely, bounds from the historical SN~1987A and detection forecasts for the next Galactic SN or a nearby NSM. We conclude with a summary in Sec.~\ref{sec:discussion}. Extended discussions of technical details are deferred to several appendices.

\section{Axion-to-photon conversion in a dipole field}\label{sec:axion-photon}

In this section, we derive simple algebraic expressions for the probability of axion-to-photon conversion in the magnetic fields of the stellar objects that produce the axions. These estimates are essential for understanding the impact of astrophysical environments on the conversion efficiency. By using simplified formulas, we can immediately discern how our results scale with each uncertain parameter that characterizes the system.

\subsection{Basic formalism}

We begin with a brief summary of axion-to-photon conversion in an external $B$-field \cite{Raffelt:1987im}, assuming the generic interaction structure of Eq.~\eqref{eq:interactionlagrangian}. The $B$-field is taken to be orthogonal to the direction of motion because its component along this direction does not spawn axion-photon conversion due to angular momentum conservation. We always consider photons with electric field orientation parallel to the magnetic field, represented by the vector potential $A$, which together with the axion forms the ``two-flavor spinor'' $\psi=(A,a)^T$. In the ultarelativistic limit, it obeys a Schr\"odinger-like equation 
\begin{equation}\label{eq:schrodinger}
    i\frac{\partial \psi}{\partial s}=-\mathcal{H}\psi,
\quad\hbox{where}\quad    
  \mathcal{H}=\begin{pmatrix}
         \Delta_B+\Delta_{\rm P} & \Delta_{a\gamma}\\
         \Delta_{a\gamma} & \Delta_a
    \end{pmatrix},
\end{equation}
and $s$ is the coordinate along the quasi-classical trajectory.

In the matrix $\mathcal{H}$, the different elements are explained as follows. The term that mixes photons with axions is
\begin{equation}
    \Delta_{a\gamma}=\frac{g_{a\gamma}B}{2}\simeq4.9\times10^{-18}~{\rm cm}^{-1}\,g_{11}\BG,
\end{equation}
where $B$ is the magnetic field component transverse to the trajectory with $\BG=B/{\rm Gauss}$ and $g_{11}=g_{a\gamma}\,10^{11}$~GeV as defined earlier. The momentum difference between a massless axion and one with mass $m_a$ is
\begin{equation}
    \Delta_{a}=-\frac{m_a^2}{2E}\simeq-2.5\times 10^{-4}\; \mathrm{cm}^{-1}\;\mev^2 E_{100}^{-1},
\end{equation}
where $m_{\rm eV}=m_a/1\;\mathrm{eV}$ and $E_{100}=E/100\;\mathrm{MeV}$. For photons, a similar expression holds when they acquire an effective plasma mass. Specifically, in a nonrelativistic plasma, relevant to interstellar space or a SN progenitor star, one finds $m_\gamma^2=\wP^2=4\pi \alpha n_e/m_e$. Numerically, $\wP\simeq 3.79\times 10^{-11}~{\rm eV}\,\sqrt{n_e/\mathrm{cm}^{-3}}$, so that
\begin{equation}
    \Delta_{\rm P}=-\frac{\wP^2}{2E}\simeq
     -3.5\times 10^{-25}\;\mathrm{cm}^{-1}\; \ncm\; E_{100}^{-1},
\end{equation}
where $n_{\rm cm^3}=n_e/1\;\mathrm{cm}^3$. In addition, a transverse $B$-field causes a refractive vacuum shift for the photons of 
\begin{equation}\label{eq: DeltaBirefringence}
    \Delta_B=\frac{7\alpha}{90\pi}\,\left(\frac{B}{B_{\rm crit}}\right)^2E
    \simeq4.7\times 10^{-19}\;\mathrm{cm}^{-1}\; 
    \BG^2 E_{100},
\end{equation}
where $B_{\rm crit}=m_e^2/e=4.41\times 10^{13}\,\rm G$. 
In all practical cases of interest to us, photon dispersion is either dominated by the $B$-field or the plasma effect, but not both in the same region.

If the conversion probability is much smaller than unity, which is always the case for the values of axion-photon coupling relevant here, it can be obtained by perturbation theory as $P_{a\gamma}(s)=|\mathcal{A}_{a\gamma}(s)|^2$ through \cite{Raffelt:1987im}
\begin{equation}\label{eq:general_conversion}
    \mathcal{A}_{a\gamma}(s)=i\int^{s}_0 ds' \Delta_{a\gamma}(s')\exp\left\{i\int_s^{s'} ds''\Bigl[\Delta_B (s'')+\Delta_{\rm P}(s'')-\Delta_a(s'')\Bigr]\right\}.
\end{equation}
This expression is obtained directly by noting that Eq.~\eqref{eq:schrodinger} has the same form as a Schr{\"o}dinger equation, to which time-dependent perturbation theory can be applied. Therefore, we can introduce the interaction basis, defined by the modified fields $\tilde{\psi}$
\begin{equation}
    \psi=\exp\left[i\int^s \mathcal{H}_0 ds'\right]\tilde{\psi},
\end{equation}
where $\mathcal{H}_0$ includes only the diagonal terms of $\mathcal{H}$. In this basis, the fields $\tilde{\psi}$ evolve only according to the perturbation $\Delta_{a\gamma}$, and their evolution can therefore be solved perturbatively, leading to Eq.~\eqref{eq:general_conversion}. This is our master formula from which we will obtain the main features of the axion-to-photon conversions in stellar magnetic fields.

\subsection{Dipole field}

Our main interest are the systematics related to axion-photon conversion in the $B$-field of a star. This might be a magnetized neutron star in the form of a pulsar, a hypermassive neutron star (HMNS) as remnant of a NSM, or the progenitor of a core-collapse SN. In all of these situations, axion-photon conversion happens outside of the star, in the region where the magnetic field is probably well-described by a dipole structure
\begin{equation}
    \bB=B_0 \left(\frac{R_0}{R}\right)^3 \left[3 \bn (\bf{m}\cdot \bn)-\bf{m}\right],
\end{equation}
where $\bf m$ is the unit vector of the dipole moment of the star, $\bn=\bR/R$ is the unit radial vector with $R=|\bR|$, and $B_0$ is the magnetic field at a reference radius $R_0$. In the equatorial plane, defined by ${\bf m}\cdot\bn=0$, and at $R=R_0$, the field is $\bB=-{\bf m} B_0$, and decreases with radius as $(R_0/R)^3$. For a radial axion trajectory with zenith angle $\theta$ relative to ${\bf m}$, the $\bB$-field components along (L) and transverse (T) to the trajectory are by modulus
\begin{equation}
    B_{\rm T}=B_0 \left(\frac{R_0}{R}\right)^3\sin\theta
    \quad\hbox{and}\quad
     B_{\rm L}=B_0 \left(\frac{R_0}{R}\right)^32\cos\theta.
\end{equation}
Notice that, in the plane transverse to the radial axion trajectory, the $B$-field orientation is fixed as a function of radius, so that our two-state assumption is justified. In practice, we will effectively consider an equatorial trajectory; an overall factor $\sin^2\theta$ in the conversion rate is trivial to include. 
The reason for the universality of the dipole structure is that we typically consider conversions in vacuum, where the plasma cannot strongly drag the magnetic field lines. Among all multipole components, the dipole decreases most slowly with radius and therefore dominates at larger distances. For each studied source, we will later provide a more detailed justification for using a dipole field. Due to this commonly assumed geometry, many properties of axion-photon conversion can be understood through simple order-of-magnitude estimates, without the need for detailed calculations. 

In this section, we explore these scaling laws. In later sections---each focused on a specific astrophysical source---we will perform more detailed numerical calculations for comparison. Here, we always neglect the potential role of the plasma density in suppressing axion-photon conversion. This approximation will be justified individually for each class of sources in the later sections.

\subsection{Massless axions}\label{sec:massless_pedestrians}

We consider first the case of massless axions ($m_a=0$). In the dipole magnetic field, we parameterize $\Delta_B=\Delta_{B,0} (R_0/R)^6$ and $\Delta_{a\gamma}=\Delta_{a\gamma,0} (R_0/R)^3$, following the scaling with the magnetic field. Therefore, up to an overall phase,
our master formula Eq.~\eqref{eq:general_conversion} yields
\begin{equation}
    \mathcal{A}_{a\gamma}=\int_{R_i}^{\infty}dR\,\frac{\Delta_{a\gamma,0}R_0^3}{R^3}\exp\left[i\frac{\Delta_{B,0}R_0^6}{5R^5}\right],
\end{equation}
where $R_i$ is the inner radius at which the axions begin converting if there are no other suppressions. Usually, $R_i$ is set by the region near the compact-star surface, where the plasma frequency drops quickly. Otherwise, we here neglect the plasma frequency; as we will see, in all of the environments that we consider efficient for conversions, this turns out to be a good approximation.

As long as the term in the exponential is large, conversions are inefficient because of its rapid oscillations, which strongly reduces the average amplitude. Therefore, a characteristic radius $R_{\rm conv}$ immediately stands out as the most relevant for conversion, defined by the condition that $\Delta_B(R_{\rm conv}) R_{\rm conv}=1$. At smaller distances, the rapid oscillations damp conversions, whereas at larger distances, the $R^{-3}$ variation of the magnetic field also reduce them, so this is the typical radial range that dominates the integral. Numerically
\begin{equation}\label{eq:conv_radius}
    R_{\rm conv}\simeq 34\;\mathrm{m}\; \BzG^{2/5} E_{100}^{1/5} R_{0,6}^{6/5},
\end{equation}
where we introduce $R_0=R_{0,6} 10^6\,\rm cm$ and $\BzG=B_0/{\rm Gauss}$. Around this region, the contribution to the integral must be of the order of $\mathcal{A}_{a\gamma}\sim \Delta_{a\gamma}(R_{\rm conv})R_{\rm conv}$. To be more precise we perform the integral explicitly. Taking $R_i=0$ we find
\begin{equation}
    \mathcal{A}_{a\gamma}=\frac{\Gamma_{2/5}}{5^{3/5}}\,
    \frac{\Delta_{a\gamma,0} R_0^3}{(-i\Delta_{B,0} R_0^6)^{2/5}},
\end{equation}
where $\Gamma_{2/5}$ is the Gamma function at argument 2/5, so that
\begin{equation}\label{eq:conv_prob_massless_strong_magnetic}
    P_{a\gamma}=\frac{\Gamma_{2/5}^2}{5^{6/5}}
    \frac{\Delta_{a\gamma,0}^2R_0^6}{(\Delta_{B,0} R_0^6)^{4/5}}\simeq 1.2\times 10^{-13} \BzG^{2/5} g_{11}^2 R_{0,6}^{6/5} E_{100}^{-4/5}.
\end{equation}

So far we have assumed $R_i\ll R_{\rm conv}$ and have taken $R_i=0$. In the opposite limit $R_i\gg R_{\rm conv}$, implying that axions are emitted directly from a region where the magnetic field can be neglected for refraction. Therefore, we take $\Delta_{B,0}\sim 0$ and obtain the conversion amplitude as $\mathcal{A}_{a\gamma}=\Delta_{a\gamma,0} R_0^3/2 R_i^2$, so that
\begin{equation}\label{eq:conv_prob_low_magnetic}
    P_{a\gamma}= \left(\frac{\Delta_{a\gamma}(R_i) R_i}{2}\right)^2\simeq 6.0\times 10^{-24}\; \BzG^2 g_{11}^2 R_{0,6}^6 R_{i,6}^{-4}.
\end{equation}
In other words, here both photon and axion dispersion was neglected, causing maximal mixing between the two flavors. Notice that even in the mass-suppressed regime, the probability does not have an oscillating behavior; its absence is due to the power-law decrease of the magnetic field, which freezes out oscillations before they can even begin, so to speak, leading to a pure power-law suppression of the conversion.

\subsection{Mass effects}\label{sec:mass_effects_pedestrians}

Accounting for the axion mass, the conversion amplitude becomes
\begin{equation}
    \mathcal{A}_{a\gamma}=\int_{R_i}^{\infty}dR\,\frac{\Delta_{a\gamma,0}R_0^3}{R^3}\exp\left[-i|\Delta_a| R+i\frac{\Delta_{B,0}R_0^6}{5R^5}\right],
\end{equation}
where both $|\Delta_a|$ and $\Delta_B$ are positive.

The axion mass decreases the conversion probability if it causes oscillations more rapid than $\Delta_B$ at the dominant conversion distance; this means that, for $R_i\ll R_{\rm conv}$, mass suppression begins when $\Delta_a\gtrsim \Delta_B(R_{\rm conv})$ or explicitly
\begin{equation}\label{eq:threshold_mass}
    m_a\gtrsim 1.1\;\mathrm{eV}\;E_{100}^{2/5}\BzG^{-1/5}R_{0,6}^{-3/5}.
\end{equation}
For larger $m_a$ values, the conversion probability decreases, providing a new characteristic scale $R_a$, defined by $\Delta_a\sim \Delta_B(R_a)$, which numerically is
\begin{equation}
    R_a\simeq 35~\mathrm{m}\; \BzG^{1/3} E_{100}^{1/3} \mev^{-1/3} R_{0,6}.
\end{equation}
At smaller distances, the large magnetic field refraction impedes efficient conversions, whereas at larger distances, the constant mass term also leads to rapid oscillations. 

The precise manner with which the probability decreases depends on how small is the axion emission radius $R_i$ compared with $R_a$. This relation is more difficult to obtain, due to the interplay between mass-induced and magnetic-induced oscillations. In Appendix~\ref{app:mass_suppression}, we study the precise behavior of the integral and find once more simple algebraic scalings. In particular, for $R_i\ll R_a$, we find that the conversion probability drops exponentially and using Eq.~\eqref{eq:conversion_small_radii} obtain
\begin{equation}\label{eq:P_asym_1}
    P_{a\gamma}\simeq 2.0\times 10^{-13} \BzG^{1/3} g_{11}^2 R_{0,6} E_{100}^{-2/3} \mev^{-1/3} \exp\left[-1.1 \BzG^{1/3} \mev^{5/3} R_{0,6} E_{100}^{-2/3}\right].
\end{equation}
The exponential decline derives from the destructive interference between the oscillations caused by $m_a$ and magnetic photon refraction. On the other hand, if $R_i$ is sufficiently close to $R_a$, the magnetic-induced oscillations play a minor role, and the conversion probability saturates to a larger value
\begin{equation}\label{eq:P_asym_2}
    P_{a\gamma}\simeq 1.1\times 10^{2}\;\frac{g_{11}^2 R_{i,6}^6}{\BzG^2 E_{100}^2 R_{0,6}^6}.
\end{equation}
The largest among Eqs.~\eqref{eq:P_asym_1} and~\eqref{eq:P_asym_2} provides the order of magnitude of the conversion probability. These expressions are valid only for $R_i\lesssim R_a$, at which point the conversion probability reaches its maximum value
\begin{equation}\label{eq:max_conversion_mass}
    P_{a\gamma}\simeq 2.1\times 10^{-13}\; g_{11}^2 \mev^{-2}.
\end{equation}
Notice that the maximum conversion probability is independent of the surface magnetic field strength, although the radius where this maximum is achieved $R_i\sim R_a$ depends of course on the axion mass and on the magnetic field.

If $R_i$ increases further, so that the axion is emitted outside of the surface where its conversion would be most efficient, the conversion probability of course decreases. Following Eq.~\eqref{eq:conversion_large_radii}, the scaling in this regime is
\begin{equation}\label{eq:conversion_large_radii_massive}
    P_{a\gamma}=\left(\frac{\Delta_{a\gamma}(R_i)}{\Delta_a}\right)^2\simeq 3.8\times 10^{-28}\; \frac{\BzG^2 E_{100}^2 g_{11}^2 R_{0,6}^6}{\mev^4 R_{i,6}^6};
\end{equation}
obviously this expression holds only for $R_i\gtrsim R_a$, at which point it smoothly merges with Eqs.~\eqref{eq:P_asym_2} and~\eqref{eq:max_conversion_mass}. By comparing Eq.~\eqref{eq:conversion_large_radii_massive} with the corresponding massless limit in Eq.~\eqref{eq:conv_prob_low_magnetic}, we see that in this low-magnetic-field limit, the mass begins to suppress the conversion when $|\Delta_a|\gtrsim 2/R_i$, that is,  Eq.~\eqref{eq:threshold_mass} is now replaced by
\begin{equation}\label{eq:threshold_mass_low_magnetic}
    m_a\gtrsim 89\;\mathrm{meV}\; \left(\frac{E_{100}}{R_{i,6}}\right)^{1/2}.
\end{equation}
Notice that in this regime of small $B$-fields, one can express the conversion amplitude completely analytically as
\begin{equation}\label{eq: SNProbPrediction}
    \mathcal{A}_{a\gamma}=\Delta_{a\gamma,0} R_0^3\left[\frac{\Delta_a^2}{2}\left[\mathrm{Ei}(-i|\Delta_a| R_0)+i\pi\right]-\frac{i e^{-i|\Delta_a| R_0} (i+|\Delta_a| R_0)}{2 R_0^2}\right];
\end{equation}
for $\Delta_a R_0\ll 1$ it becomes identical to Eq.~\eqref{eq:conv_prob_massless_strong_magnetic}, and for $\Delta_a R_0\gg 1$ to Eq.~\eqref{eq:conv_prob_low_magnetic}.

\subsection{Landscape of cosmic axion converters}

\begin{figure}[ht]
\includegraphics[width=\textwidth]{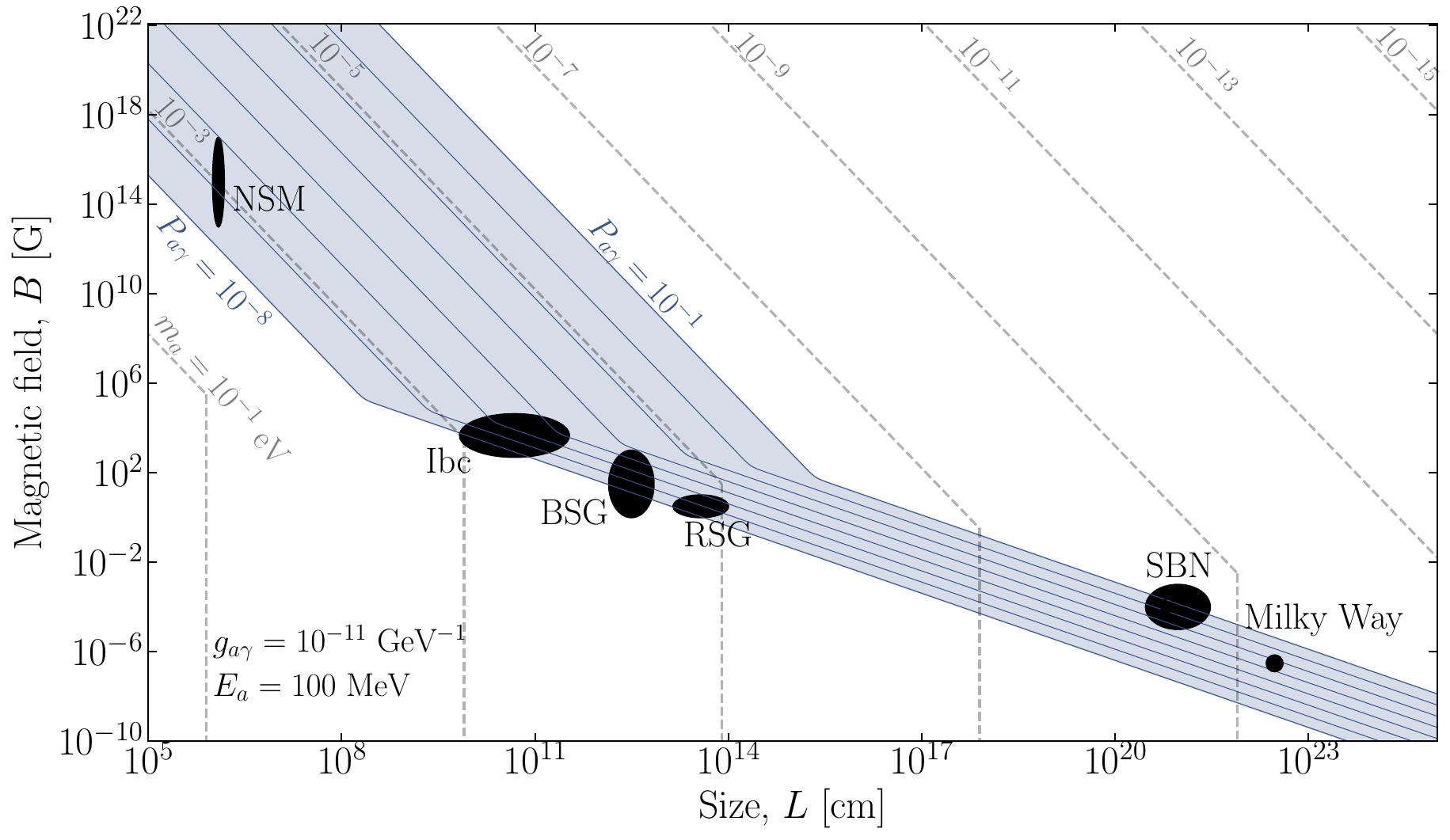}
\caption{Axion-photon conversion sites in the Hillas plane. Lines of constant conversion probability (for massless axions) are shown in blue, while the contours denoting the mass at which the axion conversion is suppressed are shown in gray. We focus on energetic axions produced in hot, compact sources, and correspondingly show the parameters of neutron star mergers (NSMs) \hbox{\cite{1992ApJ...392L...9D, Kiuchi:2014hja, Kiuchi:2015sga, Giacomazzo:2014qba, Metzger:2018qfl, Kiuchi:2023obe, Gutierrez:2025gkx}}, Type~Ibc stripped-envelope SNe~\cite{Candon:2025sdm}}, blue~\cite{Orlando:2018lbj} and red~\cite{Dorch:2004af,2010A&A...516L...2A} supergiants (BSGs and RSGs) hosting SNe, the Milky Way~\cite{Jansson:2012pc, Unger:2023lob}, and starburst nuclei (SBN)~\cite{Thompson:2006is, Lacki:2013ry, Marsh:2017yvc,2021ApJ...914...24L}. For NSMs, BSGs, and RSGs, the size $L$ denotes the surface radius of the star. For the Milky Way, we choose rough orders of magnitude for the typical magnetic field and distance, reproducing the conversion probability from the direction of SN~1987A.
\label{fig:hillas}
\end{figure}

Under the assumption of a dipole magnetic field, we now summarize the behavior of the conversion probability in a unified way in the Hillas plane, which spans size and magnetic field strength of different classes of sources. The landscape of these potential conversion sites in the Hillas plane is shown in Fig.~\ref{fig:hillas}.

At large magnetic fields, the conversion is dominated by the region around $R_{\rm conv}$, and consequently, the probability is given by Eq.~\eqref{eq:conv_prob_massless_strong_magnetic}, while at small fields, it follows Eq.~\eqref{eq:conv_prob_low_magnetic}. The separation between the two regimes happens when $R_{\rm conv}\simeq R_i$. At low magnetic fields, we recover the usual dependence of the probability on the product $BR$ that was already anticipated in the introduction.

We also highlight the axion mass at which the conversion begins being suppressed: in the large magnetic field regime, this follows from Eq.~\eqref{eq:threshold_mass}, while in the small field regime, it follows from Eq.~\eqref{eq:threshold_mass_low_magnetic} and then becomes independent of the magnetic field. Correspondingly, the contours of fixed axion mass $m_a$ turn vertical in Fig.~\ref{fig:hillas}.

This bird's eye view of possible astrophysical axion converters reveals immediately their possible significance. The largest conversion effects are associated with starburst nuclei (SBN), so that a SN in this environment would lead to a maximal conversion efficiency. The Milky Way follows immediately, confirming that for any source of high-energy massless ALPs, galactic or extragalactic, conversion in the Galactic field would dominate. We do not show the more speculative extragalactic magnetic fields, for which only upper limits exist on the order of nG over distances of hundreds of Mpc. The progenitors of core-collapse SNe, which can be either blue (BSGs) or red supergiants (RSGs) are next, with conversion probabilities between one to five orders of magnitude smaller. Finally, for neutron-star mergers (NSMs), the remnant has a very uncertain magnetic field, with a conversion probability that may be competitive or smaller than in BSGs and RSGs.

However, from the perspective of ALP masses, which can disrupt the conversion, the behavior is somewhat opposite, with NSMs reaching up to higher masses due to their small spatial scales. SBN would appear to be an excellent compromise, with their conversion probability reaching perhaps 2--3 orders of magnitude larger than the Milky Way, and extending to higher axion masses of $10^{-8}$--$10^{-9}$~eV. However, the increase in the conversion probability is offset by the geometrical flux suppression at Earth due to their larger distance; for 3~Mpc (e.g., the SBN M82 and NGC~253), compared to 50~kpc for SN~1987A, the gamma-ray flux is suppressed by nearly four orders of magnitude. In the $m_a$ range between $10^{-10}$--$10^{-8}$~eV, where Galactic conversion is suppressed by the ALP mass, a SN in a nearby SBN might become competitive with SN~1987A, since the conversion probability is four or five orders of magnitude larger than in a typical BSG progenitor, which can compensate the loss from distance. We do not investigate this question in detail, as the precise structure and magnitude of the SBN magnetic field is very uncertain.\footnote{A work dedicated to SBN appeared shortly after our paper was uploaded to the arXiv~\cite{Lecce:2025vjc}.}

Overall, Fig.~\ref{fig:hillas} shows at a glance the complementarity of different sources in testing the massless and the massive ALP regime. While the contours at low magnetic field are essentially independent of the large-scale structure of the field itself, at large field strengths, the change in slope depends on the assumed dipole structure. Thus, generalizing these results to other sources requires care, since the dipole assumption is not always realistic. For this reason, we only show sources analyzed in more detail in this work, and where axions from hot and compact transients convert. We also notice that the conversion probability depends quite sensitively on the ALP energy, and therefore this Hillas-like plot only holds for 100-MeV energies, while the conversion scenario might differ completely for lower energies, cases to be studied in a follow-up work.

\section{Axion production in dense nuclear environments} \label{sec:production}

If our axion-like particles interact exclusively with photons, they are produced in compact sources primarily through Primakoff scattering $\gamma p\to p a$. If in addition they interact with nucleons, the dominant emission processes are nuclear bremsstrahlung $NN\to NNa$. We here summarize the expression for these emission processes that we will use and defer a more detailed discussion to a series of appendices. Many authors have also used pion conversion $\pi^-p\to na$ as an additional nuclear process, which we entirely omit due to the many associated uncertainties and inconsistencies that we cannot resolve.

\subsection{Primakoff emission}

The dominant emission process provided by the axion-photon interaction is Primakoff scattering \cite{Dicus:1978fp, Raffelt:1985nk} $\gamma p\to p a$ on protons through photon exchange. Assuming static protons without recoil, and assuming massless axions as well as neglecting photon dispersion, the photon-axion conversion rate is \cite{Raffelt:1985nk}
\begin{equation}\label{eq:Primakoff-conversion-rate}
    \Gamma_{\gamma\to a}=\frac{g_{a\gamma}^2\alpha n_p}{8} 
    \left[\left(1+\frac{\kS^2}{4E^2}\right)\log\left(1+\frac{4E^2}{\kS^2}\right)-1\right].
\end{equation}
This process has a Coulomb divergence in the forward direction that is moderated by Debye screening through the scale
\begin{equation}\label{eq:Screening-Scale}
k_{\rm S}^2=\frac{4\pi\alpha\,n_p}{T}.
\end{equation}
The premise is that the background of degenerate electrons does not contribute to screening, i.e., $\kS$ accounts only for proton-proton anticorrelations caused by Coulomb repulsion. 

This simple prescription receives numerous small corrections, all at the ten percent level or less. The rate is reduced by proton degeneracy, possibly by as much as 20\%. The effective thermal photon mass, corresponding to the plasma frequency (typically around 10 MeV in a SN core), provides only a minor correction. Of course, if one were interested in the low-energy part of the axion spectrum (below tens of MeV), the photon spectral function in the medium would need to be understood in much greater detail, particularly the $\bE\cdot\bB$ fluctuations that source axion emission. The Primakoff rate is increased somewhat by muons, which also serve as essentially nondegenerate targets. In the outer regions of the source, the nuclear medium contains higher-$Z$ clusters in addition to protons, slightly increasing the overall emission. If a thermal $\pi^-$ population were present with an abundance comparable to that of muons, it would also serve as a Primakoff target. The axion spectrum is somewhat broadened by the thermal motion of the targets. In other words, many small effects act either to decrease or increase the overall rate and modify the spectrum. Some of these are discussed in more detail in Appendix~\ref{app:Primakoff}. 

In conclusion, the emission rate per unit volume will be based on Eq.~\eqref{eq:Primakoff-conversion-rate} without any corrections and therefore will be taken to be
\begin{equation}
    \frac{d\dot{N}_a}{dV}=\int \frac{2d^3\bk}{(2\pi)^3}\frac{1}{e^{E/T}-1}\,\Gamma_{\gamma\to a},
\end{equation}
whereas the differential number of axions emitted per unit energy is
\begin{equation}\label{eq:Primakoff-emission-rate}
    \frac{d\dot{N}_a}{dVdE}=\frac{E^2}{\pi^2(e^{E/T}-1)}\,\Gamma_{\gamma\to a}.
\end{equation}
Notice that, since $\Gamma_{\gamma\to a}$ depends on temperature only logarithmically through the screening scale, the total number of emitted axions per unit mass of material is nearly independent of density, and scales with the temperature roughly as $T^3$.

\subsection{Quasi-thermal axion spectrum}

The energy spectrum from Primakoff emission, but also the one from bremsstrahlung, is quasi-thermal in the sense of following roughly a Gamma distribution \cite{Tamborra:2012ac, CAST:2007jps}
\begin{equation}\label{eq:Gamma-Fit}
    \frac{d\dot N_a}{dMdE}=\frac{d\dot N_a}{dM}\,\frac{(\alpha+1)^{\alpha+1}}{\Gamma(\alpha+1)} \frac{E^\alpha}{\Eav^{\alpha+1}} \exp\left(-\frac{(1+\alpha)E}{\Eav}\right),
\end{equation}
normalized such that the right-hand side integrates to unity except for the factor $d\dot N_a/dM$. $\Eav$ is a parameter adjusted such that $\langle E\rangle=\Eav$ for a given $\alpha$. The second moment is
\begin{equation}
    \langle E^2\rangle=\frac{2+\alpha}{1+\alpha}\,\Eav^2
    \quad\hbox{implying}\quad
    \alpha=\frac{2-\langle E^2\rangle/\langle E\rangle^2}{\langle E^2\rangle/\langle E\rangle^2-1}.
\end{equation}
Therefore, the parameters $d\dot N_a/dM$, $\Eav$, and $\alpha$ follow immediately from the first moments of a given numerical or measured spectrum.

For Primakoff emission in a SN core, the screening scale is typically similar to $T$. Taking specifically $\kS=\kappa\, T$, around $\kappa=1$ a power-law approximation of the total emissivity (number of axions emitted per unit time and unit mass) is
\begin{equation}\label{eq:Na-Gamma-Primakoff}
    \frac{d\dot N_a}{dM}=\frac{g_{a\gamma}^2\alpha Y_p}{16 m_u}\,1.10\,T^3\,\kappa^{-0.70}
=\frac{g_{11}^2\,2.48\times10^{51}}{M_\odot\,{\rm s}}\,Y_p\left(\frac{T}{30\,{\rm MeV}}\right)^3\,\kappa^{-0.70},
\end{equation}
good to within a few percent for $0.5<\kappa<1.5$. Here, $Y_p$ is the average proton abundance per baryon and $m_u=1.661\times10^{-24}$~g the atomic mass unit. The spectral parameters are
\begin{equation}\label{eq:Parameters-Gamma-Primakoff}
    \Eav=3.51\,T\,\kappa^{0.09}
    \quad\mathrm{and}\quad
    \alpha=2.57\,\kappa^{0.20}.
\end{equation}
The average energy is nearly independent of the average screening scale. A Bose-Einstein spectrum at temperature $T$ provides $\Eav\simeq2.70\,T$ and $\alpha\simeq1.4$, so the Primakoff spectrum is slightly harder and more pinched than a thermal one. A Maxwell-Boltzmann spectrum has $\Eav=3T$ and $\alpha=2$.

\subsection{Nucleon-nucleon bremsstrahlung}

Nucleon-nucleon bremsstrahlung ($NN \to NNa$) is notoriously difficult to evaluate, even without a dense background medium, because modeling the nuclear interaction potential is required. The older axion literature typically used a one-pion-exchange (OPE) potential, as reviewed in Refs.~\cite{Raffelt:2006cw, Carenza:2019pxu}. However, for relevant CM energies around 100~MeV, the OPE approximation overestimates the rate by at least a factor of 4--6. This reduction has sometimes been modeled phenomenologically by including $\rho$-meson exchange~\cite{Ericson:1988wr, Carenza:2019pxu}. While the $\rho$ meson reduces the scattering amplitude in the tensor channel, it enhances it in the spin-spin channel---a phenomenon missing in  Ref.~\cite{Carenza:2019pxu}, as noted in Ref.~\cite{Bottaro:2024ugp}---and this phenomenological perturbative treatment has never been validated against actual scattering data. For axions coupling primarily to protons, as in the case of KSVZ axions, the dominant emission process is $np$ scattering. Here, flipping the proton spin alone suffices, so even non-tensor interactions can contribute to axion emission. These interactions are dominated by short-range components, which previous treatments have not accounted for. The validity of the Born approximation itself is generally not guaranteed. The only way to test these many approximations would be to directly compute the $np$ scattering cross section with the same approach and compare it with experimental data; any first-principle bremsstrahlung calculation should be validated by this test.

Even then, uncertainties remain due to static and dynamical correlations among the nucleons, their spins, and isospins \cite{Burrows:1998cg}; the vacuum interaction among them is not necessarily representative of the one in a nuclear medium. Reliable, practical forms of the dynamical structure functions are unavailable, so a simple parametric approach remains the most reasonable. Following this logic, we adopt the phenomenological treatment of Ref.~\cite{Raffelt:1991pw}, modeling axion emission as a problem of fluctuating spins. We assume a coupling only to protons
\begin{equation}\label{eq:axion-proton-Lagrangian}
    \mathcal{L}_{app}=-\frac{C_{ap}}{2f_a}\,\overline{p}\gamma^\mu \gamma^5p \,\partial_\mu a,
\end{equation}
implying that in the nonrelativistic limit, axions couple to the proton spin and are emitted by its fluctuations. In the literature, the dimensionless Yukawa coupling is often defined as $g_{ap}=C_{ap} m_p/f_a$. Focusing on the proton coupling does not significantly limit our parametric study, as the $NN$ bremsstrahlung process does not favor any particular isospin structure of the axion-nucleon couplings. Furthermore, effective couplings are substantially modified in a dense nuclear medium \cite{Springmann:2024mjp}, likely breaking any special symmetries.

Following essentially the approach taken in Ref.~\cite{Raffelt:1991pw} and discussed in more detail in Appendix~\ref{app:Bremsstrahlung}, we therefore adopt a parametric representation for the bremsstrahlung emission rate per unit mass in the form
\begin{equation}\label{eq:emission_rate_bremsstrahlung}
    \frac{d\dot{N}_a}{dE dM}=\frac{g_{ap}^2}{8\pi^2m_p^2} \,\frac{Y_p}{m_u}\,
    \frac{E}{e^{E/T}+1}\,\frac{\Gamma_\sigma}{1+(\Gamma_\sigma/2E)^2}.
\end{equation}
Here $Y_p$ is the proton fraction per nucleon and $m_u=1.661\times10^{-24}~{\rm g}$ the atomic mass unit. The emissivity is now entirely fixed by a single parameter, the spin fluctuation rate $\Gamma_\sigma$. The denominator of the last factor accounts for a reduction by multiple scattering, which however is no longer a large effect in view of the much smaller $\Gamma_\sigma$ values relative to the historical OPE assumptions.

At the CM energies relevant in a SN core, with $T\sim 30\,\mathrm{MeV}$, we may assume that the elastic $np$ scattering cross section $\sigma_{np}$ is comparable to the spin-flip cross section, implying
\begin{equation}
    \Gamma_\sigma\simeq F_{\rm deg}\,n_n v_{\rm rel}\sigma_{np},
\end{equation}
where $v_{\rm rel}\simeq 4\sqrt{T/\pi m_N}$ is the average relative thermal velocity between nondegenerate nucleons, $n_n$ is the neutron density, and $F_{\rm deg}$ a nominal correction factor for nucleon degeneracy. At a typical CM energy of 100~MeV, one finds $\sigma_{np}\simeq 40\,\mathrm{mb}$ (e.g.~Fig.~3 of Ref.~\cite{Rrapaj:2015wgs}), implying $\Gamma_\sigma/n_n\simeq 16\,\mathrm{mb}$ for $T\simeq 30\,\mathrm{MeV}$. This is surprisingly similar to the spin-flip rate obtained using the T-matrix elements for reproducing the nucleon-nucleon cross section; for example, from Fig.~6 of Ref.~\cite{Guo:2019cvs} we gather $\Gamma_\sigma/n_n\simeq 200\,\mathrm{MeV}\,\mathrm{fm}^{3}\simeq 10\,\mathrm{mb}$ for $Y_e=0.1$. Encouraged by this agreement, we model the spin fluctuation rate as
\begin{equation}\label{eq:Gamma_sigma}
    \Gamma_\sigma= 40\,\mathrm{MeV}\,\frac{\rho}{4\times 10^{14}\,\mathrm{g}/\mathrm{cm}^3}\sqrt{\frac{T}{30\,\mathrm{MeV}}}.
\end{equation}
Here we have used $\sigma_{np}=40$~mb, a neutron fraction per baryon of 0.85, and a nominal nucleon degeneracy correction of $F_{\rm deg}=0.6$, parameters inspired by our SN models. Moreover, the reference values for $\rho$ and $T$ correspond to the average values of the cold SN model to be discussed later.

In analogy to Primakoff emission, we next determine the quasi-thermal parameters describing bremsstrahlung. To study the impact of the multiple scattering factor, we here introduce a parameter $\gamma$ defined such that $\Gamma_\sigma=\gamma 4T/3$ so that for $T=30$~MeV, we find $\Gamma_\sigma=40$~MeV if $\gamma=1$. The total emission rate is
\begin{equation}
    \frac{d\dot N_a}{dM}
=\frac{g_{ap}^2\,6.24\times10^{74}}{M_\odot\,{\rm s}}\,\frac{Y_p\,\rho}{4\times 10^{14}\,\mathrm{g}/\mathrm{cm}^3}\left(\frac{T}{30\,{\rm MeV}}\right)^{5/2}
\gamma^{-0.28}.
\end{equation}
The spectral parameters are
\begin{equation}
    \Eav=2.47\,T\,\gamma^{0.12}
    \quad\mathrm{and}\quad
    \alpha=1.95\,\gamma^{0.33}.
\end{equation}
The average energy is nearly independent of the exact value of the multiple-scattering parameter.

\subsection{SN~1987A cooling bound on the axion-proton coupling}

As consistency check, we can now derive the usual SN~1987A cooling bound on $g_{ap}$ based on the schematic requirement that the energy-loss rate, calculated for $T=30$~MeV and $\rho=3\times 10^{14}\,\mathrm{g}/\mathrm{cm}^3$, should be smaller than $10^{19}~{\rm erg}~{\rm g}^{-1}~{\rm s}^{-1}$ \cite{Raffelt:1990yz, Raffelt:2006cw}. The energy-loss rate per unit mass is $(d\dot N/dM)\Eav$ and we need to evaluate it for $\rho=3\times 10^{14}\,\mathrm{g}/\mathrm{cm}^3$, implying that $\Gamma_\sigma$ is reduced by 3/4 compared with our earlier reference values and therefore $\gamma=1/4$. Assuming $Y_p=0.15$ then leads to a constraint $g_{ap}<1.52\times10^{-9}$.

In the KSVZ model, where $C_{ap}=-0.47$, this bound implies $f_a>2.9\times10^8$~GeV and the relation $m_a=5.7~{\rm meV}\,(10^9\,{\rm GeV}/f_a)$ implies $m_a<20$~meV. In Ref.~\cite{Raffelt:2006cw} a slightly more restrictive bound of 16~meV was stated, the difference caused by different choices about the spin fluctuation rate and a larger assumed $Y_p$. The latest study, based on a specific SN model and a detailed calculation of the nuclear emission rates, found \cite{Carenza:2019pxu}
\begin{equation}\label{eq:EffectiveCoupling}
    \sqrt{g_{ap}^2+1.64 g_{an}^2+0.87\,g_{ap}g_{an}}<1.16\times10^{-9}
\end{equation}
or, for the KSVZ case, $m_a<15$~meV. In view of the many differences in analysis, these constraints are identical to ours. This comparison shows that our parametric treatment of bremsstrahlung does not lead to any radical conclusions. Conversely, the treatment beyond OPE provides constraints identical with the older parametric ones \cite{Raffelt:2006cw}, although the uncanny precision of agreement (stated mass bound of 15 vs.\ 16 meV) is fortuitous. In view of these constraints, we use $g_9=g_{ap}/10^{-9}$ as a scaling factor for bremsstrahlung fluxes. As a nominal bound from SN~1987A cooling, we will adopt the simple number
\begin{equation}\label{SN1987A-cooling-bound}
    g_{ap}<1.0\times10^{-9},
\end{equation}
which we will use to multiply, for example, CAST constrains on $g_{a\gamma}$ for comparison in plots that show constraints on the product $g_{ap}\times g_{a\gamma}$. In the KSVZ model, this corresponds to a mass limit of $m_a<12.9$~meV. These adopted nominal limits are very close to those found in Ref.~\cite{Lella:2023bfb} if one omits the pion process.

\subsection{Pionic processes}

Another channel for axion emission is $\pi N\to N a$, driven by thermal pions within the dense and hot nuclear matter. This process would create axions with larger energies than bremsstrahlung and could be of particular interest for the detection of \hbox{$\gamma$-rays} produced by subsequent magnetic conversion. While the pion mass $m_\pi\simeq 140\,\mathrm{MeV}$ is somewhat large compared with typical SN temperatures $T=30$--50~MeV, the presence of pions, or that of muons (mass 105~MeV), is primarily determined by their chemical potentials, not the temperature.  The traditional focus has been on $\pi^- p\to n a$ \cite{Turner:1991ax, Raffelt:1993ix, Keil:1996ju, Fore:2019wib, Carenza:2020cis, Fischer:2021jfm, Choi:2021ign, Vonk:2022tho, Ho:2022oaw, Fore:2023gwv,Lella:2022uwi} because the $\pi^-$ population benefits from a large chemical potential $\mu_{\pi^-} = \hat\mu$. The latter is defined by $\hat\mu=\mu_n-\mu_p = \mu_e-\mu_{\nu_e} = \mu_\mu-\mu_{\nu_\mu}$, which in typical SN models can be close to $m_\pi$ and, in principle, even a $\pi^-$ condensate might form.

However, the role of pions in the nuclear matter of SN cores or in the interior of NSMs remains poorly understood. Except for a few essentially parametric studies \cite{Mayle:1993uj, Vijayan:2023qrt, Pajkos:2024iry, Fischer:2021jfm}, pions have generally been omitted without apparent justification. It is not known whether their inclusion, like that of muons \cite{Bollig:2017lki, Fischer:2020vie}, would result in a relatively benign quantitative modification of the models or require a more radical revision. 
The recent revival of pionic axion production began with Ref.~\cite{Fore:2019wib}, which estimated the $\pi^-$ abundance using a virial expansion of the equation of state, aiming to consistently include pion-nucleon and nucleon-nucleon interactions. It found a $\pi^-$ population potentially comparable to that of muons (a few percent of the baryon abundance) and rapidly increasing with density. The dispersion relation was then adjusted a posteriori to make the virial expansion result consistent with that of a thermal population at the given $\hat\mu$ and~$T$. In contrast, a recent calculation, including the same authors, used heavy-baryon chiral perturbation theory and found a zero-momentum energy (``mass'') of 200--260~MeV at nuclear density, increasing rapidly with density, though without information on the $p_\pi$ dependence~\cite{Fore:2023gwv}. This large upward shift in pion energies would suppress their abundance, preclude pion condensation, and strongly reduce the axion emissivity from thermal pions. These latest results are in agreement with the historical picture thoroughly reviewed in Ref.~\cite{Migdal:1990vm}, whereby $\pi^-$ never condense due to the $s$-wave energy shift, whereas $\pi^0$ and $\pi^+$ possess specific branches subject to condensation.

If, despite these doubts, thermal pions were present with a roughly muonic abundance, as assumed in recent studies, the resulting axion yield would be comparable to that from nucleon bremsstrahlung (not significantly larger), though with a harder spectrum. One might speculate that a high-statistics detection of such a signal could probe the properties of the nuclear medium in the SN core~\cite{Lella:2024hfk,Carenza:2025uib}. However, in the context of deriving limits or defining the detection threshold, we prefer to avoid speculations about a substantial pion contribution. Based on the current literature, we are not convinced of its predictive power; further discussion of our concerns is provided in Appendix~\ref{app:PionConversion}.

\section{Compact transient sources}\label{sec:sources}

Besides the nuclear physics of dense matter, the main other challenge for determining axion production in compact transient sources is the emission region itself. Recent studies have primarily relied on numerical simulations, followed by post-processing to extract the axion emissivity from the density, temperature, and composition profiles. While we follow a similar approach, such methods can turn to something like a ``black box,'' with limited control over the interpretation of the axion flux. To address this shortcoming, we introduce time-averaged one-zone models of such environments, guided by simulations, that provide a straightforward calculation of the axion signal and its variation with different input assumptions. Our approach stresses that higher precision is likely unattainable, making the average models a useful tool for parametric studies when only the time-integrated axion flux is important.

\subsection{Motivation}

To calculate the axion flux from a compact source, one often relies on numerical models that do not incorporate novel ingredients and are used primarily to guide expectations, assuming no strong backreaction on the models. While modern simulations focus on multi-dimensional hydrodynamics and neutrino transport, which are crucial for explosion physics, neutron-star kicks, and nucleosynthesis, axion emission from the dense core is not strongly affected by these effects. For this purpose, simpler spherically symmetric simulations are sufficient. Extensive suites of such models exist, varying in final-state neutron star (NS) masses, equations of state (EoS), and other input physics, such as the inclusion of muons or modifications to neutrino opacities. The key factors influencing axion emission are the temperatures and densities achieved within the proto-neutron star (PNS), its chemical composition, the final NS mass, and the duration of the emission.

The large variation in models and input assumptions shows that no standard supernova (SN) model exists, let alone a neutron star merger (NSM) model. As a consequence, any study of constraints or detection forecasts for novel feebly interacting particles is necessarily parametric. When deriving limits from the historical SN 1987A, the relevant parameters should be chosen conservatively to avoid overstating the constraints. In contrast, more optimistic assumptions can be used to assess detection prospects for a future event. The key question is whether, even under favorable conditions, the detection of QCD axions is entirely out of reach, or if there remains a plausible chance of discovering new physics. (We anticipate that detection is unlikely for NSMs, but reasonably plausible for the next galactic SN.) 
In particular, as we explore the SN 1987A limits or the margin of detectability in future events, we do not speculate about the opportunities that might arise from a high-statistics observation of the spectrum and time evolution of the gamma-ray signal. Since we are only interested in a signal integrated over a few seconds, we require only a few time-averaged properties of the compact sources, such as temperatures and densities, along with an understanding of the plausible range of variation for these properties. The goal of this section is therefore to construct such average compact sources and to clearly specify the input parameters and their range of variability. To examine the relationship between the axion flux from such average models and from specific numerical ones, we define a cold and a hot baseline model for SNe, and a single generic NSM model. These serve as test cases to illustrate what it means to use average models instead of detailed simulations.

\subsection{Garching supernova models}

Recent studies of SNe as sources for feebly interacting particles have often used a few models from the Garching group that include muons (but not pions) and span a representative range from ``cold'' to ``hot'' as layed out in Ref.~\cite{Bollig:2020xdr} and listed in Table~\ref{tab:SN-models}. The radial profiles of several physical parameters at different times are shown, for example, in the Supplemental Material of Ref.~\cite{Manzari:2024jns}. We use these models as reference cases, facilitating comparison with the results of other authors. These simulations were performed in spherical symmetry with the \textsc{Prometheus-Vertex} code with general-relativistic corrections and six-species neutrino transport, solving iteratively the two-moment equations for neutrino energy and momentum with a Boltzmann closure~\cite{Rampp:2002bq} and using the full set of neutrino processes listed in~Refs.~\cite{Janka:2012wk, Bollig:2017lki}. PNS convection was taken into account by a mixing-length treatment and explosions were manually triggered a few 100\,ms after bounce at the progenitor's Fe/Si or Si/O composition interface as described in~Ref.~\cite{Mirizzi:2015eza}. The models were all evolved until 10~s post bounce and they are publicly available upon request \cite{JankaWeb}.

These models were chosen to span a plausible range for SN~1987A. SFHo--18.6 has a canonical final neutron star (NS) mass and is considered representative for a typical SN.  Model SFHo--18.8 corresponds to a case near the lowest plausible NS mass, while SFHo--20.0 represents a scenario with a relatively heavy final NS mass. We recall that the largest observed gravitational NS mass is slightly above $2\,M_\odot$, but SNe forming such massive remnants must be rare. The two lighter cases are also the cooler ones, each reaching maximum temperatures of around 40~MeV. In contrast, the heavier models reach up to 60~MeV and are therefore  representative of the hotter end of the spectrum. In the model names, numbers such as 18.8 denote the progenitor star's mass in solar units.

The first letters of the model names denote the used EoS. The SFHo one \cite{Steiner:2012rk} is fully compatible with all current constraints from nuclear theory and experiment~\cite{Fischer:2013eka, Oertel:2016bki, Fischer:2017zcr} as well as astrophysics, including pulsar mass measurements \cite{Demorest:2010bx, Antoniadis:2013pzd, NANOGrav:2019jur} and the radius constraints deduced from gravitational-wave and Neutron Star Interior Composition Explorer measurements~\cite{LIGOScientific:2018cki, Bauswein:2017vtn, Essick:2020flb}. Results for the long-used LS220 EoS \cite{Lattimer:1991nc} are shown for comparison. These EoS are considerably softer than what was popular earlier, but the stiff variants used in the older literature are increasingly disfavored by the cited constraints. A softer EoS generically results in a smaller radius and higher temperature.

In Table~\ref{tab:SN-models} we show key global properties of our models, including the baryonic and gravitational mass of the final NS. 
Their difference represents the binding energy, emitted primarily as neutrinos. We also show the central density at the end of the simulation (10 s post-bounce) and the maximum temperature that is always reached within the first 1--2~s. We also show mean values for temperature and density that are averages over the NS and period of particle emission.

\begin{table}[t]
\setlength\tabcolsep{0pt}
\begin{tabular*}{\linewidth}{@{\extracolsep{\fill}} llllllllll}
\hline
\multicolumn{2}{l}{Model}&\multicolumn{2}{l}{SFHo--18.8}&\multicolumn{2}{l}{SFHo--18.6}&\multicolumn{2}{l}{LS220--20.0}&\multicolumn{2}{l}{SFHo--20.0}\\
Our name&&\multicolumn{2}{l}{Cold model}&---&&---&&\multicolumn{2}{l}{Hot model}\\
\hline
$M_{\rm NS}$ (baryon)&$M_\odot$&1.351&&1.553&&1.926&&1.947&\\
$M_{\rm NS}$ (grav.)&&1.241&&1.406&&1.707&&1.712&\\
$E_{\rm bind}$&$10^{53}~{\rm erg}$&1.98&&2.64&&3.94&&4.23&\\
\multicolumn{2}{l}{Lapse $\langle(1+z)^{-1}\rangle$}&0.77&(0.76)& 0.77 & (0.76) & 0.67 & (0.65) & 0.66 & (0.64)\\
$T_{\rm max}$     &MeV&39.4&&45.5&&60.0&&59.2&\\
$\langle T\rangle$&   &30.3&(29.4)&35.1&(34.1)&43.3&(41.1)&45.4&(44.4)\\
$\rho_{\rm max}$&$10^{14}$\,g/cm$^{3}$&7.82&&8.70&&10.2&&10.9&\\
$\langle \rho\rangle$&& 4.08&(4.73)&4.53&(5.23)&5.45&(6.33)&5.71&(6.52)\\
$\langle M t\rangle$&$M_\odot{\rm s}$& 5.28& (5.06) & 6.76 & (6.46) & 8.45 & (8.63) & 10.5 & (9.90)\\
\multicolumn{5}{l}{Average abundances per baryon}\\
\quad $\langle Y_p\rangle$&& 0.138 &(0.132)& 0.140 & (0.137) & 0.188 & (0.189) & 0.161 & (0.154)\\
\quad $\langle Y_n\rangle$&&0.853 & (0.865) & 0.849 & (0.861) & 0.811 & (0.811) & 0.834 & (0.845)\\
\quad $\langle Y_e\rangle$&&0.119 & (0.111) & 0.120 & (0.114) & 0.149 & (0.149) & 0.128 & (0.122)\\
\quad $\langle Y_\mu\rangle$&&0.022 & (0.022) & 0.025 & (0.024) & 0.039 & (0.040) & 0.035 & (0.033)\\
\multicolumn{5}{l}{Nucleon degeneracy suppression factors}\\
\quad $\langle F_{pp}\rangle$&&0.80 & (0.77) & 0.72 & (0.77) & 0.85 & (0.82) & 0.76 & (0.77)\\
\quad $\langle F_{nn}\rangle$&&0.48 & (0.42) & 0.42 & (0.44) & 0.61 & (0.55) & 0.49 & (0.47)\\
\multicolumn{5}{l}{Primakoff emission parameters}\\
\quad $\langle \kS\rangle$&MeV&27.0 & (29.4) & 26.9 & (29.2) & 31.3 & (35.2) & 28.2 & (30.1)\\
\quad $N_a$&$g_{11}^2 10^{50}$ & 15.7 && 34.0 && 100 && 133\\
\quad $\Eav$&MeV&86.0 && 94.0 && 108 && 107\\
\quad $\alpha$&&2.09 && 2.02 && 1.81 && 1.94\\
\multicolumn{5}{l}{Bremsstrahlung emission parameters}\\
\quad $\langle\Gamma_\sigma\rangle$&MeV&40.7 & (46.1) & 48.7 & (55.0) & 64.4 & (72.6) & 70.0 & (78.5)\\
\quad $N_a$&$g_{9}^2 10^{55}$ & 33.7 && 67.7 && 205 && 251\\
\quad $\Eav$&MeV& 58.5 && 65.0 && 72.5 && 76.7\\
\quad $\alpha$&& 1.55 && 1.56 && 1.42 && 1.65\\
\hline
\end{tabular*}
\caption{Muonic SN models of the Garching group \cite{Bollig:2020xdr} that serve as our reference cases and are frequently used in the axion literature. $T_{\rm max}$ is the maximum temperature reached, and $\rho_{\rm max}$ the maximum density, which is reached at the center at the end of the simulations (10~s post bounce). Average values are taken over the SN core and time using $T^3$ (in brackets using $\rho\, T^{5/2}$) as weight functions, without gravitational redshift effects. 
\label{tab:SN-models}}
\end{table}

\subsection{Average properties of the numerical models}

These averages are taken with axion emission in mind, i.e., they are taken with a weight function that emphasizes the hot and dense regions. In view of Primakoff emission, that varies essentially as $T^3$, we take the average of a quantity $Q$ in the form
\begin{equation*}
    \langle Q\rangle=\int dM dt\,Q(M,t)\,T^3(M,t)\bigg/\int dM dt\,T^3(M,t),
\end{equation*}
where $M$ is the mass coordinate. The results are not sensitive to the mass cutoff (taken as the final NS mass) or the end time (10~s for these models). Axion emission itself happens essentially in the inner half of the NS mass, moving toward the center with time, over a period of 3--4 seconds, beginning somewhat after bounce once the interior has heated up.  In view of bremsstrahlung, that varies essentially as $\rho\,T^{5/2}$, we use this quantity as an alternative weight function and show the corresponding averages in brackets. This weight function more strongly picks up the central regions with larger densities and overall larger $T$. These different averages provide us with a first impression of which overall properties of a given SN model actually matter regarding time-integrated axion emission.

In general, one could define such averages based on generic power laws of the type $\rho^m T^n$ and we denote the corresponding averages as $\langle Q\rangle_{mn}$ when it is necessary to distinguish between different cases. For instance, the $T^3$ weight corresponds to averages denoted as $\langle Q\rangle_{03}$, while the bremsstrahlung-inspired weight $\rho\,T^{5/2}$ corresponds to $\langle Q\rangle_{1\frac{5}{2}}$.

It is somewhat surprising that for all of the SFHo models, $T_{\rm max}/\langle T\rangle_{03}=1.30$, the same within 3~decimals, illustrating a strong regularity between the differnt models. For the LS220--20.0 model, with a different EoS, $T_{\rm max}/\langle T\rangle_{03}=1.39$, but still only 7\% larger. A similar observation pertains to $\rho_{\rm max}/\langle \rho\rangle_{03}=1.92$ for the two light models, 1.90 for SFHo-20.0 and 1.87 for LS20--20.0, i.e., the ratio is 1.90 for all models to better than 1.5\%. For our other weight function $\rho\,T^{5/2}$, we find $T_{\rm max}/\langle T\rangle_{1\frac{5}{2}}=1.32$--1.34 for the SFHo models and 1.46 for LS220--20.0, whereas $\rho_{\rm max}/\langle \rho\rangle_{1\frac{5}{2}}=1.65$--1.66 for the light SFHo models and 1.54 for SFHo--20.0, and 1.61 for LS220--20.0. Once more, the average and maximum densities are closely correlated on the few percent level. It would be interesting to investigate such correlations for broader classes of models, but in any case, these regularities suggest that very few basic parameters of a given SN model provide much of the critical information.

The averages listed in Table~\ref{tab:SN-models} were calculated ignoring gravitational redshift effects that are explained in Appendix~\ref{app:lapse}. A SN core with uniform $T$ would still emit radiation that reaches a distant observer with different spectra, depending on the redshift of origin. In our definition of average models, we apply an average redshift factor a posteriori to the spectrum predicted by average SN parameters that are determined by local emission.

We list several other parameters in Table~\ref{tab:SN-models} that are of interest to particle emission. If the process involves nucleons, they suffer final-state Pauli blocking. If one imagines recoil-free scattering, for example of the protons in the Primakoff process, the rate is suppressed by the factor
\begin{equation}\label{eq:Fdegpp}
    F^{\mathrm{deg}}_{pp}=\int \frac{2d^3\bp}{(2\pi)^3}f_p(1-f_p)\bigg/\int \frac{2d^3\bp}{(2\pi)^3}f_p.
\end{equation}
We show averages of this factor for both protons and neutrons. One needs to be careful to use the correct nucleon dispersion relation for the specific SN model as explained in Appendix~\ref{app:distribution}.

\subsection{Average Primakoff spectrum}
\label{sec:Primakoff-Spectrum}

The property of our emission models producing quasi-thermal spectra, with overall rates depending essentially as power laws on local parameters such as $\rho$ and $T$, suggest that, in view of all other uncertainties, one could estimate the time-integrated axion emission from average models. So instead of averaging the emission rate over a certain numerical model, one would compute the emission rate from plausible average properties of a generic model that are much easier to vary parametrically and do not depend too directly on specific, and probably unimportant, fine points of a given simulation.

\begin{figure}
    \includegraphics[width=1.\textwidth]{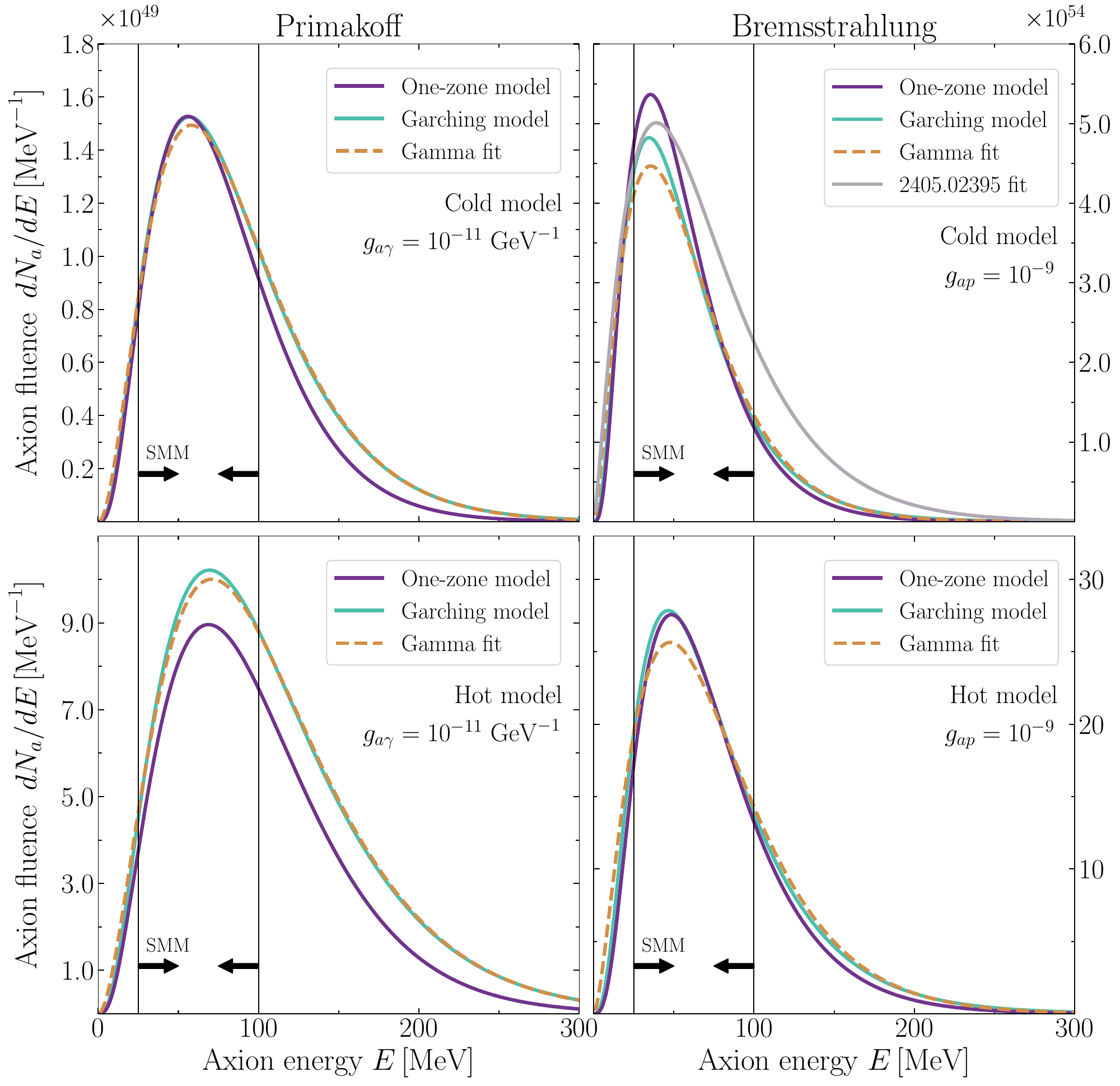}
    \caption{Primakoff and bremsstrahlung flux spectra of several SN models. The spectrum obtained after integrating Eqs.~\eqref{eq:Primakoff-emission-rate} and \eqref{eq:emission_rate_bremsstrahlung} over the cold (SFHo-18.8) and hot (SFHo-20.0) Garching models are shown in green. The best-fit Gamma distribution with the parameters given in Table~\ref{tab:SN-models} are shown in orange. In purple we show the emission considering the one-zone models with the parameters given in Table~\ref{tab:one_zone}. For bremsstrahlung (cold model), the numerical fit given in Ref.~\cite{Lella:2024hfk} is shown in gray. The vertical lines bracket the energy range covered by SMM.}
    \label{fig:Primakoff-Spectrum}
\end{figure}

To illustrate this approach, we consider Primakoff emission from our numerical models and compare it with emission from average SN properties. We compute
the flux spectrum based on Eq.~\eqref{eq:Primakoff-emission-rate} and, after integration over the numerical models, determine the Gamma-fit parameters for Eq.~\eqref{eq:Gamma-Fit}. These are the total number of axions $N_a$, their average energy $\Eav$, and pinching parameter $\alpha$, which we show in Table~\ref{tab:SN-models}.
We have included the gravitational redshift factor (lapse factor) as explained in Appendix~\ref{app:lapse}. Briefly, for an axion seen with energy $E$ by a distant observer, the emisison rate must be locally evaluated at the blue-shifted energy $E_z=(1+z) E$, where $(1+z)$ is the inverse of the lapse factor tabulated in the models. In Fig.~\ref{fig:Primakoff-Spectrum} we show the Primakoff axion spectrum with different assumptions for the cold model as an example. We show the numerical flux spectrum together with the best-fit Gamma distribution. In addition, we show the spectrum that obtains from evaluating Eq.~\eqref{eq:Primakoff-emission-rate} with the average input parameters $\langle T\rangle_{03}$, $\langle Y_p\rangle_{03}$, and $\langle \kS\rangle_{03}$ provided in Table~\ref{tab:SN-models} as well as applying the average lapse factor.

To calculate not only the spectrum from average parameters, but also the total flux, we need the ``mass exposure'' $\langle M t\rangle$, the value for mass$\times$time with which to multiply the particle emission  rate per unit mass. $\langle M t\rangle$ must be approximately a few $M_\odot {\rm s}$. One way of defining it is through
\begin{equation}
    \langle M t\rangle_{mn}=\frac{\int dM dt \rho^m T^n}{\langle\rho\rangle^m_{mn}\langle T\rangle^n_{mn}}.
\end{equation}
We provide the numerical results for the different models and weight functions in Table~\ref{tab:SN-models}. If the true SN core were a homogeneous sphere of mass $M$ with uniform $T$, $\rho$ and would radiate for a time $t$, this prescription would precisely return $M t$.

The agreement between the different curves is surprisingly good, revealing that the spectral characteristics are well described by a Gamma distribution. More surprisingly, the flux spectrum is already very well represented by very few average parameters of the SN core, i.e., the numerical averaging of the emission rate itself does not cause a large modification. For the cold model as a specific example, the total number of emitted axions is found to be $N_a=g_{11}^2 15.7\times 10^{50}$ from integrating over the numerical SN model. Conversely, inserting the average parameters for $T$, $Y_p$, $\kS$, $M t$, and lapse factor in Eq.~\eqref{eq:Na-Gamma-Primakoff} reveals $g_{11}^2 14.1\times 10^{50}$, some 10\% smaller. The numerical average energy is 86~MeV, whereas the one from Eq.~\eqref{eq:Parameters-Gamma-Primakoff}, after including the lapse factor, is 81~MeV. For the pinching parameter, numerically one finds $\alpha=2.1$ whereas the generic value for Primakoff with $\kappa=0.89$ is $\alpha=2.5$. The emission is from regions with different $T$ and one might have expected a significant broadening (a significantly smaller $\alpha$) of the distribution compared to a single-$T$ source, and indeed, the numerical spectrum is slightly less pinched than the one calculated from a single $T$, yet the difference in $\alpha$ is surprisingly small.

This exercise reveals that a useful way to compare SN models of different authors with different physical assumptions and input parameters are average properties such as those listed in Table~\ref{tab:SN-models}. For example, models with a longer cooling period due to the absence of PNS convection, but otherwise comparable, would have a larger $\langle M t\rangle$. In any event, very few average parameters provide sufficient detail on the particle emission in view of the many microphysics uncertainties of the emission process itself.

To compare with the corresponding fluxes from the same SN models found by the Berkeley group~\cite{Manzari:2024jns}, we have extracted the Gamma-fit parameters from their Fig.~S12 of the published version.\footnote{The arXiv posting v1 shows considerably different fluxes.} These authors have included proton degeneracy, but otherwise their emission rate formula is very similar to the one we have used. Therefore, we should compare our $N_a$ values multiplied with our degeneracy factor $F_{pp}$ from Table~\ref{tab:SN-models}. Depending on model, their fluxes are 10--30\% smaller than ours, a difference that we cannot explain, but small enough to make little practical difference. The average energies are very similar and their spectra are slightly more pinched (slightly larger $\alpha$). Concerning degeneracy, we have argued in Appendix~\ref{app:Primakoff} that there are several 10\% level effects going in the opposite direction so that we think the nondegenerate Primakoff rate without any corrections could be a good proxy to the true emission rate on the level of a few ten percent.

\begin{table}[ht]
\centering
\begin{tabular}{lllll}
\hline
\multicolumn{2}{l}{Model}&SFHo--18.8&SFHo--18.6&SFHo--20.0\\
\hline
\multicolumn{5}{l}{Primakoff parameters}\\
\quad $N_a$&$g_{11}^2 10^{50}$ & 10.3 & 22.1 & 76.9\\
\quad $\Eav$&MeV               & 88.3 & 96.3 & 109 \\
\quad $\alpha$&                & 2.36 & 2.47 & 2.57\\
\multicolumn{5}{l}{Bremsstrahlung parameters}\\
\quad $N_a$&$g_{11}^2 10^{50}$ & 11.7 & 21.5 & 56.3\\
\quad $\Eav$&MeV               & 76.5 & 82.5 & 93.8\\
\quad $\alpha$&                & 2.39 & 2.43 & 2.63\\
\hline
\end{tabular}
\caption{Gamma-fit parameters for the Primakoff flux spectra extracted from Fig.~S12 of Ref.~\cite{Manzari:2024jns} as well as the bremsstrahlung ones also extracted from their Fig.~S12. The latter are also proportional to the axion-photon coupling $g_{11}$ because of their assumed connection between this and 
the axion-nucleon couplings.
\label{tab:BerkeleyFits}}
\end{table}

\subsection{Average bremsstrahlung spectrum}
\label{sec:Bremsstrahlung-Spectrum}

We next repeat the same exercise for bremsstrahlung and find similar conclusions, showing in the right panels of  Fig.~\ref{fig:Primakoff-Spectrum} the flux spectra from numerical integrations over the SN models, their Gamma fits, and the spectra provided by our adopted single-zone models with the parameters listed in Table~\ref{tab:one_zone}.

In addition, for the cold model, we compare with the result of the Bari group \cite{Lella:2024hfk} as a gray line, who have provided numerical fit formulas for the bremsstrahlung fluxes for the same cold SN model. 
Specifically, they find $N_a=g_9^2\,46.0\times10^{55}$, $\Eav=69.4$~MeV, and $\alpha=1.34$. The overall agreement is quite good in view of the entirely different microphysical approach, although their spectrum is somewhat harder. In the SMM window, this makes little difference, put provides somewhat more optimistic fluxes for the next galactic SN if detected with Fermi-LAT or a similar instrument. Of course, this higher-energy  bremsstrahlung tail matters only if the putative second emission bump from the pion process is omitted.

We can also compare with the bremsstrahlung fluxes obtained in Ref.~\cite{Manzari:2024jns} that are shown in their Fig.~S12 (dark blue lines) for the different SN models. Once more, we show our Gamma-fit parameters to their curves in Table~\ref{tab:BerkeleyFits} that can be compared to our results in Table~\ref{tab:SN-models}. Their average energies are 20--30\% larger than ours and the pinching parameters are considerably larger, i.e., their spectra are more pinched (narrower) than ours. Somewhat surprisingly, their $\alpha$ values are not only much larger than ours, but also much larger than those of the Bari group, despite the analogous microphysics.

To compare the absolute fluxes (the total number $N_a$ of axions emitted), we first need to determine their assumed axion-nucleon couplings in our language. They express our $g_{a\gamma}$ as $C_{a\gamma\gamma}\alpha/(2\pi f_a)$ and our $g_{ap}$ as $C_{app} m_N/f_a$ and analogous for neutrons, so that
\begin{equation}\label{eq:gap_gagamma_connection}
    g_{ap}=\frac{C_{app}}{C_{a\gamma\gamma}}\,
    \frac{2\pi}{\alpha}\,m_N g_{a\gamma}=0.81\times10^{-12}\,g_{11},
\end{equation}
where the numerical result corresponds to their choice $C_{app}/C_{a\gamma\gamma}=10^{-4}$ and our parameter $g_{11}=g_{a\gamma}/10^{-11}\,{\rm GeV}^{-1}$. They used the same value for neutrons. Therefore, we need to estimate the enhancement factor $\xi_n$ for axion emission for the situation $g_{an}=g_{ap}$ compared with $g_{an}=0$, which is essentially the KSVZ case that we are usually considering. Of course, this factor depends on detailed conditions of chemical abundances and degeneracy. In the context of their SN~1987A cooling bounds, the Bari group \cite{Carenza:2019pxu} has provided the equivalent of our Eq.~\eqref{eq:EffectiveCoupling} which implies $\xi_n=3.51$. 
In other words, to compare with the Berkeley fluxes, we should multiply our brems\-strah\-lung results for $N_a$ with $\xi_n=3.51$ and substitute $g_9\to0.81\times10^{-3}g_{11}$. This exercise yields 7.7, 16., and 58.\ in the $N_a$ units of Table~\ref{tab:BerkeleyFits}, in much better agreement than one could have hoped, given the approximate nature of this comparison and many differences in detail.

\subsection{One-zone models}\label{sec:one_zone_models}

Motivated by the very good agreement between the emissivities obtained from average properties from the numerical post-processed model, we prefer to rely for our results on two simple average one-zone models, that condense these properties in a few parameters. The numerical values for these chosen parameters are summarized in Table~\ref{tab:one_zone}. These numbers are qualitatively representative of a generic cold and hot SN core. We also list the adopted NSM parameters to be discussed below.

\begin{table}[ht]
\centering
\begin{tabular}{llll}
\hline
\textbf{Quantity} & \textbf{Cold} & \textbf{Hot} & \textbf{NSM}\\
\hline
Density $\rho$ [$10^{14}~\mathrm{g/cm^3}$] & 4.0 & 6.0 & 4.0 \\
Temperature $T$ [MeV] & 30 & 45 &25 \\
Proton fraction $Y_p$ & 0.15 & 0.15 &0.07\\
Lapse $(1+z)^{-1}$ & 0.75 & 0.65 &0.85\\
Exposure of mass $M t$ [$M_\odot\,\mathrm{s}$] & 5.0 & 10.0 & $6.0\times10^{-3}$\\
\hline
\end{tabular}
\caption{Physical parameters adopted for our cold and hot average one-zone SN models as well as for our NSM model, which has a much shorter emission period, explaining the small $M t$ value.}\label{tab:one_zone}
\end{table}

We emphasize again the logic of the one-zone models; we are not aiming at \textit{reproducing} the results of specific numerical profiles with them. For that purpose, we could directly use the numerical profiles for our constraints. Rather, Fig.~\ref{fig:Primakoff-Spectrum} shows that the numerical profiles contain a lot of unnecessary information, since the inhomogeneous internal conditions do not change the axion emissivity by more than tens of percent, well below the systematic uncertainties in the microphysical framework used in the simulation itself. We therefore advocate for one-zone models, which distill the minimal essential information into a simple, algebraic framework. This approach not only facilitates validation by others, but also allows for straightforward parametric studies of axion production in compact transients.

\subsection{Average properties of the neutron star merger models}

A similar exercise can now be performed for the internal properties of neutron star mergers (NSMs), where it is even more crucial to define typical parameters to be adopted. Previous treatments of feebly interacting particle emission from NSMs have relied either entirely on numerical models without providing direct intuitive guidance on their internal parameters~\cite{Harris:2020qim, Diamond:2023cto, Dev:2023hax} or conversely on one-zone models with simplified parameters which however were not directly informed by numerical simulations~\cite{Diamond:2021ekg, Fiorillo:2022piv, Manzari:2024jns}. Unifying the two approaches is essential; it provides a tool to easily validate the axion emission from complicated and blackbox numerical simulations, while simultaneously putting to the test the typical values adopted in previous simplified models.

We use the four NSM remnant models considered in Refs.~\cite{Ardevol-Pulpillo:2018btx, George:2020veu, JankaWeb, phdthesisPulpillo, Diamond:2023cto} that were obtained through 3D relativistic particle hydrodynamic simulations (see Refs.~\cite{Ardevol-Pulpillo:2018btx, George:2020veu} for more details). These simulations provide less information than the SN ones; for example, we only have information on the electron abundance $Y_e$, so we use $Y_e=Y_p=1-Y_n$. The gravitational lapse is not provided, but we account for it using the gravitational potential that was determined in the Supplemental Material of Ref.~\cite{Diamond:2023cto}. The models in these simulations use two choices of EoS, SFHo and DD2, which are supposed to bracket the range from very soft to very stiff. Recall that a stiffer EoS leads to colder and less dense stars, implying less particle emission, as noted also in Ref.~\cite{Diamond:2023cto}. 

For both choices of EoS, we consider two options for the merging NS masses: an asymmetric case, with primary and secondary masses of 1.45 and $1.25\,M_\odot$, respectively, and a symmetric case with $1.35\,M_\odot$ each. These examples are motivated by the merger GW170817, whose observations suggest a primary mass of 1.36--$1.89\,M_\odot$ and secondary mass of 1.00--$1.36\ M_\odot$. While this range is fairly large, an additional hint derives from the collapse of the hypermassive remnant (HMNS) being delayed, not prompt, with the remnant likely surviving for a few hundred ms~\cite{Gill:2019bvq, Bauswein:2017vtn}. NSM simulations with high mass ratio typically result in prompt collapse~\cite{Nedora:2020hxc} and are thus disfavored. In addition, consistency with astrophysical expectations suggests a prior on the masses and spins that narrows the masses identified by LIGO~\cite{LIGOScientific:2018cki} to 90\% CL ranges of 1.15--$1.36\ M_\odot$ and 1.36--$1.62\ M_\odot$, respectively.

\begin{table}[t]
\setlength\tabcolsep{0pt}
\begin{tabular*}{\linewidth}{@{\extracolsep{\fill}} llllllllll}
\hline
\multicolumn{2}{l}{Model}&\multicolumn{2}{l}{DD2 Asym.}&\multicolumn{2}{l}{DD2 Sym.}&\multicolumn{2}{l}{SFHo Asym.}&\multicolumn{2}{l}{SFHo Sym.}\\
\hline
$M_{\rm NS}+M_{\rm NS}$ (baryon)&$M_\odot$&\multicolumn{2}{l}{1.25 + 1.45}&\multicolumn{2}{l}{1.35 + 1.35}&\multicolumn{2}{l}{1.25 + 1.45}&\multicolumn{2}{l}{1.35 + 1.35}\\
\multicolumn{2}{l}{Lapse $\langle(1+z)^{-1}\rangle$}&0.85&(0.84)& 0.82 & (0.81) & 0.88 & (0.87) & 0.82 & (0.81)\\
$T_{\rm max}$     &MeV&30.7&& 69.4 &&36.7&&73.4&\\
$\langle T\rangle$&   &19.8&(20.6)& 22.6 & (22.9) & 23.3 & (24.2) & 27.6 & (27.8)\\
$\rho_{\rm max}$&$10^{14}$\,g/cm$^{3}$&5.63&& 6.43 && 6.40 && 9.74 &\\
$\langle \rho\rangle$&& 2.58 & (3.15) & 3.78 & (4.36) & 2.70 & (3.46) & 5.46 & (6.73)\\
$\langle M t\rangle$&$10^{-3}M_\odot{\rm s}$& 7.46& (5.47) & 7.44 & (6.26) & 5.67 & (3.99) & 5.69 & (4.59)\\
\multicolumn{5}{l}{Average abundances per baryon}\\
\quad $\langle Y_e\rangle$&&0.071 & (0.069) & 0.069 & (0.069) & 0.073 & (0.067) & 0.065 & (0.062)\\
\multicolumn{5}{l}{Primakoff emission parameters}\\
\quad $\langle \kS\rangle$&MeV&18.4 & (20.4) & 21.1 & (23.1) & 17.1 & (19.2) & 21.8 & (24.6)\\
\quad $N_a$&$g_{11}^2 10^{47}$ & 3.43 && 4.72 && 5.28 && 7.21\\
\quad $\Eav$&MeV&57.6 && 66.8 && 68.8 && 80.5\\
\quad $\alpha$&&2.05 && 1.40 && 1.86 && 1.47\\
\multicolumn{5}{l}{Bremsstrahlung emission parameters}\\
\quad $\langle\Gamma_\sigma\rangle$&MeV& 21.4 & (26.4) & 33.1 & (38.3) & 24.3 & (31.5) & 52.8 & (64.9)\\
\quad $N_a$&$g_{9}^2 10^{52}$ & 6.10 && 11.2 && 7.39 && 16.3\\
\quad $\Eav$&MeV& 41.8 && 47.4 && 50.8 && 60.0\\
\quad $\alpha$&& 1.62 && 1.51 && 1.54 && 1.62\\
\hline
\end{tabular*}
\caption{Average properties of the Neutron Star Merger models in analogy to Table \ref{tab:SN-models}, also using Garching simulations \cite{Ardevol-Pulpillo:2018btx}. 
The first row shows the masses of the two merging stars. $T_{\rm max}$ and $\rho_{\rm max}$ are the maxima
reached during the available first 10~ms after the merger.}
\label{tab:NSM-models}
\end{table}

Table~\ref{tab:NSM-models} collects the parameters obtained by the two averaging procedures described in the SN context, applied here to the 10-ms time range for which the simulations are available. The typical temperatures and densities within the central HMNS are surprisingly similar to the SN case, and particularly to the cold model. The average temperature ranges from 20 to 30~MeV, and the density is lower than in SNe. In part, this reflects that particle emission favors high-$T$ regions near the surface, where the density is somewhat smaller. Notice that a temperature of 30~MeV is a factor of 3 lower than the speculated values in Ref.~\cite{Manzari:2024jns}, implying a dramatic reduction of axion emissivity. The proton fraction is a factor of~2 smaller than in SNe, a trend to be expected since matter is neutron-dominated even before the merger. The mass exposure $M t$ is about $10^{-3}$ of the SNe, primarily because of the limited time period of up to 10~ms. The remnant can survive for much longer and 
the question of how long the remnant of the historical merger GW170817 actually survived is hotly debated \cite{Gill:2019bvq,Granot:2017tbr,Shibata:2017xdx,Metzger:2018qfl,Murguia-Berthier:2020tfs,Just:2023wtj,Gottlieb:2017pju}, a discussion reviewed in Ref.~\cite{Diamond:2023cto} in the context of feebly interacting particle emission. 

The authors of Ref.~\cite{Manzari:2024jns} used the SFHo-20.0 SN model to estimate the emission from the NSM remnant, so we can compare directly the emitted number of axions that are given in Tables~\ref{tab:SN-models} and~\ref{tab:NSM-models}. Both for Primakoff and bremsstrahlung, Ref.~\cite{Manzari:2024jns} overestimated  the emitted number of axions by at least four orders of magnitude. Of course,  the emissivity in Table~\ref{tab:NSM-models} is integrated only over 10~ms, whereas Ref.~\cite{Manzari:2024jns} assumed a much longer exposure of about 1~s. However, we will see that axion-photon conversion is suppressed after 10~ms by the plasma effect of the ejecta, so this is the limiting time scale, not the HMNS lifetime. Ignoring this effect and scaling our numbers up by two orders of magnitude (1~s vs.\ 10~ms) still leaves a discrepancy of two orders of magnitude relative to Ref.~\cite{Manzari:2024jns}.

Overall, we condense the outcome of our averages in a one-zone model for the NSM core. As with SNe, we use such a generic one-zone model to estimate the emission. The properties of the adopted model are given in Table~\ref{tab:one_zone}.

\begin{figure}
    \centering
    \includegraphics[width=\textwidth]{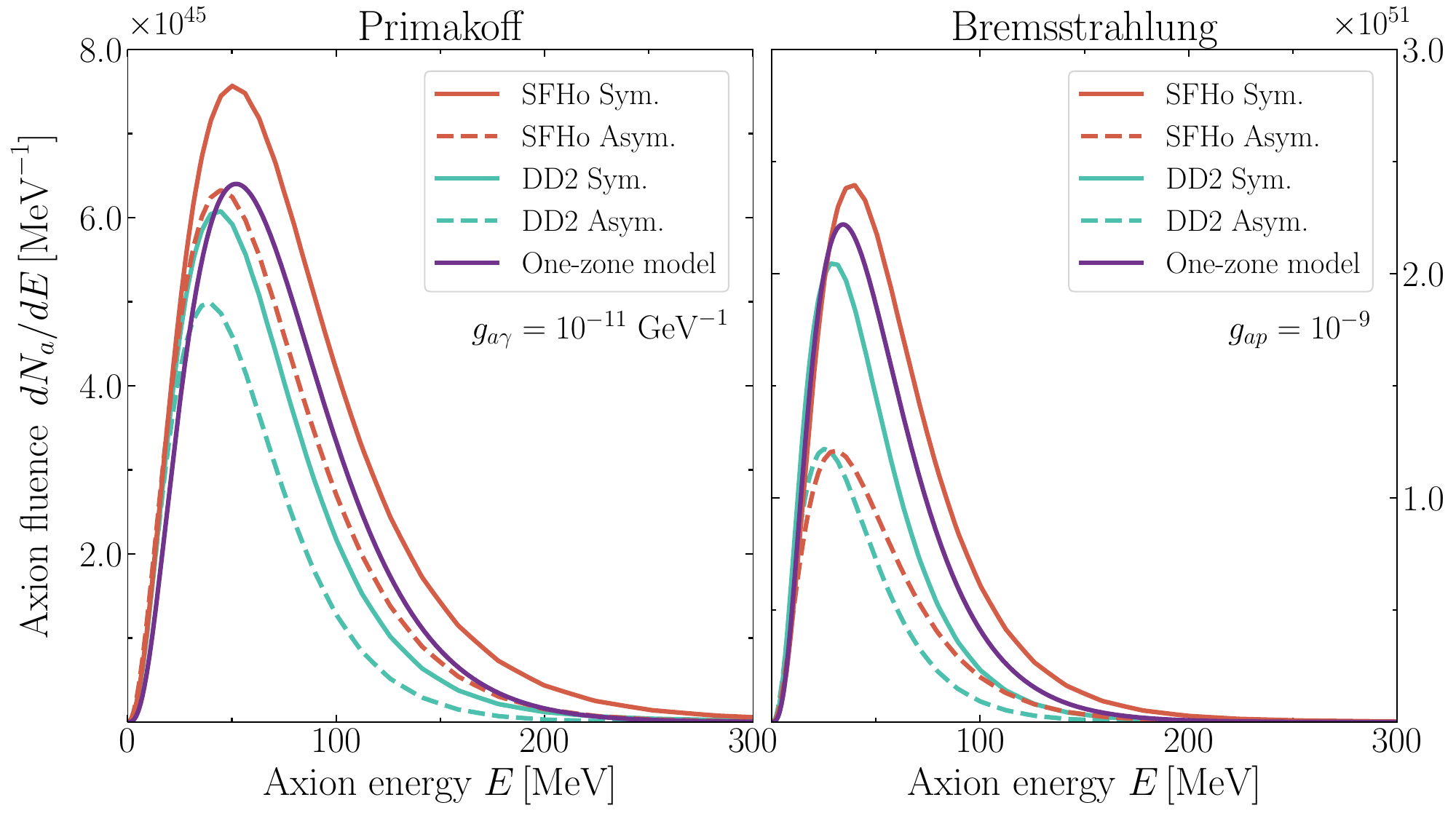}
    \caption{Primakoff and bremsstrahlung flux spectra emitted by the NSM remnant. We show the spectra obtained after integrating Eqs.~\eqref{eq:Primakoff-emission-rate} and  \eqref{eq:emission_rate_bremsstrahlung} over four Garching models listed in Table~\ref{tab:NSM-models}. In purple we show the emission by the one-zone model with the parameters given in Table~\ref{tab:one_zone}. The  time integration period is 10 ms.}
    \label{fig:NSM-flux}
\end{figure}

In Fig.~\ref{fig:NSM-flux}, we show the flux spectra from numerical integrations over the Garching NSM models, together with our one-zone model. The Garching models provide information from 0 to 10 ms, so the axion fluences shown in Fig.~\ref{fig:NSM-flux} have been integrated over this time period, which agrees with the early period before axion-photon conversion is suppressed by the ejecta. Our choice is on the optimistic side of the four numerical models as behooves a study of future detection opportunities.

\section{Conversions in astrophysical magnetic fields}
\label{sec:conversion-in-astro-B-field}

In this section, we examine the general properties of magnetically induced axion-photon conversion in the environments of SN progenitors and the magnetic field of NSMs. In addition, we also discuss the axion-photon conversion in the Galactic magnetic field.

\subsection{Conversions in the progenitor magnetic field}\label{sec:progenitor_conversion}

We first examine axions produced in SN cores, propagating as a short burst through the progenitor magnetic field that survives for several hours after collapse. Within the star, conversion is impeded primarily by photon absorption. For 100~MeV energies, the opacity is dominated by Compton scattering, and the total cross section on electrons at rest is $\sigma_{\rm C}=(\pi\alpha^2/m_e E)\,[1/2+\log(2E/m_e)]\simeq6\pi\alpha^2/m_e E$, where we have assumed $E\simeq 100$--200~MeV, providing an optical depth of $\tau\sim\sigma_{\rm C}n_e R\simeq(6\pi\alpha^2/m_e E)n_e R$. For typical internal densities of the star $\rho\sim 10^{-8}\,\mathrm{g/cm}^3$, leading to $n_e\sim 10^{16}\,\mathrm{cm}^{-3}$, this leads to an optical depth $\tau\sim 100$ at the stellar radius $R\sim 10^{12}\,\mathrm{cm}$. Close to the photospheric surface, the density drops, allowing in principle for axion-photon conversion below the photosphere. However, the corresponding uncertainty on the onset radius of conversion is not large, due to the rapidly dropping density; for a radiative envelope, $\rho\propto (R_*-R)^3$ holds (see, e.g., Ref.~\cite{Matzner:1998mg}), so that at a radius $R\sim R_*/2$ the density has already reached more than $10\%$ of its internal value. An uncertainty by a factor of 2 in the onset radius is in itself comparable to the overall uncertainty on the progenitor radius itself. Moreover, within the photosphere, the field structure would not be dipolar, so there is no reliable way to model it. The order of magnitude of the conversion is captured by the region outside of the photosphere, even if conversion begins at somewhat smaller radii.

In the region outside the photosphere, plasma refraction can be neglected, as follows from a self-consistency argument: the Thomson optical depth for visual photons, which is about $\sim 100 \tau$ due to the lack of Klein-Nishina suppression, must be smaller than 1.  We now relate $\tau$ to the effect of plasma refraction; since $\wP^2 R/E\lesssim1$ with  $\wP^2=4\pi\alpha n_e/m_e$, we have $\wP^2 R/E\simeq 4\tau/6\alpha\simeq 100\,\tau\lesssim1$.  In other words, at the stellar surface, defined by the visual photo sphere, both photon absorption and plasma refraction can be neglected. For stellar $B$-fields below the kG range, magnetic refraction is also completely irrelevant.

Thus, for SN~1987A, we can consider axion conversion to  start at the surface of the star (radius $R=45\pm 15\, R_\odot$) in the dipolar magnetic field.\footnote{Recent binary-merger models that best match observational constraints for the SN light curve and for the progenitor star of SN~1987A possess radii of $\sim$37\,$R_\odot$~\cite{Menon+2019,Utrobin+2021}.} Stellar magnetic fields are usually measured through Zeeman splitting at the photosphere, but for highly evolved blue supergiants (BSGs) such as Sanduleak $-$69$^\circ$202, the progenitor of SN 1987A, no measurements exist. In Ref.~\cite{Manzari:2024jns}, the authors consider as expected values for such stars a range from $100\,\rm G$ to $10\,\rm kG$, based on \cite{Donati_2009}. However, although the aforementioned values are quoted in \cite{Donati_2009}, they are associated to massive main-sequence stars and not highly evolved BSGs. The authors of Ref.~\cite{Orlando:2018lbj} relate the SN~1987A progenitor magnetic field with the nebular one measured through the polarization of radio waves \cite{Zanardo:2018kss} and obtain an estimate at the surface of $B_0\sim 3\,\rm kG$. However, these measurements carry significant uncertainties---the Faraday Rotation Measure (RM) are taken at their upper bounds, and several measurements in the relevant wavelength range are compatible with zero. In addition, relating the nebular field with the progenitor one requires some assumptions on the intermediate evolution. In Ref.~\cite{Orlando:2018lbj}, this is assumed to follow the Parker spiral model in the limit of strong dragging by the stellar wind. 

Viewed from an unbiased perspective, $B$-fields of BSGs at the stage of their collapse are basically unknown. Existing measurements of supergiant surface fields are constrained to main-sequence and early post-main-sequence stars (just because of the longer stellar evolution times during this phase) and are found to be typically on the order of tens of Gauss \cite{Shultz+2014,Grunhut+2017}. One way to estimate the fields for highly evolved massive stars is by projection of the measured fields to late stages under the assumption of magnetic flux conservation~\cite{Petermann+2015}. Because of the growing stellar radii, this leads to lower field strengths, yielding estimated values of only a few G. However, applying such simple scaling arguments in the opposite direction to extrapolate from measured RSG surface fields on the order of 1\,G \cite{Tessore+2017,Plachinda+2022} to the more compact BSGs, whose radii are only several 10 $R_\odot$ instead of hundreds to thousands of solar radii, one gets $B_0(\mathrm{BSG}) \sim (1000\,R_\odot/(30...40)\,R_\odot)^2\times 1\,\mathrm{G} \sim 1000\,\mathrm{G}$. Such arguments, however, ignore possible field-amplifying processes during the post-main-sequence evolution in the form of dynamos and binary mergers events~\cite{Schneider+2019}. The latter are particularly relevant for SN~1987A, because its BSG progenitor had probably experienced a late-time binary merger some 20,000 to 30,000 years before its collapse and explosion \cite{Podsiadlowski+2007,Morris+2009,Menon+2017}. Such an event could have caused a substantial growth of the $B$-field in the hydrogen-helium envelope of the progenitor. For all these reasons BSG surface fields on the order of 1000\,G or even higher cannot be excluded, although $B_0\sim 100$\,G may be a safer, more conservative assumption. Overall, such estimates, while within the range of possible values for BSGs, cannot be trusted without actual measurements in order to derive robust axion constraints.

Rather than in BSGs, very strong magnetic fields might be more common in Type~Ibc SNe, which possess stripped-envelope progenitors and are therefore a much more interesting target \cite{Candon:2025sdm} (see Note Added). However, we will follow Ref.~\cite{Manzari:2024jns} in their assumptions of BSG magnetic fields to execute the comparison with their work as directly as possible, and also because those $B$-fields have the required strengths to probe QCD-axion models. In this context, we will consider $B_0\sim 100\,\rm G$ as a reasonably conservative assumption, but stress that this value is not based on a direct measurement for Sanduleak $-$69$^\circ$202. Regarding the assumed dipole structure, Ref.~\cite{Manzari:2024jns} also mentions the possibility of a Parker spiral structure, where the transverse component would be mostly toroidal, driven by field dragging by the rotating plasma ejected from the star. The $B_\phi\propto R^{-1}$ variation might look more optimistic for axion-photon conversion. However, the amplitude of the toroidal component strongly depends on the angular velocity of the rotating star, and most importantly, the spiral structure cannot form immediately outside the photosphere. Rather, the distance must be so large that the wind velocity has dropped below the Alfv\`en velocity, and thus efficient at field dragging. For a typical mass loss rate $\dot{M}\sim 6\times 10^{-6}\; M_\odot/$yr, surface field $B_0\sim 100\,\rm G$, and wind velocity $v_w\sim 550\,\rm km/s$~\cite{1991ApJ...380..575L}, this radius is on the order of $100\;R_\odot$, significantly larger than the progenitor radius. Since conversion occurs mostly just outside the stellar surface, the less rapidly decreasing field strength is presumably inessential compared to the dipole structure.

\begin{figure*}
\includegraphics[width=\textwidth]{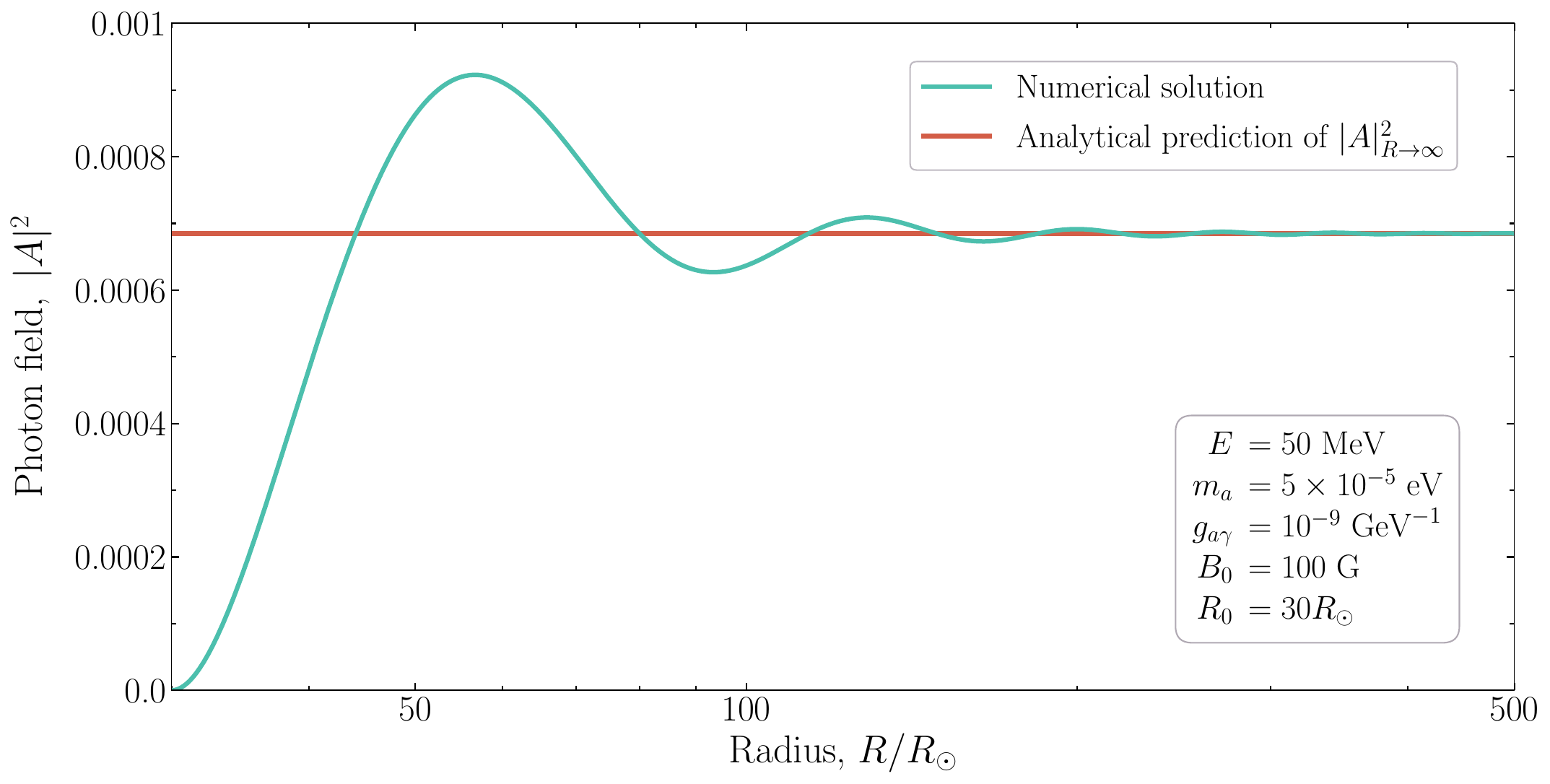}
\caption{Numerical solution of the axion-photon mixing system in the SN case, assuming $a(0)=1$ and $A(0)=0$, and for the reference values of axion energy $E$, mass $m_a$, axion-photon coupling $g_{a\gamma}$, surface magnetic field $B_0$, and progenitor radius $R_0$ as shown in the legend. The contributions from magnetic refraction and plasma density are negligible. The asymptotic value of $|A|^2$ matches with its analytical prediction found in Eq.~\eqref{eq: SNProbPrediction}.}
\label{fig:sn_conversion}
\end{figure*}

Motivated by this discussion, we set SN~1987A constraints using a dipole field with strength $B_0$ at the stellar surface at radius $R_0$. We consider the range $B_0=30$--$100\,\rm G$ with fixed $R_0=30\;R_\odot$ (the lower end of the $1\sigma$ interval). For such small fields, magnetic refraction is inessential, and so we can use the algebraic probabilities derived in Sec.~\ref{sec:axion-photon} for $R_i=R_0\gg R_{\rm conv}$. Thus, in the massless limit, Eq.~\eqref{eq:conv_prob_low_magnetic} implies \begin{equation}\label{eq:sn_probability_massless}
    P_{a\gamma}\simeq 2.6\times 10^{-7}\; g_{11}^2 B_{100}^2 R_{30}^2,
\end{equation}
with $B_{100}=B_0/100\,\rm G$ and $R_{30}=R_0/30\;R_\odot$. Instead, according to Eq.~\eqref{eq:threshold_mass_low_magnetic}, the axion mass becomes relevant for
\begin{equation}\label{eq:maximum_mass_supernovae}
   m_a\gtrsim 62\;\mu\mathrm{eV}\;(E_{100}/R_{30})^{1/2}, 
\end{equation}
and according to Eq.~\eqref{eq:conversion_large_radii_massive} the conversion probability is suppressed to 
\begin{equation}\label{eq:sn_probability_massive}
    P_{a\gamma}\simeq 3.8\times 10^{-8}\; B_{100}^2 E_{100}^2 g_{11}^2/m_{-4}^4.
\end{equation}
It is convenient here to introduce a new reference scale for the mass of $m_{-4}=m_a/10^{-4}\;\mathrm{eV}$. These values are confirmed by our numerical solution of Eq.~\eqref{eq:schrodinger}, which we show in Fig.~\ref{fig:sn_conversion} for the massless case.

\subsection{Conversions in the merger magnetic field}\label{sec:nsm_conversions}

For axions produced in the merger, conversions occur already close to its surface, due to the absence of a stellar mantle surrounding the hot core. This much more compact region implies much larger fields, as expected from magnetic flux conservation alone, perhaps reaching up to $10^{16}\, \mathrm{G}$. These fields would be highly turbulent, but at slightly larger distances, the dipole structure is likely established in the absence of plasma. Following Ref.~\cite{Manzari:2024jns}, we then use a characteristic value of $B_0\sim 10^{14}\,\mathrm{G}$ at the inner radius.

In these strong fields, photons are strongly refracted, conversions are suppressed, and massless axions start to efficiently convert only at $R_{\rm conv}$ defined by Eq.~\eqref{eq:conv_radius}, which is $R_{\rm conv}\simeq 1.4\times 10^{4}\,\mathrm{km}\; B_{14}^{2/5} E_{100}^{1/5} R_{0,6}^{6/5}$, now using $B_{14}=B_0/10^{14}\,\rm G$. The conversion probability for massless axions is then estimated from Eq.~\eqref{eq:conv_prob_massless_strong_magnetic} as 
\begin{equation}\label{eq:nsm_probability_massless}
P_{a\gamma}\simeq 7.2\times 10^{-8}\; B_{14}^{2/5} g_{11}^2 R_{0,6}^{6/5} E_{100}^{-4/5}.
\end{equation}
However, this estimate is too optimistic because it ignores plasma suppression. It is true that the matter surrounding the merger is probably dilute enough that plasma refraction is small---compare with our earlier observation that, in the Klein-Nishina regime, if plasma refraction was relevant, it would imply a Compton optical depth of 0.01. However, the merger itself ejects a sizable amount of material, thus polluting the environment.

To estimate this effect, we observe that the ejected mass can exceed even $M_{\rm ej}\sim 10^{-3}\;M_\odot$ for a merger that is not very symmetric \cite{Bauswein:2013yna, Shibata:2019wef}. Typical velocities are around $V\sim 0.25\,c$, but a small fraction can be much faster with $V\sim 0.8\,c$ \cite{Metzger:2014yda}. The density at the conversion radius is estimated to be $\rho\sim M_{\rm ej}/(4\pi R_{\rm conv}^2 V \delta t)$, where $\delta t\sim 100\,\rm ms$ is a reasonable timescale for the bulk of axion emission (the precise value will be irrelevant), so $\rho\sim 100\,\rm g/cm^3$. For a photon with $E\sim 100\,\rm MeV$, the optical depth for Compton scattering is $\tau_{\rm C}\sim (\rho/m_N) \sigma_{\rm T} R_{\rm conv} m_e/E$, where $m_N$ and $m_e$ are the nucleon and electron masses, and the factor $m_e/E$ accounts for the Klein-Nishina suppression of the Thomson cross section, so one easily finds $\tau_{\rm C}\sim 10^8$. Even if the fraction of fast ejecta ($V\gtrsim 0.8\,c$) was suppressed by 6--7 orders of magnitude (actually much larger ejecta masses are found in simulations), they would still absorb the photons, for which $\rho\sim 10^{-6}\,\rm g/cm^3$ is enough. Even smaller densities interrupt axion-photon conversion, provided that $|\Delta_{\rm P}(R_{\rm conv})| R_{\rm conv}\gg 1$; for $R_{\rm conv}\sim 10^4\,\rm km$, this occurs already at $\rho \sim 10^{-9}\,\rm g/cm^3$. So, even as little ejected mass as $\sim 10^{-14}\, M_{\odot}$ expelled over $100\,\rm ms$ through $R_{\rm conv}\sim 10^4\,\rm km$ can shut down conversion. Usually, the mass ejected with relativistic velocities is much larger, at least some $10^{-7}\, M_\odot$. In our estimates of refraction, we have assumed nonrelativistic matter at rest, whereas the temperature of the ejecta can rival $m_e$, introducing corrections of order unity to the plasma frequency. In addition, $Y_e$ of the neutron-dominated ejecta can be as low as 10\%, proportionally reducing $|\Delta_{\rm P}|$, while the ALP energy in the ejecta rest frame is reduced by $\sqrt{(1-V)/(1+V)}\sim 0.3$, correspondingly increasing $|\Delta_{\rm P}|$. These numerical factors do not change the conclusion that fast ejecta can quench conversions altogether.

Within the ejecta, refraction is largely determined by the plasma, while outside, the magnetic field dominates, suggesting the intriguing possibility of enhanced conversion at the resonance point where the two effects cancel. However, in practice no robust modeling is possible, since the transition from ejecta to vacuum happens in a region of small ejecta mass which is challenging to resolve numerically. In addition, one cannot expect the conversion to be near-adiabatic, since the transition to the circumstellar medium must be discontinuous, with a sharp drop in density over the dissipative length scale determined by viscosity, which is much shorter than the hydrodynamical length scales. Therefore, we do not consider this possibility, and simply model the ejecta as having a sharply discontinuous surface.

We briefly comment on the relation of this argument with the constraints that some of us have derived~\cite{Diamond:2023cto} on massive ALPs decaying into photons around NSMs, in particular GW170817, forming a fireball. The nonobservation of X-rays was used to constrain $g_{a\gamma}$. While the ejecta can also absorb the photons, affecting a portion of the excluded parameter space, these constraints are less sensitive to a fast ejecta component because decays are not impeded by matter. However, if the absorbed energy $\mathcal{E}_\gamma$ exceeds the mass of the fast component of the ejecta $M_{\rm ej}$, the fireball will survive, simply carrying a baryon loading from the matter enraptured in its expansion. Using Eq.~(5) of Ref.~\cite{Diamond:2023cto}, $g_{a\gamma}=10^{-10}\,\rm GeV^{-1}$ and $m_a=200\,\rm MeV$ implies $\mathcal{E}_\gamma=9\times 10^{46}\,\rm erg$, so that $M_{\rm ej}\lesssim 5\times 10^{-8}\, M_\odot$ could not disrupt the constraints. Therefore, a small fraction of fast ejecta would not significantly impact the constraints. In contrast, even a minimal amount of matter inhibits axion-photon conversion, provided that $|\Delta_{\rm P}(R_{\rm conv})|R_{\rm conv}\gg 1$, as argued above.

Thus, we can safely assume that, as soon as the ejecta reach $R_{\rm conv}$, the conversion abruptly ends. Since axions move essentially with the speed of light, while the ejecta with $V\simeq 0.8$, means that only axions emitted at the source within a time span of
\begin{equation}
    \Delta t\simeq \frac{R_{\rm conv}}{V}-R_{\rm conv}\sim 12\; \mathrm{ms}\; B_{14}^{2/5} R_{0,6}^{6/5} E_{100}^{1/5}
\end{equation}
convert efficiently. Massless axions emitted at a later time $t_{\rm em}$ convert just outside the ejecta, at $R\sim V t_{\rm em}/(1-V)$; since $\Delta_{a\gamma}\propto R^{-3}$, the probability reduces for later emission times as $P_{a\gamma}\propto t_{\rm em}^{-4}$, making their overall contribution negligible. Thus, essentially only the axions emitted within the first 10~ms contribute to the observable gamma-ray flux.

\begin{figure}[b]
    \centering
    \includegraphics[width=1.\linewidth]{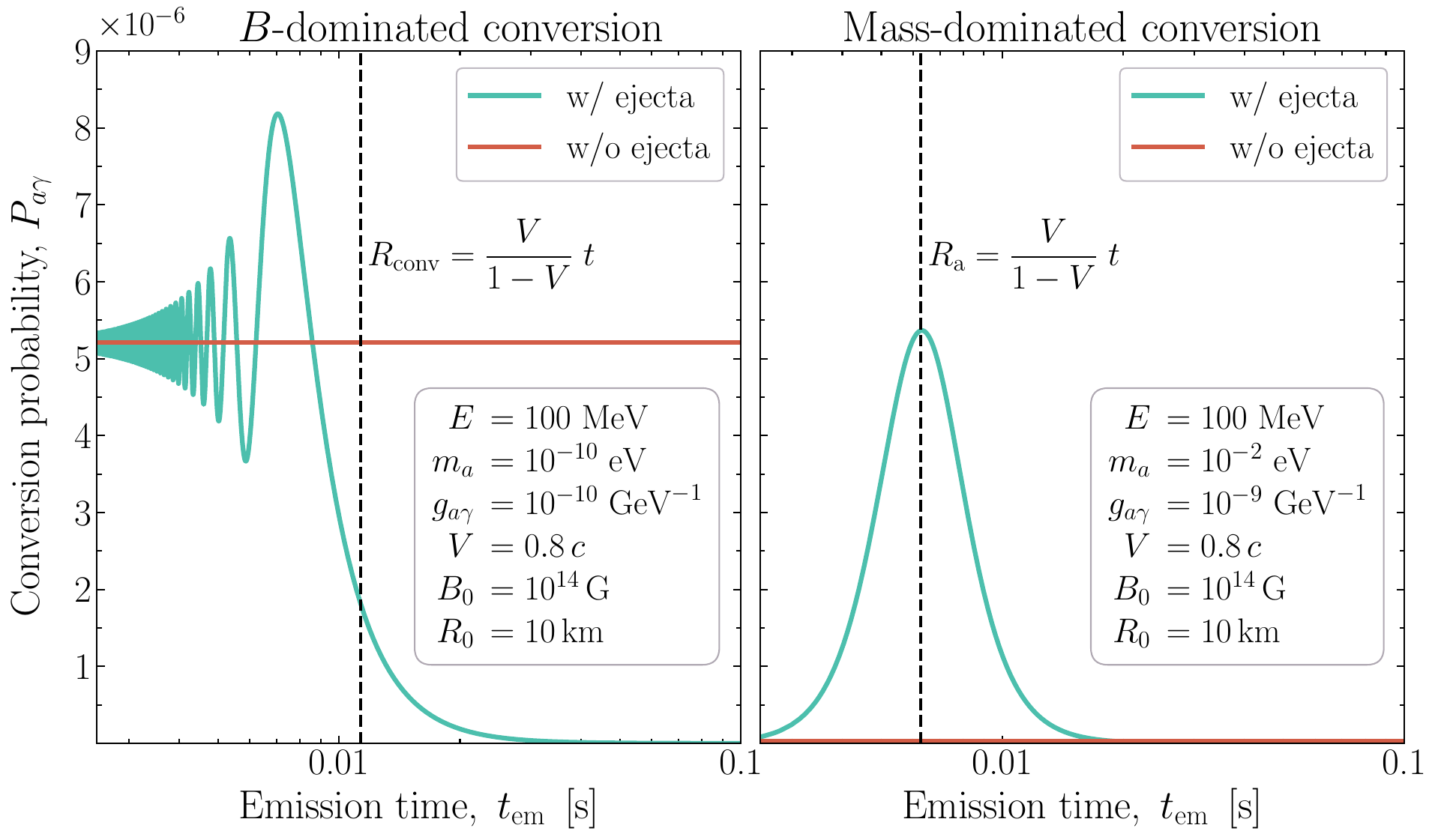}
    \caption{Axion-photon conversion probability as a function of emission time for two scenarios. \textit{Left:} quasi-massless axion ($m_a=10^{-10}$~eV), where conversion happens in the region dominated by magnetic-induced refraction, and \textit{Right:} a massive case ($m_a=10^{-2}$ eV), where instead the mass term $\Delta_a$ dominates. The case including ejecta (blue) is compared with their absence (red), using the same choices of axion energy $E$, axion-photon coupling $g_{a\gamma}$, ejecta velocity $V$, magnetic field at the HMNS surface $B_0$, and radius $R_0$. The black dashed line marks the instant when the ejecta reach the radius where most of the conversion takes place.
    }
    \label{fig: NSMConversion}
\end{figure}

For massive axions, as discussed in Section~\ref{sec:mass_effects_pedestrians}, conversions at $R_{\rm conv}$ are suppressed by mass refraction when
\begin{equation}\label{eq:maximum_mass_nsm}
    m_a\gtrsim 1.7\;\mathrm{meV}\; E_{100}^{2/5} B_{14}^{-1/5} R_{0,6}^{-3/5}.
\end{equation} 
Now $R_a\simeq 1.6 \times 10^{4}\;\mathrm{km}\; B_{14}^{1/3} E_{100}^{1/3} m_{-3}^{-1/3} R_{0,6}$, with $m_a=m_{-3}\,\rm meV$, is the characteristic scale at which conversion is most efficient. For $R_i\ll R_a$, the conversions are actually exponentially suppressed by the interference between mass-induced and magnetic-induced refraction. Axions emitted at $t_{\rm em}\gg R_a (V^{-1}-1)$ are still converted less efficiently than in the absence of ejecta, as in the massless case. However, for $t_{\rm em}\lesssim R_a (V^{-1}-1)$, the conversion is actually \textit{more} efficient than for emission from the center, as seen from Eq.~\eqref{eq:P_asym_2} which shows the probability increasing with the emission radius $R_i$. We will indeed see that this effect impacts the constraints.

Overall, we model the axion-photon conversion by assuming a dipole field with a strength $B_0$ at the surface radius $R_0=10^6\,\rm cm$, and solve the mixing equation Eq.~\eqref{eq:schrodinger} accounting for the magnetic field refraction effect. The line of sight is chosen such that the $B$-field is completely transverse to the axion direction. For the impact of the ejecta, following our earlier arguments, we model them as a hard surface, such that the starting point for conversions of an axion emitted at a time $t_{\rm em}$ is at a radius $V t_{\rm em}/(1-V)$.

Figure~\ref{fig: NSMConversion} shows the conversion probability for ALPs emitted at $t_{\rm em}$. In the absence of ejecta, this probability is constant, since axions always convert primarily at the conversion radius $R_{\rm conv}$, which is independent of time. However, when the ejecta are included, the conversion probability depends on time, since axions begin their conversion at a radius which changes with $t_{\rm em}$. In particular, for ALPs emitted after about 10~ms, the conversion probability drops rapidly, as discussed above,

For massive ALPs, the conversion probability still drops rapidly when conversion begins outside of the critical radius $R_a$, but the peak probability is much larger compared to the case without ejecta, due to the exponential suppression of the conversion for ALPs emitted from the center (Eq.~\eqref{eq:P_asym_1}). Therefore, the ejecta here increase, rather than decrease, the produced gamma-ray fluence.

\subsection{Conversions in the Galactic magnetic field}

For small masses, conversion is dominated by the Galactic magnetic field. We use the regular field model of Unger and Farrar~\cite{Unger:2023lob}, and in particular the ``\texttt{base}'' model. We do not test different choices among the suite identified in Ref.~\cite{Unger:2023lob}, since these are expected to yield comparable results, an expectation validated by Fig.~S14 of Ref.~\cite{Manzari:2024jns}.

\begin{figure}
    \centering
    \includegraphics[width=\textwidth]{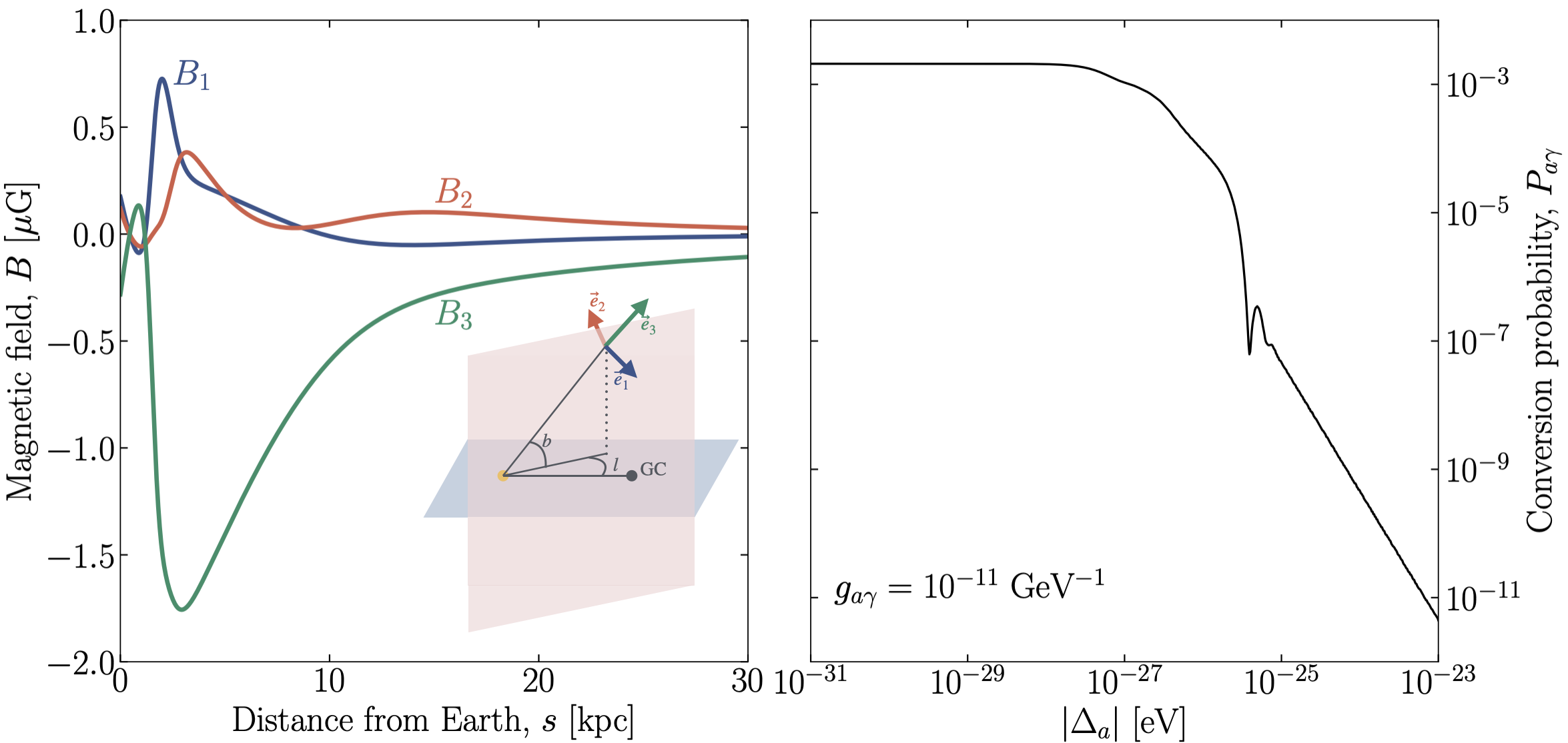}
    \caption{ \textit{Left:} Galactic magnetic field along the line of sight to SN~1987A, which has Galactic coordinates $b_{\rm SN}=-31.8^\circ$ and $l_{\rm SN}=279.7^\circ$. The system of axes defining the components is shown as an inset, with GC the Galactic Center, and the orange point from which coordinates are measured is the Solar System. \textit{Right:} Conversion probability as a function of $|\Delta_a|=m_a^2/2E$ and fixed coupling $g_{a\gamma}=10^{-11}\,\rm GeV^{-1}$.}
    \label{fig:galactic}
\end{figure}

We show in Fig.~\ref{fig:galactic} the components of the Galactic magnetic field along the line of sight connecting Earth with SN~1987A. $B_3$ is the component along the line of sight, which however does not cause axion-photon conversion, whereas the other two components are in the orthogonal plane. We define $B_1$ as the component in the plane orthogonal to the Galactic plane passing through the Solar System and SN~1987A (see the sketch in Fig.~\ref{fig:galactic}). For this field profile, we obtain the conversion probability by solving Eq.~\eqref{eq:schrodinger} along the line of sight. Since we are safely in the perturbative regime, in which the mixing $\Delta_{a\gamma}$ is a small parameter, the conversion probability increases in proportion to $g_{a\gamma}^2$. The corresponding conversion probability for $g_{a\gamma}=10^{-11}\,\rm GeV^{-1}$ is shown in the right panel of Fig.~\ref{fig:galactic}, as a function of the mass-induced refraction term.

For comparison with previous studies of axion-photon conversion in the Galactic magnetic field, we mention explicitly our value in the massless limit of
\begin{equation}\label{eq:GalacticConversionProbability}
    P_{a\gamma}=g_{11}^2\,2.1\times 10^{-3}.
\end{equation}
For the same UF base model, the Berkeley group~\cite{Manzari:2024jns} found $P_{a\gamma}=g_{11}^2\,1.4\times 10^{-3}$ (private communication). From Fig.~S14 of Ref.~\cite{Manzari:2024jns}, we infer that for the suite of UF models \cite{Unger:2023lob}, their $g_{a\gamma}$ constraints cover an approximate range of $7/6\simeq1.17$ (the width of the blue band at small mass). As the $g_{a\gamma}$ bound scales as $P_{a\gamma}^{1/4}$, the spread in conversion rate is approximately $(7/6)^4\simeq2$. The limit from the older JF model \cite{Jansson:2012pc}, shown as a green line in that figure, is shifted roughly by a factor 4/3 upwards relative to the average blue band, implying that $P_{a\gamma}$ is reduced by roughly 1/3 relative to UF. An explicit value for the JF  conversion rate was given in Ref.~\cite{Payez:2014xsa} at the end of their Sec.~3 as $P_{a\gamma}=g_{11}^2\,9\times10^{-4}$, a factor 2.3 smaller than Eq.~\eqref{eq:GalacticConversionProbability}. However, this conversion rate was critiqued in Ref.~\cite{Hoof:2022xbe} as being a factor of~1.6 too large for the same $B$-field model. An even larger value of $P_{a\gamma}=g_{11}^2 1.5\times 10^{-3}$ was reported in Ref.~\cite{Calore:2020tjw}, based on the  updated JF 12c model~\cite{Planck:2016gdp}.

Overall, while the final $g_{a\gamma}$ limit is not strongly affected by the exact $B$-field model, $P_{a\gamma}$ itself shows a significant spread between different cases. Also the variation with $m_a$ depends significantly on the model. Specifically, the relative $P_{a\gamma}$ between small and large $m_a$ depends on the model as highlighted by the green curve and blue band in Fig.~S14 of Ref.~\cite{Manzari:2024jns} crossing as a function of $m_a$. The JF model shows a much larger $P_{a\gamma}$ than UF for large $m_a$, the opposite as for small $m_a$.

For the next galactic SN or a future NSM, we do not include the effect of the Galactic $B$-field, that would depend on the source location in the sky. Moreover, the Galactic field is relevant for very small masses, far away from the QCD axion band, which is of primary interest for a detection perspective. For SN~1987A, with a known source location, and the purpose of deriving constraints, the Galactic field extends the explored range.

\section{Gamma-ray satellites}\label{sec:satellites}

In this section we describe the gamma-ray satellites used in our analysis and the adopted effective areas and flux limits.

\subsection{Solar Maximum Mission (SMM)}

The conversion of axions into photons should have led to a gamma-ray flux from SN~1987A in the tens to hundreds of MeV range. The primary active experiment at the time was the Solar Maximum Mission (SMM) satellite, launched by NASA in 1980. It was primarily designed to study solar flares and their high-energy emissions. However, it also included the Gamma-Ray Spectrometer (GRS), which allowed it to detect and analyze high-energy gamma-rays in the range from a few to about a hundred MeV. No excess above background was observed in association with the neutrino burst, constraining neutrino radiative decays, as reported in two early papers that we will refer to as GRS-89 \cite{Chupp:1989kx} and GRS-93
\cite{Oberauer:1993yr}. The shown information and data were later used by many authors to set limits on various feebly interacting particles that could produce gamma-rays by decay or conversion. We explicitly cite those papers that, like our work, consider magnetically induced ALP-photon conversion \cite{Brockway:1996yr, Grifols:1996id, Payez:2014xsa, Hoof:2022xbe, Manzari:2024jns}. 

The GRS data were reported in the energy intervals 4.1--6.4, 10--25, and 25--100~MeV, of which only the last one is relevant for our study. GRS-89 reported the counts for 10.24~s bins around the time of neutrino detection, whereas GRS-93 shownd 2.048~s bins, the smallest ones available in the GRS data. Concerning the 25--100~MeV channel, another difference is that GRS-89 only included counts from the CsI detectors, whereas GRS-93 showed the sum of the CsI and NaI counters. Correspondingly, they reported different effective areas of 63 vs.\ $90~{\rm cm}^2$, meaning that in GRS-93, 70\% of the counts come from CsI, another 30\% from NaI. Moreover, the two papers reported average background rates of 6.3~Hz vs.\ 8.8~Hz, reflecting the same ratio as that of the effective areas. In the lower-energy channels, both papers show the same data, except for the different bin widths, and report the same effective areas in those channels. The difference of the reported data in the 25--100~MeV channel has caused considerable confusion in the recent literature on which we comment in Appendix~\ref{app:SMM}.

No excess counts were observed around the time of the neutrino burst,
tagged by the arrival time $t_\nu$ of the first neutrino in the IMB detector, which was the only detector with a reliable clock. Axion emission, in our models, lasts at most for some 3--5~s. Therefore, based on GRS-89, only the first 10.24~s bin after $t_\nu$ matters, for which we extract a number of 61 measured counts in the 25--100~MeV channel, to be compared with an average background of 65 counts per bin. As shown in Appendix~\ref{app:SMM}, one then finds a $3\sigma$ upper limit of
\begin{equation}\label{eq:SMM-fluence-limit}
    \Phi_{\gamma}^{\rm SMM}<0.35~{\rm cm}^{-2},
\end{equation}
to be compared with the corresponding limit of $0.6~{\rm cm}^{-2}$ stated in GRS-89, who used a different formula for the relation between background and $3\sigma$ limit.

For the more finely binned GRS-93 data, only the first three 2.048~s bins after $t_\nu$ are important, for which we extract a total of 60 counts, to be compared with the average expectation for three bins of 54 counts. Fortuitously, one finds the same bound as Eq.~\eqref{eq:SMM-fluence-limit} to be derived in Appendix~\ref{app:SMM}. As the observed data are close to the average background, without any particular upward or downward fluctuation, the limits are very similar to what one would have expected in view of the background rates.

We have also explicitly tested that binning the data in energy and time provides negligible changes in the constraints obtained from the integrated fluence. The energy binning is of course irrelevant; the bin 25--100~MeV contains the bulk of the signal, so all the others are irrelevant in practice. Time binning is similarly not very relevant; using the total counts in the three time bins around $t_\nu$ as we do provides a constraint on the fluence that differs by less than $10\%$ from doing a binned analysis on these bins. Overall, the precise alignment of $t_\nu$ with the GRS-93 data cannot be determined accurately, certainly not to better than hundreds of ms, since the first neutrino event does not necessarily coincide with the collapse. Moreover, the time structure of the gamma-ray signal depends sensitively on the equation of state and PNS mass of the simulation (see, e.g., Fig.~15 of Ref.~\cite{Fiorillo:2023frv}). Therefore, a time-binned analysis would not provide any reliable benefit to the constraints obtained here.

\subsection{Fermi-LAT and future satellites}
\label{eq:FermiLat}

\begin{figure}
    \centering
    \includegraphics[width=\textwidth]{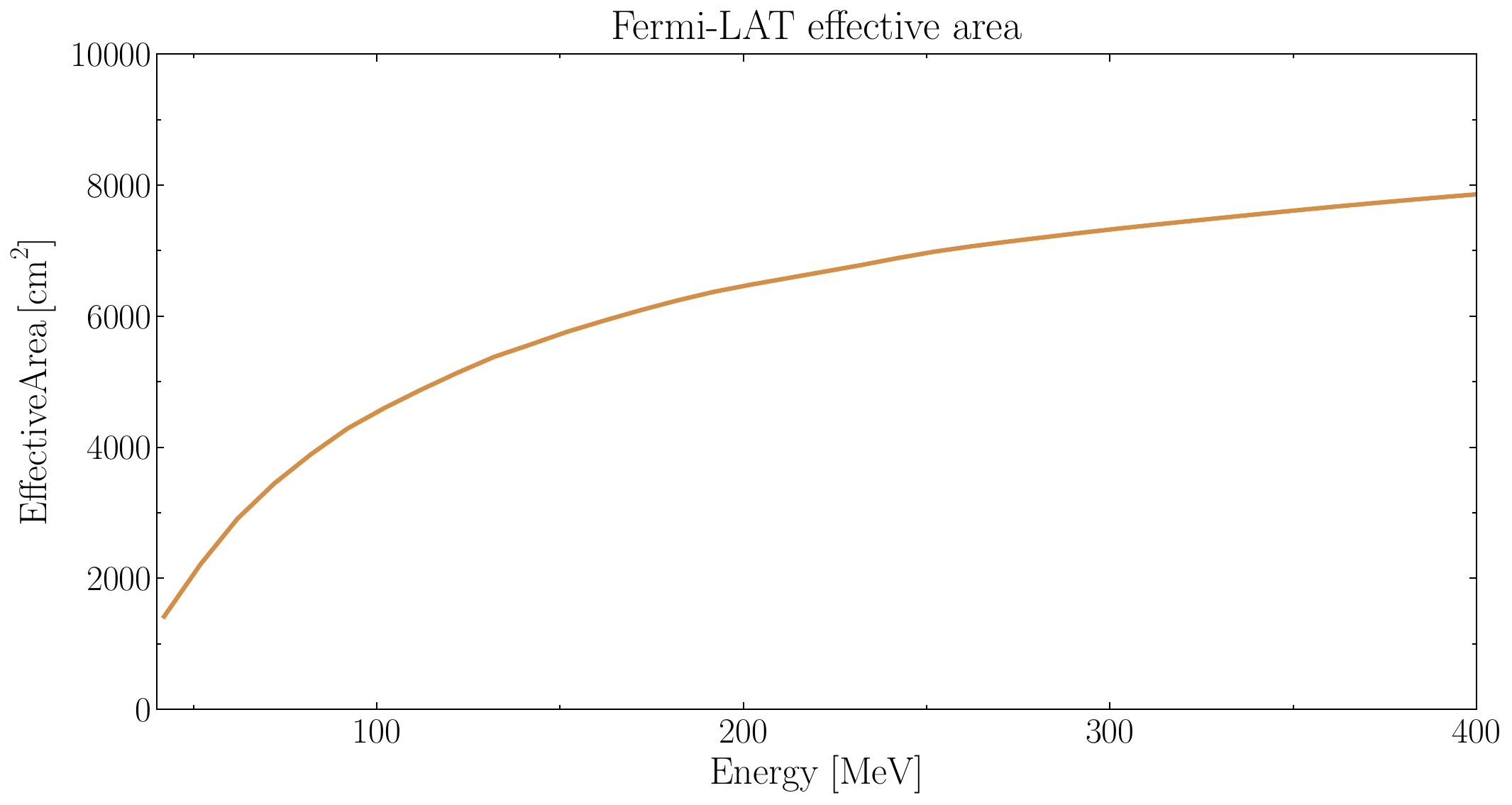}
    \caption{Effective area of the Fermi-LAT satellite for the transient analysis class (dotted curve in Fig.~14 of Ref.~\cite{Fermi-LAT:2009ihh}).}
    \label{fig:AeffFermiLAT}
\end{figure}

The Fermi Large Area Telescope (Fermi-LAT) is a high-energy gamma-ray observatory designed for the 20~MeV to 300~GeV range with high sensitivity, large effective area, and excellent background rejection. At 100 MeV, Fermi-LAT has an effective area of up to $2000\,\mathrm{cm}^2$, a dramatic improvement over SMM, where GRS had an effective area of only a few $\mathrm{cm}^2$ at these energies. In addition, Fermi-LAT provides superior energy resolution (10–15\%), angular resolution ($5^\circ$ at 100 MeV, improving at higher energies), and timing precision, along with continuous all-sky monitoring due to its wide field of view (2.4 sr). Its onboard tracker and calorimeter allow for precise reconstruction of photon direction and energy, while rejecting charged-particle backgrounds.  

In our investigation, we model Fermi-LAT through its effective area extracted from Ref.~\cite{Fermi-LAT:2009ihh} and shown in Fig.~\ref{fig:AeffFermiLAT}. Given the large uncertainties on the axion emission itself, and the unpredictability of which instrument will dominate gamma-ray observations at the time of the next Galactic SN, a more refined modeling of detector response would be moot. In other words, our forecasts assume that the observation is made with an instrument having the response shown in Fig.~\ref{fig:AeffFermiLAT}, which defines our ``Fermi-LAT-like'' instrument.

We rely on a background-free method, since within a few seconds and within the angular resolution of the instrument, far less than one background event is expected~\cite{Lecce:2025dbz}, allowing us to neglect background altogether. As a threshold for detection, we may use the number of expected signal events leading to at least one detected event with more than $95\%$ probability; for Poisson statistics this is $s=3$.

We mention in passing a difference to the procedure used in the recent Ref.~\cite{Lecce:2025dbz} for the NSM detection forecast. Their criterion was that the signal should exceed background, which for Fermi-LAT within the appropriate observation time and angular window is about 0.08 events, but still effectively no signal. In a signal-limited, not background-limited, observation, one needs a minimal number of events, not a minimal signal-to-noise ratio.

\section{Results}
\label{sec:results}

In this section, we assemble all the previous elements, and derive our SN~1987A constraints and the sensitivities for a future SN or NSM. We discuss separately ALPs, in our definition with a coupling to photons alone, and the more general axion case, inspired by the KSVZ model, where we consider the coupling to protons for bremsstrahlung emission and the one to photons for conversion. For the sources, we use the representative one-zone models described in Table~\ref{tab:one_zone}, allowing us to derive the scaling of our results with assumed input parameters. While we use the full numerical computation of the conversion probability, the scaling of the detection rates for massless and massive axions can be analytically confirmed with the formulae provided in Sec.~\ref{sec:axion-photon}.

\subsection[SN~1987A constraint on \texorpdfstring{$g_{a\gamma}$}{}]{SN~1987A constraints on \texorpdfstring{\boldmath$g_{a\gamma}$}{}}

\label{sec:gag-limits}

\begin{figure}
\includegraphics[width=\textwidth]{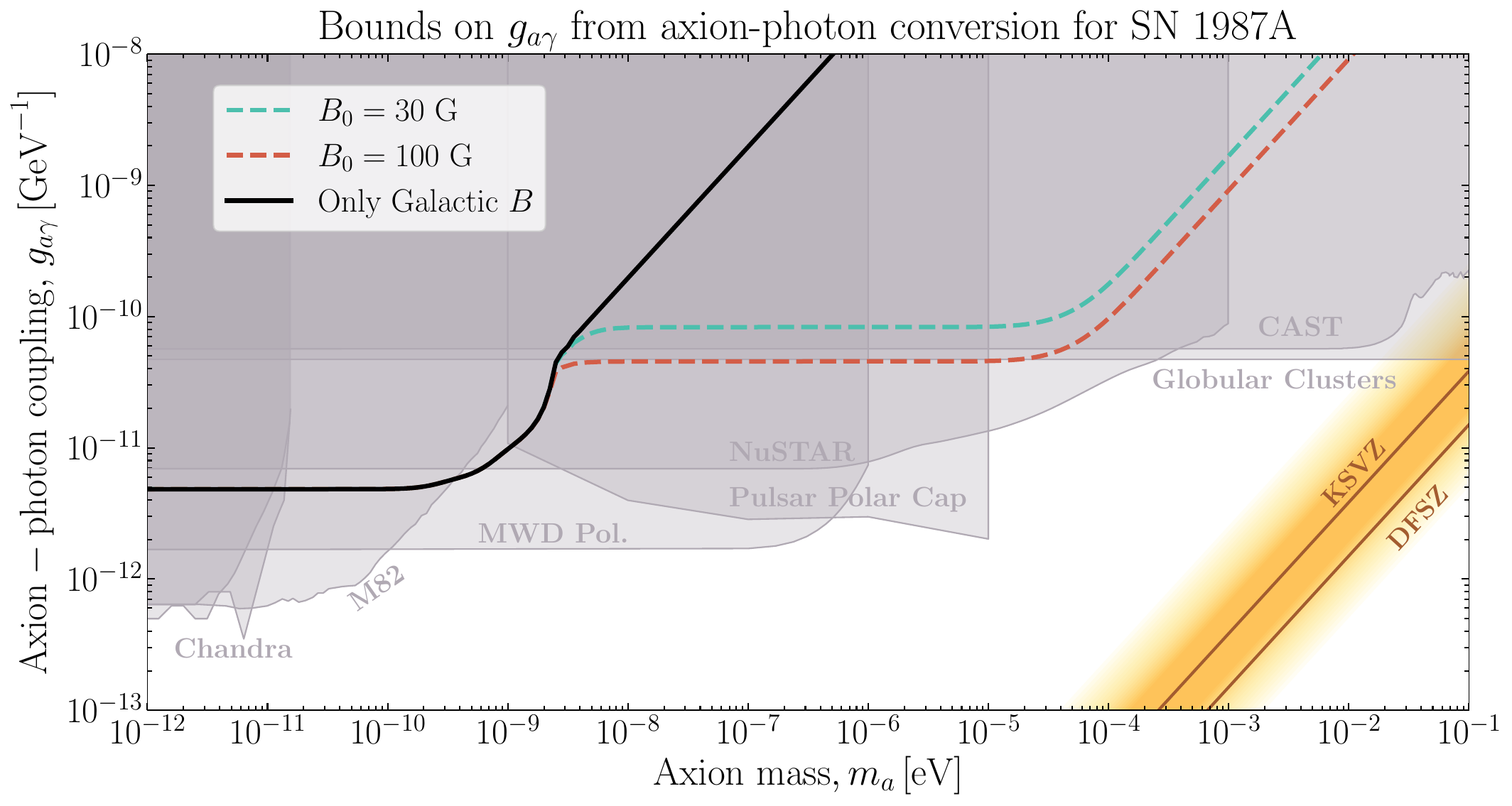}
\caption{Constraints on the axion-photon coupling $g_{a\gamma}$ from the SMM non-detection of $\gamma$-rays from SN~1987A, assuming Primakoff production alone and using the cold one-zone model to describe the  SN core. For $m_a\lesssim 10^{-9}$~eV, the conversion is mainly in the Galactic $B$-field (black line). For higher masses, the exclusion range is extended by the stellar $B$-field (dashed red and blue curves), assumed to have dipole structure, with strength $B_0=30$ or 100~G, while the progenitor radius is fixed to $R_0=30~R_\odot$. The yellow band represents QCD axions and the limits from other sources (grey shaded) are taken from the compilation of Ref.~\cite{OHare}; specifically, we show the bounds from Chandra~\cite{Reynolds:2019uqt, Reynes:2021bpe}, M82~\cite{Ning:2024eky}, polarized light from magnetic white dwarfs~\cite{Benabou:2025jcv}, pulsar polar-cap cascades~\cite{Noordhuis:2022ljw}, NuSTAR~\cite{Ruz:2024gkl}, globular clusters~\cite{Dolan:2022kul}, and CAST~\cite{CAST:2024eil}.}\label{fig:sn1987a}
\end{figure}

Using the emissivity from the cold one-zone model introduced in Sec.~\ref{sec:one_zone_models}, we now derive constraints from the SMM non-observation of gamma-rays from SN~1987A. The results for $g_{a\gamma}$ alone are shown in Fig.~\ref{fig:sn1987a} and those for the product $g_{ap}\times g_{a\gamma}$ in Fig.~\ref{fig:sn_1987A_nucleons}. However, besides this graphical illustration of numerical results, it is quantitatively more instructive to obtain them purely algebraically, highlighting the parametric dependence on the systematic uncertainties. We begin with the $g_{a\gamma}$ constraint, based on Primakoff emission alone.

The SMM satellite has reported constraints in three energy windows, where for our expected signal, the 25--100~MeV range dominates by far as illustrated in  Fig.~\ref{fig:Primakoff-Spectrum}. Therefore, we focus on that window alone, ignoring a small contribution from photons with smaller energies. We thus need the integrated number of photons in the 25--100~MeV range
\begin{equation}
    N_{\gamma
    }=\int_{25\,\mathrm{MeV}}^{100\,\mathrm{MeV}}\frac{dN_a}{dE} P_{a\gamma}\,dE.
\end{equation}
For small axion masses, $P_{a\gamma}$ is given by Eq.~\eqref{eq:conv_prob_low_magnetic} and is energy-independent. The resulting number of axions is obtained for our one-zone model accounting for Primakoff emission by direct integration. In addition, we perform a power-law expansion for parameters close to those of the one-zone model to highlight the parametric dependence on input parameters. For compactness of notation, we introduce parameters measured relative to the reference values
\begin{eqnarray}
     (Mt)_c&=&\frac{Mt}{5\, M_\odot\, \mathrm{s}},\quad Y_{p,c}=\frac{Y_p}{0.15}, \quad
    \rho_{c}=\frac{ \rho}{4\times 10^{14}\, \mathrm{g/cm}^3},\nonumber\\ T_c&=&\frac{T}{30\,\mathrm{MeV}},\quad \left[(1+z)^{-1}\right]_c=\frac{(1+z)^{-1}}{0.75}.
\end{eqnarray}
As in Sec.~\ref{sec:progenitor_conversion}, the parameters of the progenitor are expressed in terms of the dimensionless quantities $R_{30}$ and $B_{100}$.

With these parameters, we find the total number of axions emitted and converted in the stellar $B$-field to photons to be
\begin{equation}\label{eq:MasslessPrimakoffBound_number}
    N_{\gamma    
    }=2.6\times 10^{44}\,g_{11}^4 B_{100}^2 R_{30}^2 (Mt)_c Y_{p,c}(Y_{p,c}\rho_c)^{-0.36} T_c^{3.33} \left[(1+z)^{-1}\right]_c^{0.26}.
\end{equation}
The photon fluence at Earth is then $\Phi_\gamma=N_{\gamma, 25-100\,\mathrm{MeV}}/(4\pi d^2)$, where $d=51.4\,\mathrm{kpc}$ is the distance to SN~1987A, so that
\begin{equation}
    \Phi_\gamma=8.2\times 10^{-4}\,\mathrm{cm}^{-2}\,g_{11}^4 B_{c}^2 R_{c}^2 (Mt)_c Y_{p,c}(Y_{p,c}\rho_c)^{-0.36} T_c^{3.33} \left[(1+z)^{-1}\right]_c^{0.26}.
\end{equation}
Comparing this prediction with the SMM $3\sigma$ upper limit
of $0.35~{\rm cm}^{-2}$, given in Eq.~\eqref{eq:SMM-fluence-limit}, we finally find
\begin{equation}\label{eq:MasslessPrimakoffBound}
    g_{11}<4.5 \, \left(B_{c}R_{c}\right)^{-1/2} \left[(Mt)_c Y_{p,c}\right]^{-1/4}(Y_{p,c}\rho_c)^{0.09} T_c^{-0.83} \left[(1+z)^{-1}\right]_c^{-0.07}
\end{equation}
as a nominal $3\sigma$ upper limit for quasi-massless ALPs.

When the axion mass $m_a$ is large enough, the conversion probability $P_{a\gamma}$ decreases. In the SN case, the magnetic field refraction and the plasma density outside the photosphere are negligible, so $P_{a\gamma}$ is given by Eq.~\eqref{eq:conversion_large_radii_massive} with $R_i=R_0$, meaning that the conversion starts right outside the object. The mass value at which this suppression starts is given by Eq.~\eqref{eq:threshold_mass_low_magnetic}. The probability is now energy-dependent, and we find
\begin{equation}\label{eq:MassivePrimakoffBound_number}
    N_{\gamma
    }=1.6\times 10^{27}\,g_{11}^4 B_{c}^2 m_\text{eV}^{-4} (Mt)_c Y_{p,c}(Y_{p,c}\rho_c)^{-0.32} T_c^{3.75} \left[(1+z)^{-1}\right]_c^{0.79},
\end{equation}
where $m_\text{eV}$ is $m_a/\text{eV}$. Note that $N_\gamma$ does not depend directly on the progenitor radius
because, when $P_{a\gamma}$ is suppressed by the axion mass, the conversion probability is independent of $R_0$, although the boundary between the massless and massive regimes depends on this radius (see Eq.~\eqref{eq:threshold_mass_low_magnetic}). The final photon fluence is
\begin{equation}
    \Phi_\gamma=5.0\times 10^{-21}\,\mathrm{cm}^{-2}\,g_{11}^4 B_{c}^2 m_\text{eV}^{-4} (Mt)_c Y_{p,c}(Y_{p,c}\rho_c)^{-0.32} T_c^{3.75} \left[(1+z)^{-1}\right]_c^{0.79},
\end{equation}
leading to our constraint for massive axions of
\begin{equation}\label{eq:MassivePrimakoffBound}
    g_{11}<9.2\times 10^{4} \, B_{c}^{-1/2} m_\text{eV} \left[(Mt)_c Y_{p,c}\right]^{-1/4}(Y_{p,c}\rho_c)^{0.08} T_c^{-0.94} \left[(1+z)^{-1}\right]_c^{-0.20}.
\end{equation}
For our reference parameters, Eqs.~\eqref{eq:MasslessPrimakoffBound} and \eqref{eq:MassivePrimakoffBound} cross at $m_a=49\,\mu$eV. The algebraic results in Eqs.~\eqref{eq:MasslessPrimakoffBound} and \eqref{eq:MassivePrimakoffBound} agree with those shown in Fig.~\ref{fig:sn1987a} that were obtained with a numerical integration of the mixing equations. The variation with assumed $B$-field strength is also borne out of the algebraic scalings.

For large masses, the $g_{a\gamma}$ constraint grows proportionally to $m_a$, matching Eq.~\eqref{eq:MassivePrimakoffBound}. This ``diagonal part'' of the exclusion line in Fig.~\ref{fig:sn1987a} is parallel to the axion band, an effect analogous to the mass suppression in CAST or the future IAXO. It is this effect that makes it hard to extend sensitivity to realistic QCD axions. We also find that, with our parameter choices, Sanduleak's field does not improve the sensitivity to smaller $g_{a\gamma}$ values than already excluded by CAST and globular clusters. Due to the unknown strength and orientation of the stellar field, the dashed lines in Fig.~\ref{fig:sn1987a} are anyway not limits, they only provide tentative orientation in parameter space.

The previous expressions are valid for axion masses $m_a\gtrsim 10^{-9}\,\rm eV$. For masses below this value, the constraints derive primarily from axion-photon conversion in the Galactic magnetic field (black line in Fig.~\ref{fig:sn1987a}). Repeating the previous exercise with our fixed Galactic field and using Eq.~\eqref{eq:GalacticConversionProbability}, we find
\begin{equation}\label{eq:GalacticPrimakoffBound_number}
    N_{\gamma
    }=2.0\times 10^{48}\,g_{11}^4  (Mt)_c Y_{p,c}(Y_{p,c}\rho_c)^{-0.36} T_c^{3.33} \left[(1+z)^{-1}\right]_c^{0.26}.
\end{equation}
Of course, in this case there is no dependence on the progenitor radius or magnetic field. The dependence on the other parameters, involved only in the axion emission, is the same as in the previous massless case. The photon fluence at Earth is
\begin{equation}
    \Phi_\gamma=6.4\,\mathrm{cm}^{-2}\,g_{11}^4 (Mt)_c Y_{p,c}(Y_{p,c}\rho_c)^{-0.36} T_c^{3.33} \left[(1+z)^{-1}\right]_c^{0.26}.
\end{equation}
Therefore, we find 
\begin{equation}\label{eq:GalacticPrimakoffBound}
    g_{11}<0.48 \,  \left[(Mt)_c Y_{p,c}\right]^{-1/4}(Y_{p,c}\rho_c)^{0.09} T_c^{-0.83} \left[(1+z)^{-1}\right]_c^{-0.07}
\end{equation}
as our SMM bound from SN~1987A
for axion masses below $10^{-9}\,\rm eV$.

Over the years, similar SN~1987A bounds on $g_{a\gamma}$ have been derived by several groups \cite{Brockway:1996yr, Grifols:1996id, Payez:2014xsa, Hoof:2022xbe, Manzari:2024jns}. In an early paper, Brockway et al.\ \cite{Brockway:1996yr} found $g_{11}<0.76$, but allowing for large uncertainties, they stated $g_{11}<1$ as a conservative bound. Around the same time, Grifols et al.\ \cite{Grifols:1996id} found $g_{11}<0.3$, assuming a rather hot SN model with $T=60$~MeV.

Twenty years later, a completely fresh analysis by Payez et al.\ \cite{Payez:2014xsa} found $g_{11}<0.53$. These authors used a cold numerical SN model with typical $T\simeq30$~MeV. They reported the Gamma-fit parameters of the time-integrated axion flux spectrum, in particular $\Eav=102.3$~MeV and $\alpha=2.25$, but they did not include gravitational lapse. If the latter is taken to be 0.8, their average energy is entirely in line with our $\Eav=86.0$~MeV for the SFHo-18.8 model and our $\alpha=2.09$. For the total number of emitted axions, they found $N_a=g_{11}^2\,5.3\times10^{51}$, but even accounting for the missing lapse factor, it is around 3 times larger than our corresponding $N_a$. The difference for sure derives from their very long cooling time, where axion emission is still significant at 10~s post bounce, a very long SN cooling time, that must be caused by the absence of PNS convection in their model. This simple example shows that it is not important to use a specific numerical SN model, it is more important to worry about the critical overall parameters. It is clear that what we have called $M t$ is much larger in their model, caused by a specific physical difference, not a numerical detail.

Still, their $g_{a\gamma}$ bound is practically identical to ours, probably caused by their conversion probability being a factor 2.3 smaller as discussed after Eq.~\eqref{eq:GalacticConversionProbability}, so the expected gamma-ray flux would be comparable. They also used the less restrictive SMM limit of $0.6~{\rm cm}^{-2}$ of Ref.~\cite{Chupp:1989kx}, compared to our $0.35~{\rm cm}^{-2}$ given in Eq.~\eqref{eq:SMM-fluence-limit}. Overall, one can understand how these factors can conspire to make their bound similar to ours, despite these relatively large differences in the various factors.

A few years ago, this topic was revisited by Hoof and Schulz
\cite{Hoof:2022xbe} who found a $3\sigma$ limit of $g_{11}<0.52$, based on the same SN model as Payez et al.~\cite{Payez:2014xsa}. (Hoof and Schulz never explicitly state their limit as a number, so we refer to their Fig.~3 and the uploaded limit data on O'Hare's github page \cite{OHare}.) Their limit is practically identical to the one found by Payez et al.~\cite{Payez:2014xsa}, although there are many differences in the detailed analysis, including the conversion probability in the Galactic $B$-field, using the time structure of the expected signal, and revising the SMM data (see the comments in our Appendix~\ref{app:SMM}). 

The latest SN~1987A limit on $g_{a\gamma}$ was derived by the Berkeley group \cite{Manzari:2024jns}. They assumed axion production by both the Primakoff process and bremsstrahlung, the latter based on loop-induced nucleon couplings. In this way, both reactions depend on $g_{11}$ and the total number of produced axions by both reactions are shown in our Table~\ref{tab:BerkeleyFits}. In agreement with Figs.~2 and~S12 of their published version, both channels produce nearly equal values for $N_a$. Therefore, to compare with our Primakoff-only assumption, and in agreement with the usual ALP constraints, their bound on $g_{11}$ should be relaxed (multiplied) by an approximate factor of $2^{1/4}=1.19$.

In their published version, they showed a $2\sigma$ bound of $g_{11}<0.27$, a number extracted from the dark blue line in their Fig.~1, whereas in v2 of their arXiv posting, that supersedes the published paper, they show 0.36 instead,\footnote{The difference arises from correcting a coding error in their computation of the photon flux that we discovered after they had made their files available to us and led to the update of their PRL version in the form of v2 of their arXiv posting. Unfortunately, they still use the scaled SMM data (see our Appendix~\ref{app:SMM}) with an impact on their analysis that is hard to estimate.} a seemingly small change, which however corresponds to a flux difference of $(0.36/0.27)^4=3.2$. After multiplying with $2^{1/4}$ as explained in the previous paragraph, this bound becomes 0.43, slightly more restrictive than our value of 0.48. Actually, this difference could be explained by them using a $2\sigma$ limit, whereas ours is $3\sigma$ (see end of Appendix~\ref{app:SMM}). On the other hand, their Primakoff flux is roughly 2/3 of the one we have used, and their conversion probability is at least a factor 2/3 smaller than ours (see the comment after Eq.~\ref{eq:GalacticConversionProbability}), two effects that should weaken their bound compared to ours, and using the scaled SMM data goes in the same direction. Of course, on the level of $g_{11}$, due to the 1/4 power between flux limit and coupling-strength limit, such differences look minimal on a panoramic logarithmic~plot, although these are clear differences that we cannot explain.

\subsection[SN~1987A constraints on \texorpdfstring{$g_{ap}\times g_{a\gamma}$}{}]{SN~1987A constraints on \texorpdfstring{\boldmath$g_{ap}\times g_{a\gamma}$}{}}

\begin{figure}[ht]
\includegraphics[width=\textwidth]{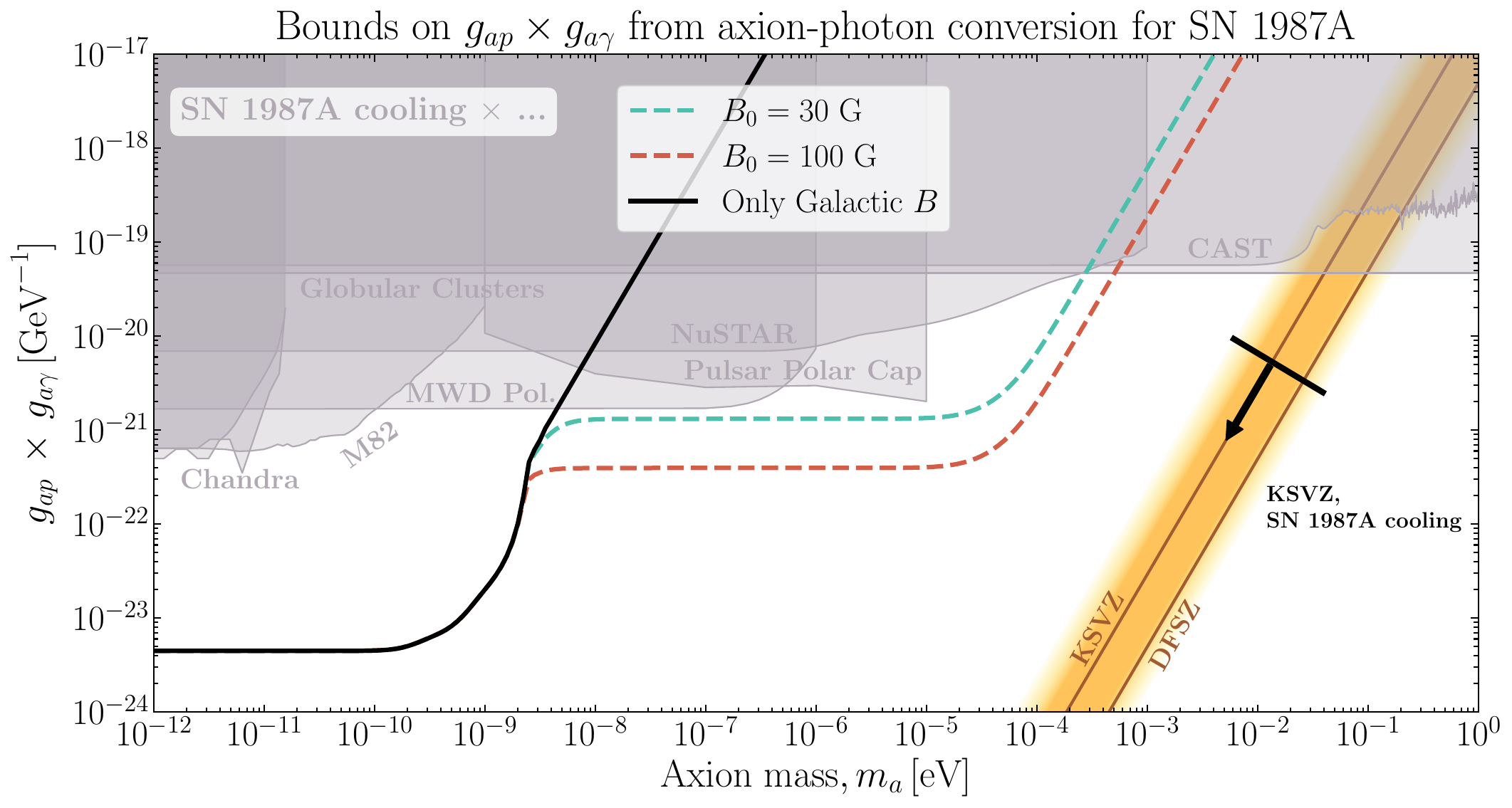}
\caption{Same as Fig.~\ref{fig:sn1987a}, now for the product of the axion-proton and axion-photon couplings $g_{a p}\times g_{a\gamma}$. Constraints from other sources (grey shaded) are those of Fig.~\ref{fig:sn1987a}, multiplied with the SN~1987A cooling bound on $g_{ap}$ of Eq.~\eqref{SN1987A-cooling-bound}. The yellow band represents QCD axions, and the black arrow indicates the SN~1987A cooling constraint on KSVZ axions of $m_a<13$~meV, corresponding to the $g_{ap}$ limit. 
}\label{fig:sn_1987A_nucleons}
\end{figure}

We next turn to constraining the product $g_{ap}\times g_{a\gamma}$ and show our results in Fig.~\ref{fig:sn_1987A_nucleons}. To compare with the constraints from other sources that assume only the $g_{a\gamma}$ coupling, we have multiplied these limits with that on $g_{ap}$ from SN~1987A cooling stated in Eq.~\eqref{SN1987A-cooling-bound}. Algebraically, to constrain $g_{ap}\times g_{a\gamma}$, we repeat the previous exercise for Primakoff emission, now for bremsstrahlung, while $g_{a\gamma}$ remains responsible for axion-photon conversion. The number of axions emitted and converted to photons in the 25--100~MeV window becomes
\begin{equation}\label{eq:MasslessBremsBound_number}
    N_{\gamma
    }=7.1\times 10^{49}\,g_{20}^2 B_{c}^2 R_{c}^2 (Mt)_c Y_{p,c}\rho_c^{0.80} T_c^{2.60} \left[(1+z)^{-1}\right]_c^{1.00},
\end{equation}
where
\begin{equation}
    g_{20}=\frac{g_{ap}\times g_{a\gamma}}{10^{-9}\times 10^{-11} \text{ GeV}^{-1}}.
\end{equation}
The scaling of the lapse factor turns out to be almost linear, but this is only approximate, unlike the exact linear scaling for $Mt$, for example. The photon fluence at Earth is
\begin{equation}
    \Phi_\gamma=2.2\times 10^{2}\,\mathrm{cm}^{-2}\,g_{20}^2 B_{c}^2 R_{c}^2 (Mt)_c Y_{p,c}\rho_c^{0.80} T_c^{2.60} \left[(1+z)^{-1}\right]_c^{1.00},
\end{equation}
leading to the constraint
\begin{equation}\label{eq:MasslessBremsBound-87A}
    g_{20}<4.0\times 10^{-2} \, \left(B_{c}\,  R_{c}\right)^{-1} \left[(Mt)_c Y_{p,c}\right]^{-1/2}\rho_c^{-0.40} T_c^{-1.30} \left[(1+z)^{-1}\right]_c^{-0.50}
\end{equation}
for massless axions.

Finally, for large axion masses, with corresponding suppression of the conversion probability, we find
\begin{equation}\label{eq:MassiveBremsBound_number}
    N_{\gamma
    }=3.3\times 10^{32}\,g_{20}^2 B_{c}^2 m_\text{eV}^{-4} (Mt)_c Y_{p,c}\rho_c^{0.88} T_c^{3.28} \left[(1+z)^{-1}\right]_c^{1.72}.
\end{equation}
The final photon fluence becomes
\begin{equation}
    \Phi_\gamma=1.0\times 10^{-15}\,\mathrm{cm}^{-2}\,g_{20}^2 B_{c}^2 m_\text{eV}^{-4} (Mt)_c Y_{p,c}\rho_c^{0.88} T_c^{3.28} \left[(1+z)^{-1}\right]_c^{1.72},
\end{equation}
leading to the constraint for massive axions of
\begin{equation}\label{eq:MassiveBremsBound}
    g_{20}<1.8\times 10^{7} \, B_{c}^{-1} m_\text{eV}^2 \left[(Mt)_c Y_{p,c}\right]^{-1/2}\rho_c^{-0.44} T_c^{-1.64} \left[(1+z)^{-1}\right]_c^{-0.86}.
\end{equation}
As explained earlier, this scaling applies when the axion mass exceeds the value specified in Eq.~\eqref{eq:threshold_mass_low_magnetic}. Numerically, Eqs.~\eqref{eq:MasslessBremsBound} and \eqref{eq:MassiveBremsBound} cross at $m_a=47~\mu$eV.
In analogy to the Primakoff-only source, the results obtained from numerically integrating the mixing equations agree with the algebraic ones from Eqs.~\eqref{eq:MasslessBremsBound} and \eqref{eq:MassiveBremsBound}. The conversion probability is, of course, the same as for Primakoff, only the source process is now bremsstrahlung. The bounds are now more restrictive than by other arguments (shown in gray in Fig.~\ref{fig:sn_1987A_nucleons}), but once more we show them only dashed because the field strength and orientation is speculative, not measured.

Once more, for $m_a\lesssim 10^{-9}$ eV, the conversion is dominated by the Galactic magnetic field, which provides the only true limit from SN~1987A as discussed in the Primakoff case. Using Eq.~\eqref{eq:GalacticConversionProbability}, we now find
\begin{equation}
    N_{\gamma
    }=5.5\times 10^{53}\,g_{20}^2 (Mt)_c Y_{p,c}\rho_c^{0.80} T_c^{2.60} \left[(1+z)^{-1}\right]_c^{1.00}.
\end{equation}
Again, when considering the Galactic magnetic field, there is no dependence on the progenitor radius or magnetic field. The photon fluence at Earth is
\begin{equation}
    \Phi_\gamma=1.8\times 10^{6}\,\mathrm{cm}^{-2}\,g_{20}^2 (Mt)_c Y_{p,c}\rho_c^{0.80} T_c^{2.60} \left[(1+z)^{-1}\right]_c^{1.00},
\end{equation}
leading to the constraint
\begin{equation}\label{eq:MasslessBremsBound}
    g_{20}<4.5\times 10^{-4} \, \left[(Mt)_c Y_{p,c}\right]^{-1/2}\rho_c^{-0.40} T_c^{-1.30} \left[(1+z)^{-1}\right]_c^{-0.50}
\end{equation}
We can compare this result with the constraints reported by Ref.~\cite{Manzari:2024jns} on the KSVZ-like axion, which is reported in their Fig.~1 to be $g_{a\gamma}<10^{-13}\,\mathrm{GeV}^{-1}$. Since for the KSVZ-like axion we have $C_{ap}=-0.47$ and $C_{a\gamma}=1.92$, from Eq.~\eqref{eq:gap_gagamma_connection} we find $g_{ap}<2.0\times 10^{-11}$ and therefore $g_{20}<2.0\times 10^{-4}$, which is more than a factor 2 more restrictive than our constraint, implying a constraint on the overall photon signal more restrictive than a factor~4. 
Some of the discrepancy probably can be attributed to the same issues that we discussed earlier at the end of Sec.~\ref{sec:gag-limits}.

There is a tantalizing hard X-ray excess detected from the Magnificent Seven NSs~\cite{Dessert:2019dos} that is consistent with ALPs that couple to protons and photons. These could be produced in NS cores by bremsstrahlung and converted into X-rays in the magnetospheres \cite{Buschmann:2019pfp,Fan:2025ixw}. However, even our most conservative results exclude this hypothesis. Likewise, the recent proposal to explore axion production from nuclear de-excitations in Galactic stellar populations (see, e.g., Refs.~\cite{Ning:2025kyu,Haxton:2025xqz}) can only probe parameters that are already excluded by our constraints.

\subsection{Sensitivity from a future Galactic supernova}\label{sec:results_sensitivities_sne}

The observation of a future Galactic SN could be accompanied by the detection of gamma-rays with Fermi-LAT or even the proposed future satellite network GALAXIS \cite{Manzari:2024jns}, thus providing evidence for axions. The instant for the putative signal would be set by the neutrino burst, assumed to be observed in the existing network of SN neutrino detectors \cite{SNEWS:2020tbu}. Moreover, the SN location in the sky likely will be determined immediately or later by astronomical means, so one is looking for a signal tightly constrained in time and space. Therefore, a few 100-MeV-range photons would be enough as a detection threshold. As explained in Sec.~\ref{eq:FermiLat}, we consider a ``Fermi-LAT-like'' experiment, defined by the effective area shown in Fig.~\ref{fig:AeffFermiLAT}, and we consider 3 expected events as our detection threshold.

Moreover, since we are now exploring future detection opportunities, not past or future constraints, we can be optimistic in our choice of source parameters. Our goal here is to explore if there is even a small chance to detect QCD axions with this technique, whereas the constraints that would arise from a non-observation are a different matter that we do not study here.
For the description of the SN core, we will bracket our forecast with our cold and hot one-zone models (Table~\ref{tab:one_zone}). Regarding the distance to the SN and the magnetic field around the progenitor, the results will be presented for a pessimistic case, assuming a distance $d=20\ \mathrm{kpc}$ and a surface magnetic field $B_0=100\ \mathrm{G}$, and for a more optimistic case, with $d=2\ \mathrm{kpc}$ and $B_0=1\ \mathrm{kG}$. For the progenitor radius we will consider a fixed value $30~R_\odot$, corresponding to the typical radius of a BSG. One should notice that CCSNe featuring a BSG as a progenitor are extremely rare, with a rate of 1-7\% of the hydrogen-rich Type II SNe~\cite{Pastorello:2011mn,Shivvers:2016bnc,Schneider:2025ixo}, making detection prospects somewhat daunting. CCSNe from RSGs are more common, though the magnetic field is expected to be smaller and likely turbulent; they would generally not allow to probe the QCD axion through axion-photon conversion in their stellar magnetic field.

\begin{figure}
\includegraphics[width=\textwidth]{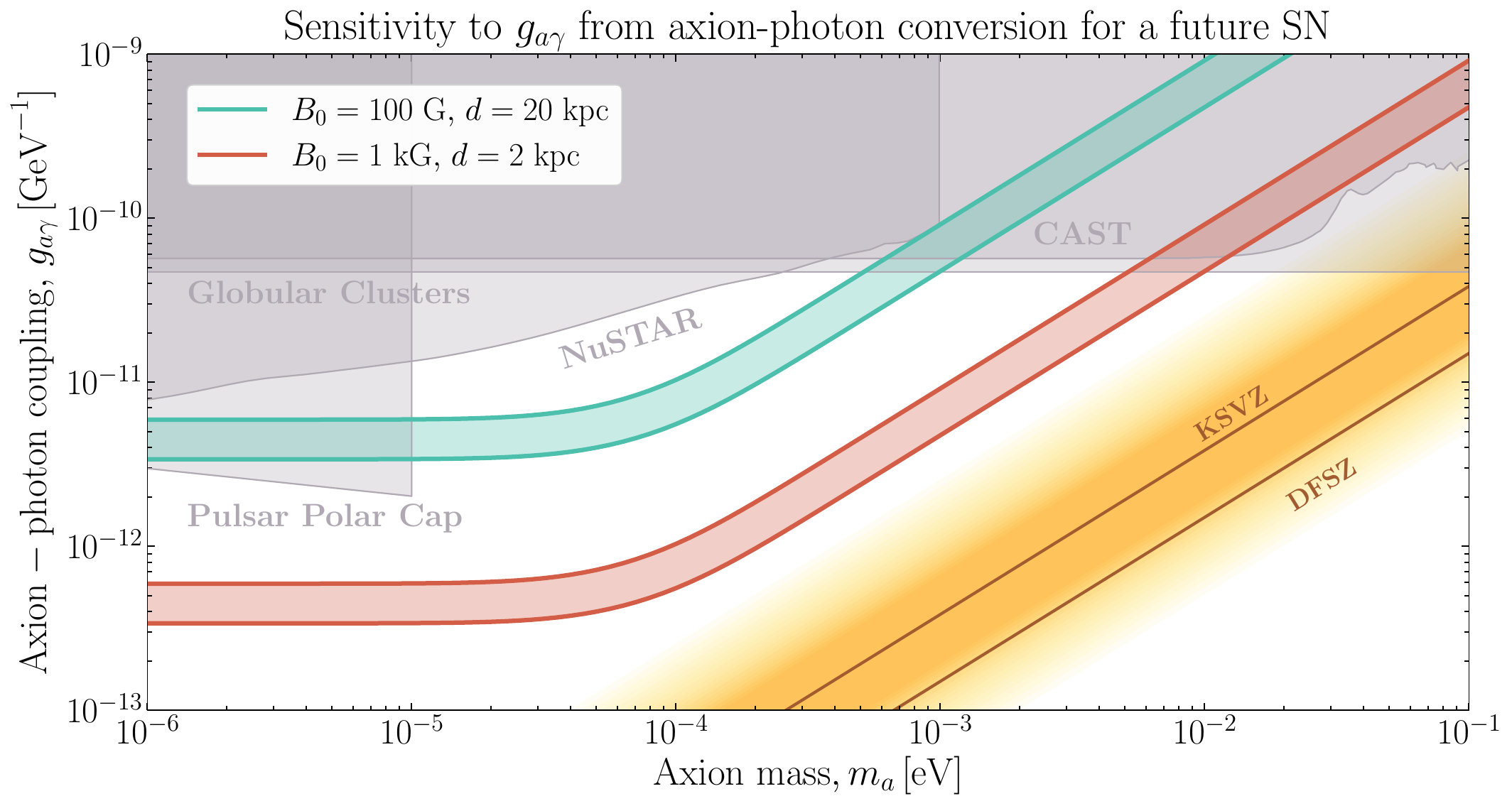}
\vskip-6pt
\caption{Projected reach on the axion-photon coupling $g_{a\gamma}$ as a function of the axion mass, assuming that Fermi-LAT or a similar satellite observes 100-MeV-range gamma-rays coming from a future SN. The blue and red sensitivity bands are defined by the pessimistic and optimistic choices for $B_0$ and distance shown in the legend.
The edges of the shaded bands are defined by the hot and cold one-zone models.
The progenitor radius is fixed to be $R_0=30R_\odot$.
}\label{fig:sn_projected}

\vskip24pt
\includegraphics[width=\textwidth]{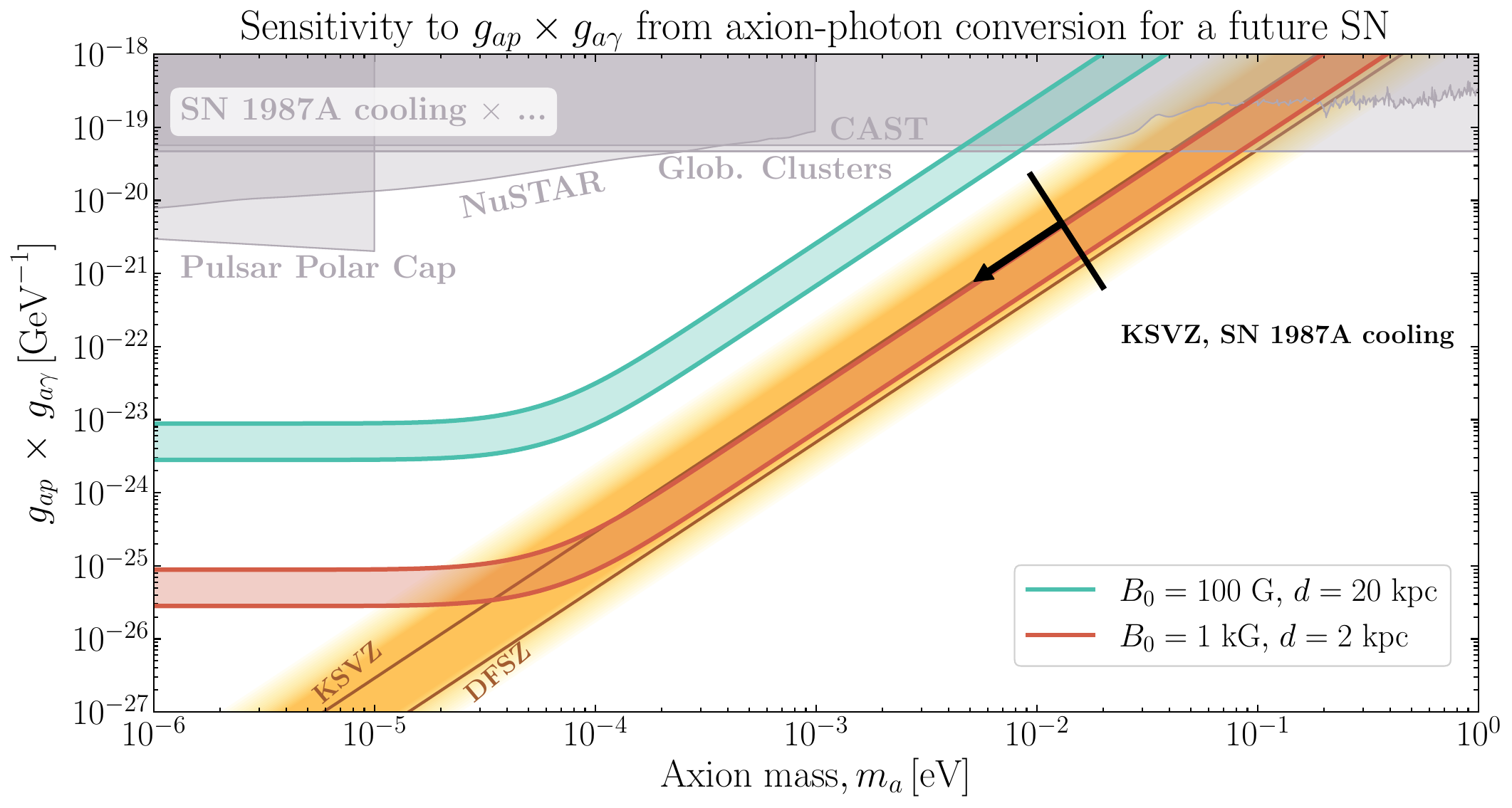}
\vskip-6pt
\caption{Same as Fig.~\ref{fig:sn_projected}, now for $g_{a p}\times g_{a\gamma}$, relying on bremsstrahlung instead of Primakoff emission.
Previous constraints (grey shaded) are obtained from those of Fig.~\ref{fig:sn_projected} by multiplying with the SN~1987A cooling bound on $g_{ap}$ of Eq.~\eqref{SN1987A-cooling-bound}.
The black arrow indicates the SN~1987A cooling bound on KSVZ axions.
}\label{fig:sn_projected_nucleons}
\end{figure}

Figure~\ref{fig:sn_projected} shows the projected sensitivity for pure ALPs, which in our definition couple only to photons, relying on Primakoff emission alone.
Even in the more pessimistic case, the sensitivity exceeds the recent NuSTAR constraints~\cite{Ruz:2024gkl}, opening up parameter space in an unexplored region, and discovering an axion would test the pulsar polar cap scenario~\cite{Noordhuis:2022ljw}. The optimistic choices for $B_0$ and distance improve the reach on $g_{a\gamma}$ by a factor of 10. In photon flux, one gains a factor of 100 from the ten times larger $B_0$-field and another factor of 100 from the distance reduced by a factor of 10, so four orders of magnitude in gamma-ray flux lead to a factor of 10 improvement in coupling-constant sensititivity, as for any production$\times$detection axion experiment, which corresponds to the challenge for IAXO to improve on the CAST sensitivity.
There is a factor of~1.74 sensitivity difference between the hot and cold models (the width of the shaded bands). This factor can be confirmed by comparing with Eqs.~\eqref{eq:MasslessPrimakoffBound_number} and \eqref{eq:MassivePrimakoffBound_number} which predict that our hot model emits $\sim 7$ times more axions through Primakoff process than our cold model, so the predicted sensitivity on the coupling by the hot model is $7^{1/4}=1.6\sim1.74$ times better than the cold model.

The projected sensitivity on $g_{a p}\times g_{a\gamma}$ is shown in Fig.~\ref{fig:sn_projected_nucleons}, using bremsstrahlung as a source. We find that under the optimistic assumptions---a SN exploding at a distance of 2 kpc with a progenitor with $B_0=1\;\mathrm{kG}$---the high-energy gamma-ray flux might be detectable by a ``Fermi-LAT-like'' experiment if the axion has a coupling structure as suggested by the KSVZ model and a mass $m_a\gtrsim 10^{-4}\,\rm eV$. For slightly more optimistic conditions, even the DFSZ axion model could be probed. In this sense, we confirm that the QCD axion might be testable by the detection of high-energy gamma-rays, as proposed in Ref.~\cite{Manzari:2024jns}.
However, to reach QCD axions, our estimate requires an optimistic distance of 2~kpc, whereas they used a typical distance of 10~kpc. Both estimates consider a relatively small stellar radius appropriate for a BSG. In other words, our gamma-ray flux estimate, for the same conditions, is significantly smaller. The main reason for this large difference is that we omit the pionic processes; as we argue in Appendix~\ref{app:PionConversion}, our current understanding of pions in SNe rests on too thin basis to have any reliable estimate of how large this signal could be, with different recent papers contradicting each other. From an optimistic perspective, of course, it is not excluded that the axion flux could be much larger than our conservative estimate.
 
Concerning our projected difference between hot and cold model, the reach to the product $g_{a p}\times g_{a\gamma}$ differs by about a factor of 3. Here, Eqs.~\eqref{eq:MasslessBremsBound_number} and \eqref{eq:MassiveBremsBound_number} predict that the hot model emits $\sim8$ times more axions than the cold one, so the predicted sensitivity is $\sqrt{8}\sim3$ times better.

\subsection{Sensitivity from future neutron-star mergers}

In this last section we present our results for the case of a future NSM. We show our projected reach for a future event, similar to GW170817, from which gamma-ray observations were not available from Fermi-LAT. We follow a strategy similar to the projected reach for SN events in Section~\ref{sec:results_sensitivities_sne}; the gamma-ray fluence at Earth is the time-integrated gamma-ray flux, obtained as the product of the axion emissivity and the conversion probability, which depends on the emission time as discussed in Section~\ref{sec:nsm_conversions}. We assume a NSM happening at 30~Mpc from Earth (for comparison, the distance of GW170817 was around 40~Mpc). 
Although first estimates of the merger rate were, at $90\%$~C.L., were as large as $110-3840 \;\rm Gpc^{-3}y^{-1}$~\cite{LIGOScientific:2018mvr}, more recent results finds $10-1700\;\rm  Gpc^{-3}y^{-1}$~\cite{KAGRA:2021duu}.
We model the response of the instrument using the effective area from Fig.~\ref{fig:AeffFermiLAT}, and require the number of detected events to be above 3 for a detection, following the Feldman-Cousins prescription already used in Section~\ref{sec:results_sensitivities_sne}.

The gamma rays originating from axion conversion would be visible at Earth as a 100-MeV gamma-ray burst. Since a standard gamma-ray burst is also associated with the merger itself, one might be concerned that the search for events is not really background-free. However, the gamma rays from the event, in the standard scenario without axions, have completely different properties from the axion-induced ones: they arrive with a significant delay of about 2 seconds, as observed from GW170817---a delay caused by the time for the formation of the central black hole and subsequent breakout of the jet---and they typically peak in the MeV region. On the other hand, the detection of the gamma ray arising from the axion-photon conversion should take place immediately after the detection of the merger, when the HMNS is formed. Therefore, they can be easily discriminated from the non-standard signal associated with axion conversion.

We use our one-zone NSM model described in Table~\ref{tab:one_zone} to obtain the axion flux spectrum. When building Tables~\ref{tab:one_zone}, \ref{tab:NSM-models} and, more specifically, when finding our estimates for the mass exposure $Mt$, we integrate the Garching models over time, so our one-zone models give a prediction for the time-integrated axion fluence $dN/dE$. However, to find the photon fluence at Earth for the NSM case we need $dN/(dt\ dE)$, since in this case the axion-photon conversion is time-dependent. For that purpose we assume that the axion-fluence emitted by the HMNS is constant in time, and since the time-integration period for the Garching models is 10~ms, then $dN/(dt\ dE)=(10\ \mathrm{ms})^{-1}\times dN/dE$. We will assume that the remnant HMNS lives for one second; however, as we will explain, this assumption is only relevant in the non-physical case where we ignore the presence of the ejected material. Once we take this material into account, the relevant timescale is when the ejecta reaches the conversion radius and shuts down the axion-photon conversion.

\begin{figure}
    \centering
    \includegraphics[width=1.\linewidth]{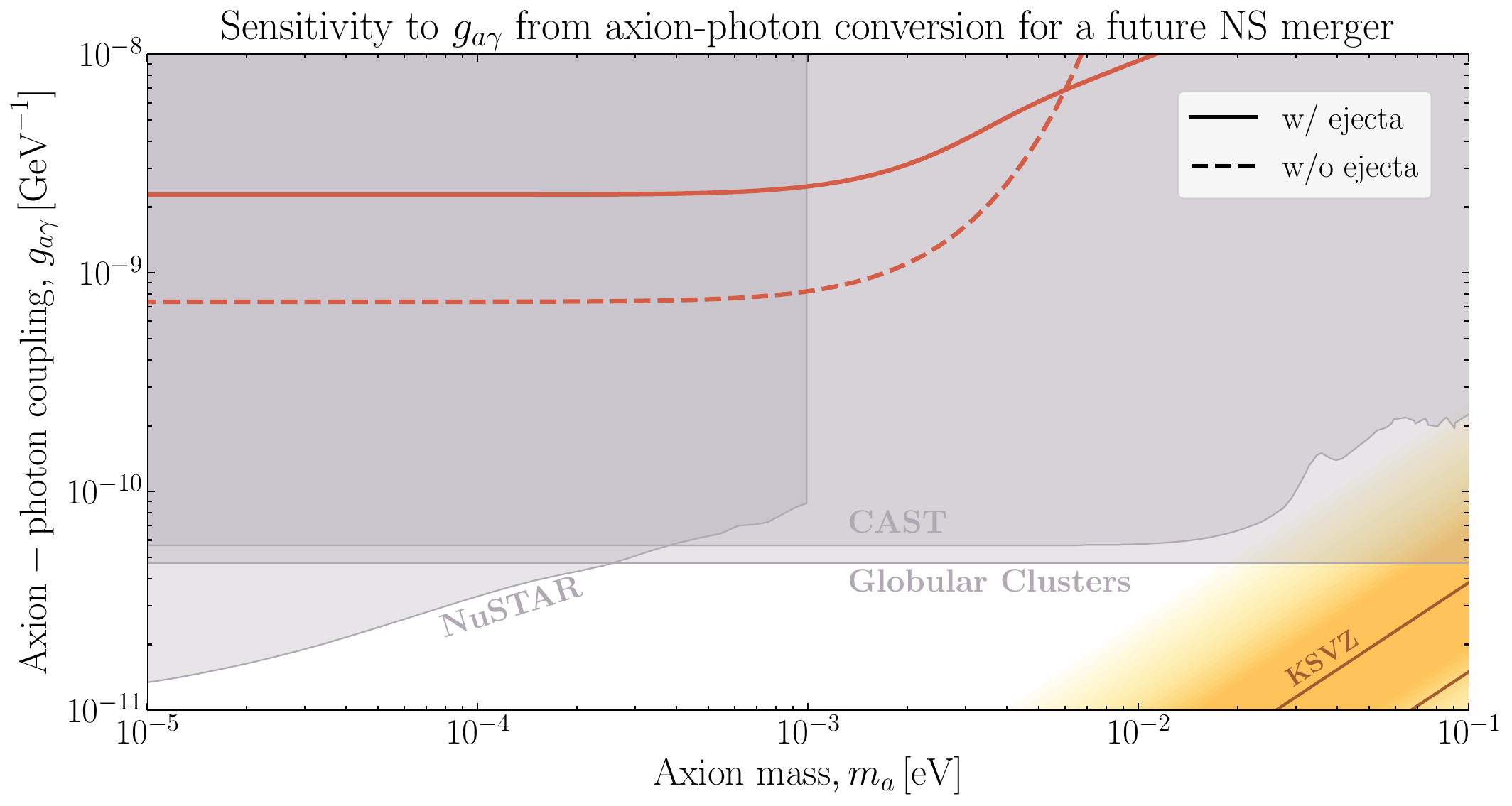}
    \caption{Projected reach on the axion-photon coupling as a function of the axion mass, assuming a NSM at 30~Mpc from Earth. The yellow band represents the QCD axion band. The solid line corresponds to the bound obtained including the ejecta, while the dashed line corresponds to the case where the ejected material is not taken into account. The HMNS is described by the one-zone model determined in Table~\ref{tab:one_zone}. We assume the HMNS to have a radius $R_0=10$~km and a surface magnetic field $B_0=10^{14}$~G, and we fix the velocity of the ejected material to be $V=0.8 c$.}
    \label{fig: NSMBounds}
\end{figure}

Figure~\ref{fig: NSMBounds} shows our projected reach for $g_{a\gamma}$, considering only Primakoff emission of axions. Unfortunately, we find that even under very optimistic conditions, the sensitivity to the ALP-photon coupling is significantly above the CAST exclusion region. In the absence of ejecta, the sensitivity curve is flat at low masses, and weakens exponentially at the threshold mass anticipated by Eq.~\eqref{eq:maximum_mass_nsm}. This exponential trend is also noticeable in the numerical curves of Ref.~\cite{Manzari:2024jns}, and we now understand it explicitly based on our earlier estimates. However, these curves without accounting for ejecta are overall unphysical. The impact of the ejecta, previously neglected in this context, is quite noticeable, and can weaken the projected reach by even a factor 3 for massless axions; since the gamma-ray flux grows in proportion to $g_{a\gamma}^4$, this means that ejecta can reduce the gamma-ray fluence by as much as two orders of magnitude. This is expected, since ejecta can shut off entirely axion-photon conversions from 10~ms to 1~s, therefore reducing the amount of time in which the NSM is actively emitting gamma-rays. At large masses, the projected reach in the presence of ejecta turns out to be better than without the ejecta, due to the positive effect they have on the conversions of ALPs emitted close to the dominant conversion radius $R_a$, as we have previously pointed out. The results depend on the choice of the surface magnetic field, but, as we have seen in Section~\ref{sec:nsm_conversions}, the conversion probability grows as $P_{a\gamma}\propto B^{2/5} g_{a\gamma}^2$, so that the gamma-ray signal at Earth scales in proportion to $B^{2/5} g_{a\gamma}^4$. Therefore, the sensitivity reach for the coupling scales as $g\propto B^{-1/10}$. Thus, choosing a magnetic field of $10^{16}$~G, rather than $10^{14}$~G as we have done here, would improve the sensitivity reach by about $60\%$. On the other hand, increasing the magnetic field also reduces the maximum axion mass that can be probed, according to Eq.~\eqref{eq:maximum_mass_nsm}.

\begin{figure}
    \centering
    \includegraphics[width=1.\linewidth]{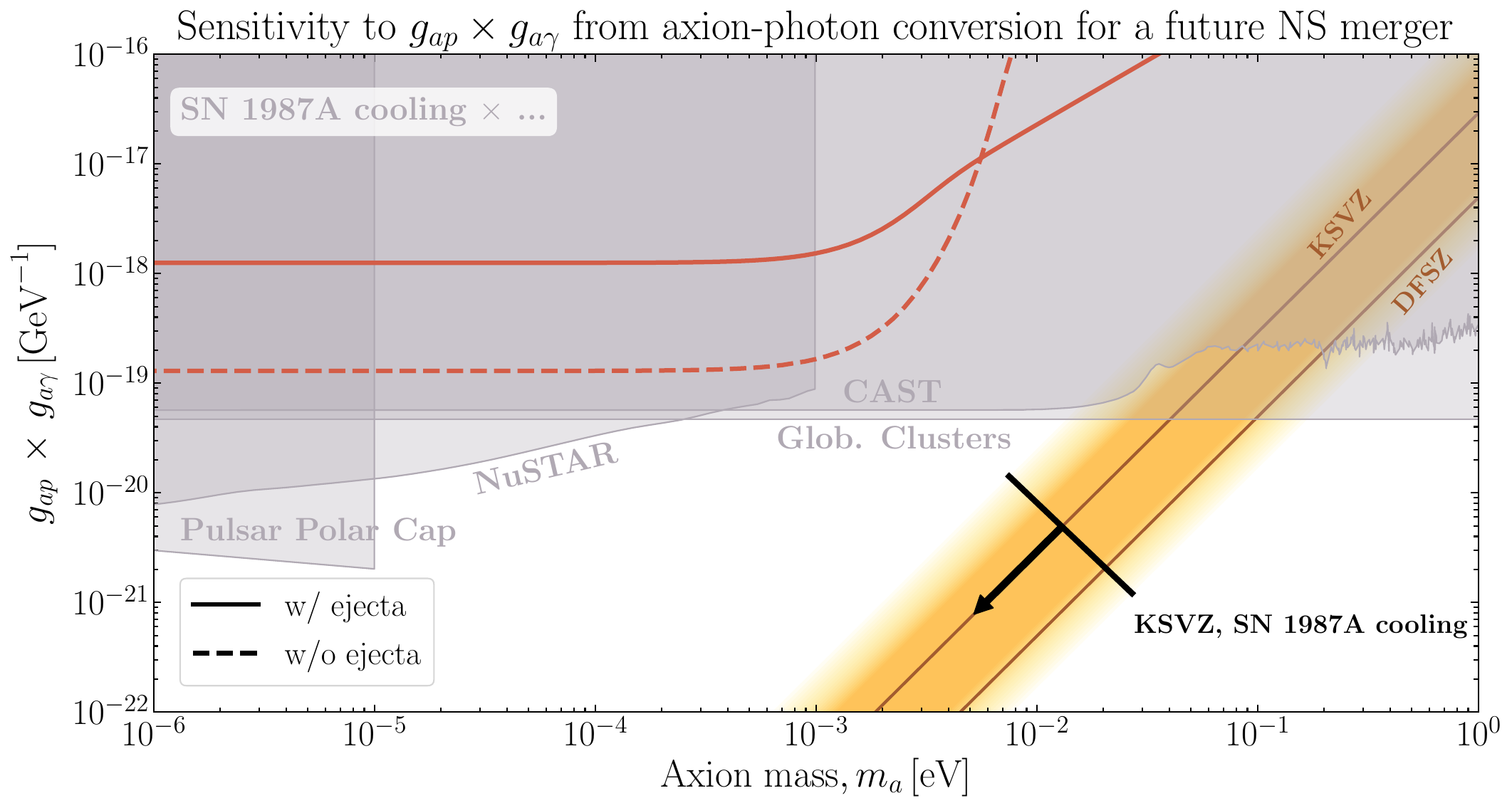}
    \caption{Projected reach on the product of the axion-photon and axion-proton couplings $g_{ap}\times g_{a\gamma}$ as a function of the axion mass, assuming a NSM at 30~Mpc from Earth. The solid line corresponds to the bound obtained including the ejecta, while the dashed line corresponds to the case where the presence of the ejected material is ignored. The HMNS is described by a one-zone model which parameteres are specified in Table~\ref{tab:one_zone}. We assume a surface magnetic field $B_0=10^{14}$~G, a HMNS radius of $R_0=10$~km, and the velocity of the ejecta to be $V=0.8 c$. The existing constraints, shown in grey, are obtained by multiplying the current constraints on $g_{a\gamma}$ times the bound on $g_{ap}$ obtained from the cooling of SN~1987A (see Eq.~\eqref{SN1987A-cooling-bound}). The yellow band represents the QCD axion band, and the black line indicates the restriction on KSVZ QCD axion found from the cooling of SN~1987A.}
    \label{fig: NSMNucleonBounds}
\end{figure}

Finally, let us comment on the possibility of probing the QCD axion via nucleon-induced processes. A coupling to nucleons would induce nucleon-nucleon bremsstrahlung, which is a much more efficient production channel than the Primakoff process previously considered. Can the sensitivity reach in this case the couplings associated with the QCD axion band? 
We believe that NSMs are unlikely to probe the QCD axion via this methodology. This conclusion seems relatively robust, as can be seen by the following argument.

NSMs can potentially compete with SNe only at large axion masses, since the SN sensitivity weakens at masses $m_a\gtrsim 10^{-4}$~eV (see Eq.~\eqref{eq:maximum_mass_supernovae}, as well as Figs.~\ref{fig:sn1987a}, \ref{fig:sn_1987A_nucleons}, \ref{fig:sn_projected}, \ref{fig:sn_projected_nucleons}), while the NSM one weakens at masses $m_a\gtrsim 10^{-3}$~eV (see Eq.~\eqref{eq:maximum_mass_nsm}, as well as Fig.~\ref{fig: NSMBounds}). At these larger masses, the coupling combination $g_{ap}\times g_{a\gamma}$ of the QCD axion is two orders of magnitude larger, since both $g_{ap}$ and $g_{a\gamma}$ are proportional to the QCD axion mass. For NSMs to probe the QCD axion band, it is therefore sufficient if the sensitivity is about two orders of magnitude weaker than the range of future SN observations on $g_{ap}$ and $g_{a\gamma}$ assuming massless axions. However, this seems unfeasible. The gamma-ray fluence at Earth can schematically be estimated for both sources in order of magnitude as the product $\Phi_\gamma\sim N_a  P_{a\gamma}/4\pi d^2$, where $N_a$ is the total number of axions emitted (for NSMs this should be computed only within the first 10~ms, since axions emitted later do not convert efficiently due to the ejecta), $d$ is the source distance from the Earth. Let us consider each factor separately.

The number of axions $N_a$ for NSMs is about 3-4 orders of magnitude lower than in SNe; since the internal temperature and density conditions are similar, but only the axions emitted within 10~ms from NSMs can convert into photons compared to the 10-seconds signal of SNe, we expect a difference of about 3 orders of magnitude in the emission from the two sources.
This is confirmed by a comparison of Tables~\ref{tab:SN-models} and~\ref{tab:NSM-models}; our one-zone NSM model is close to the SFHo-asymmetric NSM model (see Fig.~\ref{fig:NSM-flux}), which is the most optimistic among the suite we consider. Even for this model, $N_a$ is about 3-4 orders of magnitude lower than the cold and the hot SN models in Table~\ref{tab:SN-models}. Even assuming that NSMs could reach temperatures much higher than our one-zone model, and perhaps much higher than in SNe---although this is not a general finding of NSM simulations---one can still not expect a time-integrated number of axions larger than in SNe. 

The conversion probability $P_{a\gamma}$ is about two orders of magnitude lower for NSMs than for next galactic SN, as can be gathered from Eqs.~\eqref{eq:sn_probability_massless} and~\eqref{eq:nsm_probability_massless}; for the next galactic SN, one could assume optimistically a magnetic field $B_0\sim 1$~kG. On the other hand, the typical distance of a NSM from the Earth is of the order of tens of Mpc, while for our prospects for the QCD axion from a next galactic SN we have adopted a distance of 2~kpc for the optimistic case. Therefore, the factor $1/d^{2}$ introduces an additional geometrical suppression by about 8~orders of magnitude in NSMs compared to the optimistic scenario for SNe. Overall, the gamma-ray fluence at Earth is expected to be about 14~orders of magnitude lower from a typical NSMs than from a next galactic SN. Since the gamma-ray fluence is proportional to $(g_{ap}\times g_{a\gamma})^2$, we expect a corresponding worsening in the sensitivity by about 7 orders of magnitude, which, according to our previous estimate, is too much to reach the QCD axion band.

Putting everything together we get the results shown in Fig.~\ref{fig: NSMNucleonBounds}. The constraints on the combination $g_{ap}\times g_{a\gamma}$ are about 7 orders of magnitude worse than those obtained for SNe; they are not competitive with already existing constraints, not even if we ignore the presence of the ejected material, and they do not probe the QCD axion band. The impact of including such ejecta is stronger compared to the case of pure axion-photon coupling, since in this case the bound on $g_{ap}\times g_{a\gamma}$ depends on the square root of the fluence; thus, since the fluence is reduced by a factor of 100 due to the ejecta, the bound worsens by an order of magnitude.

Our projected reach is about a factor 4 weaker than the recent estimate in Ref.~\cite{Lecce:2025dbz}. The origin of this discrepancy is in their procedure to obtain constraints, namely to require a number of signal events larger than the background ones, which for Fermi-LAT within the observation time and angular window is about 0.08. In reality, in this background-free regime, the number of signal events should rather exceed a threshold of order unity; our threshold is set at 3 events, which roughly explain the discrepancy.

\section{Discussion}\label{sec:discussion}

Motivated by the recent idea, advanced by the Berkeley group \cite{Manzari:2024jns},
of axion-photon conversion in stellar magnetic fields that either persist for some time after a SN collapse or arise in a neutron star merger, we have studied this scenario from multiple perspectives. We have essentially reexamined the entire pipeline, from axion production in a hot and dense nuclear environment to potential detection by gamma-ray satellites. 
Along the way, we have identified numerous issues, both major and minor, that influence this scenario and modify previous constraints. However, the main conclusion persists: a satellite like Fermi-LAT could detect a distinctive signature of QCD axions in a mass range not yet excluded by other methods (Fig.~\ref{fig:sn_projected_nucleons}). This includes axion masses below 100~meV, which are especially difficult to probe with current techniques, although the future IAXO, searching for solar axions, may partly cover this range.

On the other hand, a future neutron-star merger is far less promising. For one, typical distances are too large. Second, the most important conversion region, close to the compact remnant's surface, is rapidly contaminated by ejecta. This suppresses magnetically induced axion-photon conversion, reducing the potential reach for axion couplings by as much as a factor of 100. Although this early quenching of the conversion efficiency was not accounted for in a recent parallel study \cite{Lecce:2025dbz}, the overall conclusion remains the same: NSMs are disfavored as targets for axion searches by this method.

A bird's-eye view of the prospects for axion-photon conversions in astrophysical $B$-fields, for MeV photons, is provided by the axion Hillas plot introduced in Fig.~\ref{fig:hillas}. It illustrates graphically the interplay between $B$-field strength and characteristic length scale in determining the conversion efficiency. This schematic landscape presents a compact perspective on the efficiency across different axion sources, though it does not capture their complementarity or the actual detection prospects, which depend on distance.

Regarding the SN~1987A sensitivity to the axion-photon coupling (production and detection by this interaction alone), we find that conversion in the progenitor's $B$-field does not extend the excluded parameter space, in contrast with earlier findings~\cite{Manzari:2024jns}. The traditional approach relies on the Galactic $B$-field, which we also reexamine, with results broadly consistent with previous studies (Fig.~\ref{fig:sn1987a}). The stellar $B$-field, due to its shorter length scale, shifts sensitivity to higher axion masses. However, its strength and orientation are not known, so one cannot derive a limit, only a putative sensitivity region under generic assumptions.
Even then, the hypothetical sensitivity on $g_{a\gamma}$ is at best comparable to those from CAST and globular cluster observations. All excluded $m_a$--$g_{a\gamma}$--parameters are also covered by axion-photon conversion constraints from other sources. In constrast to the Berkeley group, we have not included a putative loop-induced nucleon interaction, i.e., we have stuck to the traditional definition of an ALP as coupling to photons alone, arguing in Appendix~\ref{app:axion_nucleon} that such couplings are not generic or model-independent. Anyway, even assuming their loop-induced nucleon coupling only doubles the axion flux and improves the limit only minimally.

Concerning axion production in dense nuclear environments, we have omitted the pionic process $\pi^-p\to na$ because we have identified several issues and open questions in the previous approaches. We are particularly concerned with the putative $\pi^-$ abundance and how it relates to the $\pi^-$ dispersion relation, which is inextricably linked with the possible condensation of other pionic states (specifically $\pi^+_s$ and $\pi^0_s$). 
Of course, one can be optimistic and speculate that pions might produce more axions and improve detectability at a future~SN beyond our estimates based on bremsstrahlung emission.

For the latter, we have used a parametric estimate using the picture of fluctuating proton spins, scaled with the measured $np$ cross section, which however may receive large corrections in a nuclear medium. Once more, detailed calculations would be favored only if these were validated against measured quantities, notably elastic $np$ scattering, which factors from bremsstrahlung in the soft limit. Recent calculations use a nuclear potential that fails to reproduce the dominant $np$ interaction channels and rely on a Born treatment. Going reliably beyond an educated dimensional analysis, and in addition include nuclear medium effects, appears to be impossible with simple techniques. Pragmatically, our overall bremsstrahlung emission is surprisingly similar to that of other authors, as we have seen by comparing it for identical numerical SN models.

We have devoted substantial effort on assessing whether detailed numerical SN models provide significantly more information about axion emission than simple estimates based on average properties---particularly when there is no standard model for the proto-neutron star (PNS) due to uncertain input physics and unknown mass. Our analysis shows that only a few parameters are crucial for axion emission. In fact, the distinction between a ``cold'' and a ``hot'' model---spanning a plausible physical range---is already sufficient to capture the essential behavior. Therefore, when detailed integrations over SN models are performed, a good basis for comparison would be similar average quantities to those we have introduced. It would be interesting to examine how these averages vary across a broader suite of numerical models than we have studied here.

A significant latent source of systematic uncertainty lies in the $B$-field conversion between compact source and detector. Except for conversion in the galaxy, we have assumed a dipole structure for the $B$-fields of compact sources or progenitor stars. Conversion essentially begins at a well-defined radius, where the matter density---and consequently the plasma frequency---drops rapidly, lifting the suppression of axion-photon conversion. At larger distances, the $B$-field decreases smoothly as a power law, and the axion-photon conversion probability asymptotically approaches a constant value. This behavior shows no energy-dependent oscillations, in contrast to the CAST experiment, where two sharp $B$-field boundaries lead to such features. We have developed simple power-law scaling formulas for the conversion efficiency, applicable across different scenarios and limits.

Likewise, we have developed simple scaling laws for our limits and sensitivities to the main input assumptions, most notably the inner SN temperature, which shows the scaling of our results with input assumptions.

The idea of using ALP conversion in stellar $B$-fields from compact transient sources is an intriguing approach to searching for QCD axions. However, the devil is in the details, and we have aimed to develop a better understanding of many relevant issues. We were primarily guided by simplicity: substantial previous numerical effort was condensed to a few algebraic estimates that capture parametrically the dependence on the uncertain physics of compact transients. Seemingly more refined treatments often gloss over input uncertainties, offering no real benefit. Our parametric approach makes these dependencies explicit and points to several questions that should be examined in future work.

\section*{Note Added}

While this paper was in the review process, in an extended collaboration, we have come to the conclusion that stripped-envelope supernovae, the spectral types Ibc, are actually the optimal targets for a future search with this method. They have smaller surface radii and large surface $B$-fields and comprise at least one third of all core-collapse events~\cite{Candon:2025sdm}.

\section*{Acknowledgments}

We acknowledge useful correspondence with Sanjay Reddy, and we thank Alessandro Lella, Giuseppe Lucente, Claudio  Manzari, Alessandro Mirizzi, and Ben Safdi for useful comments.
DFGF is supported by the Alexander von Humboldt Foundation (Germany). In Munich, we acknowledge partial support by the German Research Foundation (DFG) through the Collaborative Research Centre ``Neutrinos and Dark Matter in Astro- and Particle Physics (NDM)'', Grant SFB-1258-283604770, and under Germany’s Excellence Strategy through the Cluster of Excellence ORIGINS EXC-2094-390783311. In Padova, we acknowledge support by the Italian MUR Departments of Excellence grant 2023--2027 ``Quantum Frontier'' and by Istituto Nazionale di Fisica Nucleare (INFN) through the Theoretical Astroparticle Physics (TAsP) project. AGM acknowledges the support of FCT - Fundação para a
Ciência e Tecnologia, I.P., with DOI identifiers 10.54499/2023.11681.PEX, and the project 10.54499/2024.00249.CERN. EV acknowledges support from the Italian Ministero dell'Universit\`a e della
Ricerca through the FIS 2 project FIS-2023-01577 (DD n. 23314 10-12-2024, CUP C53C24001460001). This article is based upon work from COST Action COSMIC WISPers (CA21106), supported by COST (European Cooperation in Science and Technology).

\appendix

\section{Is an ALP-nucleon coupling generic?}\label{app:axion_nucleon}

ALP constraints or detection opportunities are discussed under varying assumptions about their interaction channels. Traditionally, an ALP (unlike a QCD axion) has been defined as coupling only to photons, as described by Eq.~\eqref{eq:interactionlagrangian}, with no other interactions. Limits are typically summarized as constraints on $g_{a\gamma}$ as a function of $m_a$ \cite{OHare}. The authors of Ref.~\cite{Manzari:2024jns} argue that this framework is overly conservative, or even inconsistent, since any ALP will generically also couple to nucleons via loop effects. The constraints on $g_{a\gamma}$, shown in their Fig.~1, then actually depend on the implied nucleon coupling.

The observation that loop-induced processes can rival tree-level ones is intriguing, but the quantitative impact depends on the specifics of the UV completion. The traditional notion of a ``minimal ALP'' coupling only to photons concerns the IR behavior, not the UV theory; it assumes interactions with other species vanish at low, not high, energies. It is essential to distinguish whether the photon-only assumption is meant to apply in the IR or the UV. Conventional constraints target IR behavior without assumptions about the UV. Therefore, we consider it phenomenologically consistent, as done in this paper, to derive ALP constraints based solely on the well-defined IR interaction structure.

This approach raises the question of whether UV completions can allow ALPs to avoid sizeable IR nuclear interactions without fine-tuning. But what does this even mean when the theory is already fine-tuned to produce ALP masses much smaller than the QCD axion at the same $f_a$ scale? Moreover, loop-induced couplings are not renormalizable, as they depend logarithmically on the scale at which they are defined.

In any event, loop-induced nuclear processes that remain subdominant in the dense SN environment do not require special fine-tuning. The opposite conclusion was reached in Ref.~\cite{Manzari:2024jns} primarily because they assumed that, above the electroweak scale, the axion would couple only, and with equal strength, to $W^\mu$ and $B^\mu$, the mediators of SU(2) and U(1). The effective couplings to protons and neutrons are then found to be
\begin{eqnarray}
    C_{app}&\sim& +3.5\times 10^{-5} C_{aBB} + 2.1\times 10^{-4} C_{aWW},
    \nonumber\\ 
    C_{ann}&\sim& -5.6\times 10^{-6} C_{aBB}+2.1\times 10^{-4} C_{aWW},
\end{eqnarray}
implying a photon coupling of $C_{a\gamma\gamma}=C_{aBB}+C_{aWW}$. For $C_{aWW}\sim C_{aBB}$, the nucleon coupling is correctly identified as $C_{app}\sim C_{ann}\sim 10^{-4} C_{a\gamma\gamma}$, and this small coupling is enough for nuclear processes to dominate over photonic ones in a SN core.

However, the assumption $C_{aWW}\sim C_{aBB}$ is arbitrary. An equally motivated choice, $C_{aWW}=0$, reduces $C_{app}$ by one and $C_{ann}$ by even two orders of magnitude. Since the emissivity scales quadratically with the couplings, the nucleon-induced processes drop by two to four orders of magnitude, while leaving the photon ones nearly unchanged. As a result, an axion that couples primarily to $B^\mu$ in the UV is dominated by photon-induced processes in the IR. This simple counterexample illustrates the difficulty of applying general naturalness arguments to IR phenomenology without a well-specified UV sector.

\section{Mass suppression of the conversion in a dipolar magnetic field}\label{app:mass_suppression}

Most of the configurations we have considered in the text involve a dipolar magnetic field. In the massless limit, it is straightforward to understand how the conversion works in practice, using the intuitive arguments of Section~\ref{sec:massless_pedestrians}. When the axion has a mass, however, the interference between refraction caused by the mass and by the magnetic field
can produce non-trivial behavior of the conversions, which we now discuss.

We parameterize $\Delta_B=\Delta_{B,0} (R_0/R)^6$ and $\Delta_{a\gamma}=\Delta_{a\gamma,0} (R_0/R)^3$, following the scaling with the magnetic field. In the perturbative regime, the conversion probability is then $P_{a\gamma}=|\mathcal{A}_{a\gamma}|^2$, with
\begin{equation}
    \mathcal{A}_{a\gamma}=\int_{R_i}^{+\infty}dR\, \frac{\Delta_{a\gamma,0}R_0^3}{R^3}\exp\left[-i|\Delta_a| R+i\frac{\Delta_{B,0}R_0^6}{5R^5}\right],
\end{equation}
up to an unimportant phase factor. Here $R_i$ is the radius at which conversion starts; in the absence of ejecta, this can be taken to be 0, since the merger radius is much smaller than the typical distance at which conversion happens, except at very large masses. 

We can make the spatial coordinate dimensionless by $R=(\Delta_{B,0} R_0^6)^{1/5} \eta$, so that the transition amplitude becomes
\begin{equation}\label{eq:amplitude_conversion}
    \mathcal{A}_{a\gamma}=\Delta_{a\gamma,0}R_0^3(\Delta_{B,0}R_0^6)^{-2/5}\int_{\eta_i}^{+\infty}\frac{d\eta}{\eta^3}\exp\left[-i\delta_a \eta+\frac{i}{5\eta^5}\right],
\end{equation}
with $\delta_a=|\Delta_a| (\Delta_{B,0}R_0^6)^{1/5}$. This expression can be additionally brought to the form
\begin{equation}
    \mathcal{A}_{a\gamma}=\frac{\Delta_{a\gamma,0}R_0^3 (\Delta_{B,0}R_0^6)^{-2/5} \delta_a^{1/3}}{2}F\left[\frac{(\Delta_{B,0}R_0^6)^{2/5}}{\delta_a^{1/3}R_i^2},\delta_a\right],
\end{equation}
where
\begin{equation}
    F(x,\delta_a)=\int_0^x \exp\left[-i\delta_a^{5/6}\left(\frac{1}{\sqrt{y}}-\frac{y^{5/2}}{5}\right)\right]dy.
\end{equation}
For $x\to 0$, an integration by parts reveals that $F(x,\delta_a)\to -2i x^{3/2} e^{-i\delta_a^{5/6}/\sqrt{x}}/\delta_a^{5/6}$; for $x\to +\infty$, by a subtraction we find that $F(x,\delta_a)\to F(+\infty,\delta_a)-2i x^{-3/2}e^{i\delta_a^{5/6} x^{5/2}/5}/\delta_a^{5/6}$. The value of $F(+\infty,\delta_a)$ can be approximately obtained using the steepest descent method; to do so, one must first fix a branch for the square-root function in the integrand. It is convenient to choose it such that the branch cut is along the line $\mathrm{Re}(x)<0$, $\mathrm{Im}(x)=0$. The exponent of the integrand function has three points of stationary phase; the point $y=-1$ lies on the branch cut, so we consider the other two points $y=e^{-i\pi/3}$ and $y=e^{i\pi/3}$. The point $y=e^{-i\pi/3}$ is a local maximum along the direction of the line $y=e^{-i\pi/3}+s e^{i\pi/3}$, with $s$ from $-\infty$ to $+\infty$. In principle, we could deform the integration contour to match this line; however, for $s\to -\infty$, the exponent of the integrand function is dominated by the term $i \delta_a^{5/6} y^{5/2}/5$ and diverges. Therefore, we consider the other point $y=e^{i\pi/3}$, which is a maximum along the direction of the line $y=e^{i\pi/3}+s e^{i\pi/6}$; we easily see that the integrand function drops to 0 for $s\to \pm \infty$, so we can deform the contour (previously coinciding with the positive real axis) into this line. By expanding around this point and performing the saddle point integration, we finally find
\begin{equation}
    F(+\infty,\delta_a)\simeq \exp\left[\frac{6\delta_a^{5/6}}{5} e^{2i\pi/3}+\frac{i\pi}{6}\right]\sqrt{\frac{4\pi}{3 \delta_a^{5/6}}}.
\end{equation}
This expression approximates very precisely the numerical result of the integral. 

Thus, overall we find that for $R_i\to\infty$ ($x\to 0$) the conversion probability is approximated by
\begin{equation}\label{eq:conversion_large_radii}
    P_{a\gamma}\simeq \frac{\Delta_{a\gamma,0}^2 R_0^6}{\Delta_a^2 R_i^6}.
\end{equation}
Instead, for $R_i\to 0$ ($x\to \infty$), the function $F(x,\delta_a)\to F(+\infty,\delta_a)-2ix^{-3/2} e^{i\delta_a^{5/6} x^{5/2}/5}/\delta_a^{5/6}$ can be dominated either by the first or the second term. Therefore,
\begin{equation}\label{eq:conversion_small_radii}
    P_{a\gamma}\simeq \frac{\Delta_{a\gamma,0}^2R_i^6}{\Delta_{B,0} R_0^6}\mathrm{max}\left[1,\frac{\pi R_0^7 \Delta_{B,0}^2\exp\left[-\frac{6}{5}R_0 \left(|\Delta_a| \Delta_{B,0}^5\right)^{1/6}\right]}{3R_i^6 \left(|\Delta_a| \Delta_{B,0}^5\right)^{1/6}}\right],
\end{equation}
if we assume that approximately only the largest of the two is relevant.

\section{Some practical details of numerical supernova models}

\subsection{Nucleon distribution functions}

\label{app:distribution}

When using numerical models to evaluate emission rates of feebly interacting particles, and especially to estimate degeneracy effects, one needs to make sure that the assumptions are consistent with those made within the given SN model. For example, to evaluate Eq.~\eqref{eq:Fdegpp}, one needs nucleon phase-space densities that are consistent with the used EOS. In the Garching models, the necessary information is tabulated in the form of the effective nucleon mass $m_N^*$ and a scalar potential $U_N$, caused by the background medium, for $N=n$ or $p$. The phase-space density takes the form 
\begin{equation}
    f_N(p)=\left(\exp\left[\frac{U_N+K_N(p)-\mu_N^*}{T}\right]+1\right)^{-1},
\end{equation}
where $K_N(p)$ is the nucleon kinetic energy as a function of momentum $p$, whereas $\mu_N^*$ is the tabulated nucleon chemical potential that does not include the mass. For other particles, notably muons, the relativistic chemical potentials are tabulated that include the mass.

The SFHo equation of state (after Steiner, Fischer, and Hempel \cite{Steiner:2012rk}) is based on a relativistic mean field approach, where the kinetic energy is
\begin{equation}
    K_N(p)=\sqrt{(m_N^*)^2+p^2}-m_N^*
\end{equation}
with the tabulated values of $m_N^*$. For the effective nucleon masses,
$m_n^*-m_p^*=\Delta m_N=1.2933$~MeV, i.e., their effective mass difference corresponds to the vacuum one, but the effective masses themselves are much smaller than in vacuum and depend on density. Beta equilibrium implies the relation $\hat\mu= \mu_e-\mu_{\nu_e}= \mu_\mu-\mu_{\nu_\mu}= \mu_n^*-\mu_p^*+\Delta m_N$ between the chemical potentials and is reflected in this form in the tabulated values.

On the other hand, the Lattimer and Swesty (LS) EOS \cite{Lattimer:1991nc} uses the vacuum nucleon masses and specifically $m_n=939.57$~MeV and $m_p=938.27$~MeV. Moreover, it uses the nonrelativistic expression for the kinetic energy
\begin{equation}
    K_N(p)=\frac{p^2}{2m_N}
\end{equation}
with the vacuum mass for each nucleon. In the LS220-20.0 model, nucleon degeneracies are smaller both because this more massive NS becomes hotter, but also because the used effective nucleon masses, being the vacuum ones, are larger.

\subsection{Gravitational lapse factor}

\label{app:lapse}

Axions reaching us from the interior of a SN core suffer a considerable redshift. In a one-zone model of homogeneous density $3\times10^{14}~{\rm g}~{\rm cm}^{-3}$ and mass $1.4\,M_\odot$, the NS radius is $R_{\rm NS}=13.06~{\rm km}$ and the Newtonian gravitational potential for $R<R_{\rm NS}$ is
\begin{equation}
     U(R)=U_{\rm NS}\,\frac{3-(R/R_{\rm NS})^2}{2}
     \quad\hbox{with}\quad
     U_{\rm NS}=-\frac{G_{\rm N}M_{\rm NS}}{R_{\rm NS}}=-0.159.
\end{equation}
The gravitational redshift is approximately $z(R)=-U(R)$ so that the reduction of axion energies is a 20\% effect. Moreover, the locally calculated emission rate, viewed from a distant observed, is likewise reduced so that also the total number of emitted axions, calculated from local conditions, is reduced by the same factor.

In the Garching models, general relativistic effects are accounted for in an approximate way. The practical consequences for our calculation is that the spatial metric is that of flat Minkowski space, whereas temporal effects are included in the ``gravitational lapse'' factor that is tabulated for every radial shell and corresponds to the inverse redshift factor $({\rm lapse})=(1+z)^{-1}$ for both energies and emission rates. The time sequence of models is represented by time stamps for a distant observer. Let us assume that the differential axion emission rate $\dot N'_a(t,E_z)=d\dot N_a(t,E_z)/dE_zdV$ based on local conditions is given by a formula of the type Eq.~\eqref{eq:Primakoff-emission-rate}, where $E_z$ is the local energy at a place with redshift $z$, whereas $t$ is the clock time of the distant observer. Then the emission rate relevant for a distant observer is
\begin{equation}
    \frac{d\dot N_a}{dEdV}=\dot N'_a\bigl[(1+z)E\bigr],
\end{equation}
that is, for an observed energy $E$ we need to evaluate the local emission rate at the blue-shifted energy $E_z=(1+z)E$. The energy-integrated emission rate is then found to be $\dot N_a=\int_0^\infty dE N'_a\bigl[(1+z)E\bigr]=(1+z)^{-1}\int_0^\infty d E_z N'_a(E_z)$. Therefore, the energy-integrated number of axions per proton emitted per unit time observed from far away is $(1+z)^{-1}$ times the one calculated from local parameters. The axion energy loss rate suffers two lapse factors in that the total number of emitted axions is reduced by one lapse factor and the average energy by another.

\section{Primakoff emission}
\label{app:Primakoff}

\subsection{More details on the Primakoff process}

The photon-to-axion Primakoff conversion rate, assuming recoil-free scattering on protons, was given in Eq.~\eqref{eq:Primakoff-conversion-rate} together with the screening scale in Eq.~\eqref{eq:Screening-Scale} that represents the proton-proton spatial anticorrelation caused by their Coulomb repulsion. In this form, Primakoff production has been used for a long time in the literature. We always consider axions that are essentially massless relative to thermal energy scales in the medium. Otherwise, the axion-photon mass difference and concomitant momentum difference also moderates the Coulomb divergence; the corresponding expression for the term in square brackets in Eq.~\eqref{eq:Primakoff-conversion-rate} was provided, for example, in Refs.~\cite{DiLella:2000dn, Caputo:2022mah}. 

Photons suffer dispersion and are not kinematically  massless. The relevant scale is the plasma frequency, here dominated by relativistic degenerate electrons, so that $\wP^2=(4\alpha/3\pi)\,\mu^2$ in terms of the electron chemical potential \cite{Braaten:1993jw}. Assuming that the number of protons essentially equals that of electrons implies $n_p=n_e=p_{\rm F}^3/3\pi^2\simeq \mu^3/3\pi^2 $ and therefore \begin{equation}\label{eq:plasma-screening-ratio}
    \wP^2/k_{\rm S}^2\simeq T/\mu.
\end{equation}
Due to this hierarchy, the screening scale is indeed far more important than the photon plasma mass for the moderation of the Coulomb divergence. Typical estimates in the relevant regions of a SN core are $\wP\simeq10$~MeV and $k_{\rm S}\simeq 30$~MeV.

Photon energies are on average $3T$, so the smallness of $\wP$ relative to $T$ implies that photon dispersion is a small effect that matters only for the soft part of the spectrum. We will be mostly interested in $E>10~{\rm MeV}$, so it is safe to ignore photon dispersion. On the other hand, including it consistently would entail a significant number of complications, such as the role of longitudinal plasmons, the appropriate wave function renormalization, and the generally nontrivial plasmon dispersion relation. Notice, for example, that in a relativistic plasma, transverse excitations have $m_\gamma^2=\wP^2$ for small $k$, and $m_\gamma^2=\frac{3}{2}\wP^2$ asymptotically for large $k$. Moreover, for the soft part of the spectrum, the proton velocity distribution and its collective nature in the form of plasma oscillations cannot be ignored. For momentum transfers small compared with $k_{\rm S}$, Primakoff conversion can also be pictured as the interaction of a transverse photon with a longitudinal plasmon \cite{Raffelt:1987np}, imprinting itself on the low-energy axion distribution. In the massless photon and axion limit, the expression for $\Gamma_{\gamma\to a}$ given in Eq.~(S5) of Ref.~\cite{Manzari:2024jns} agrees with our Eq.~\eqref{eq:Primakoff-conversion-rate}. However, their expansion for small masses in terms of the parameter $\xi=(\wP^2+m_a^2)/E^2$ is not selfconsistent, since if one expands the momenta to lowest order in $m_{a,\gamma}^2$, also the overall final expression makes sense only up to this order. In any case, if one wishes to include the masses, an expansion is not useful because the main effect occurs precisely for soft $E$ near $\wP$, where the expansion fails, and the effects mentioned above should be accounted for. In view of their smallness throughout the majority of the relevant ranges, we neglect photon plasma mass effects.

There are several 10\%-level corrections to our simplified Primakoff prescription. We ignore proton recoils that anyway would be small for near-forward scattering. The final-state proton is Pauli blocked and, in the recoil-free approximation, has the same occupation number as the initial one. Therefore, the average rate is reduced by an approximate factor given in Eq.~\eqref{eq:Fdegpp}, which for typical SN conditions implies a reduction by some 20\%, as revealed by the numerical results shown in Table~\ref{tab:SN-models}. 

On the other hand, additional contributions come from other charged particles. Besides protons, there are higher-$Z$ nuclear clusters which make a larger contribution because of a $Z^2$ coherent enhancement. Based on the Garching models, the average proton and neutron abundances do not add up exactly to one (Table~\ref{tab:SN-models}), providing an average enhancement of perhaps a few percent, but the details depend on the equation of state and how it deals with such clusters. In Ref.~\cite{Caputo:2021rux} it was argued that for the effective charged nucleon density one should use $1-Y_n$ from the tabulated SN models instead of $Y_p$, roughly correcting for the ``proton deficit'' revealed by $Y_p<1-Y_n$. However, we here do not worry about this small difference and use the tabulated $Y_p$.

Concerning negatively charged particles, we have already argued that the contribution of electrons is suppressed by their degeneracy, but an exact estimate is not available. In addition, muons exist with a considerable abundance, typically with 15--20\% of the proton abundance according to Table~\ref{tab:SN-models}. Degeneracy effects are small and so muons would contribute somewhat to screening and as Primakoff targets, although the impact of their semi-relativistic kinematics would have to be determined. Overall, they might provide a 10\%-level enhancement of the Primakoff rate.

A similar upward correction could derive from a thermal $\pi^-$ population that is not included in most numerical SN models and that we will also omit. However, their abundance likely is not much larger, and probably much smaller, than the muon abundance. Therefore, optimistically, they might contribute another 10\%-level enhancement of Primakoff emission. The inclusion of pions would also cause an increase of the proton fraction  \cite{Fore:2019wib} and in this way indirectly enhance the Primakoff rate, again on the 10\% level.

Of course, in view of all other uncertainties, these are minor issues. Since proton degeneracy causes a reduction by some 20\%, whereas the muon and possibly pion contributions together might provide a similar enhancement, we finally resolve to using the elementary prescription of Eq.~\eqref{eq:Primakoff-conversion-rate} without degeneracy and without negatively charged particles. The result should be a reasonable proxy for the true Primakoff emission rate on perhaps the overall 20\% level of accuracy.

\subsection{Electro-Primakoff emission}

\label{sec:Electro-Primakoff}

In Primakoff production, ALPs are sourced by $\bE\cdot\bB$ fluctuations provided by the magnetic fields of propagating electromagnetic fields (photons) and the electric fields of charged particles in the medium. However, there are also $B$-fields associated with moving charges, or, in the Weizs\"acker-Williams picture, the electromagnetic fields of moving charges can be interpreted in terms of a virtual photon density. In other words, axions can be emitted by two interacting virtual photons. For solar conditions, this electro-Primakoff effect was shown to be vastly subdominant due to the small electron velocities that imply small $B$ fields associated with moving charges \cite{Raffelt:1985nk}.

In a SN core, on the other hand, electrons are relativistic and much more abundant than photons, so the Weizs\"acker-Williams photons need not be less abundant than real ones. Taking nuclear density at a temperature of 30~MeV as an example, the number of thermal photons per nucleon is $Y_\gamma= 4.7\times10^{-3}$, to be compared with electrons of perhaps $Y_e=0.12$, meaning that electrons are some 25 times more abundant than photons. Since the abundance of virtual photons suffers from $\alpha=1/137$, being sourced by relativistic electrons, one may expect that their contribution remains subdominant, not to mention electron degeneracy. As this question has not been previously addressed, we determine the corresponding emission rate here, even though the final result will be small.

We denote by $P_1^\mu$ and $P_2^\mu$ the four-momenta of the initial and final electron, respectively, by $K^\mu$ the axion four-momentum, and by $Q^\mu$ the four-momentum exchanged with the proton field, which is purely space-like in the laboratory frame. The corresponding spatial momenta are denoted by bold-faced non-capital letters (e.g.\ $\bq$ for $Q^\mu$) and their modules are denoted by non-capital letters (e.g.\ $q=|\bq|$). Other than protons, which are taken to be static, all particles are assumed to be ultra-relativistic. We still need to account for the corrections induced by the electron thermal mass (around 10~MeV in a SN core), since massless electrons can radiate photons on-shell collinearly, which can then convert via Primakoff in the proton field. This would lead to a divergence of the electro-Primakoff rate, in turn caused by the unphysical double-counting of the on-shell photons produced by the electrons. Therefore, the lowest-order corrections due to the electron effective mass must be kept, but only in the denominators of the internal propagators, which lead to strongly forward-peaked angular distributions. Similarly, also the thermal mass of the photon $m_\gamma\simeq\wP$ must be kept, which is also around 10~MeV.

The squared matrix element for the electro-Primakoff process, summed over the spins of the electrons, is
\begin{equation}
|\mathcal{M}|^2=\frac{16\pi^2\alpha^2 g_{a\gamma}^2}{q^2 (q^2+\kS^2)}\varepsilon_{\mu\nu\alpha 0}\varepsilon_{\theta\sigma\rho 0}K^\mu Q^\alpha K^\theta Q^\rho \frac{\mathrm{Tr}\bigl[\gamma^\nu \slashed{P}_1\gamma^\sigma \slashed{P}_2\bigr]}{\left[(P_2-P_1)^2-m_\gamma^2\right]}.
\end{equation}
Here $\kS$ is the screening scale, defined in Eq.~\eqref{eq:Screening-Scale}, representing the proton anticorrelations caused by their Coulomb repulsion. In terms of the spatial momenta, we find
\begin{equation}
    |\mathcal{M}|^2=\frac{16\pi^2\alpha^2 g_{a\gamma}^2}{q^2(q^2+\kS^2)\left[(P_2-P_1)^2-m_\gamma^2\right]}
    \bigl[2\bk\cdot(\bp_1\times\bq)\,\bk\cdot(\bp_2\times \bq)+(p_1\cdot p_2)(k^2 q^2-(\bk\cdot\bq)^2)\bigr].
\end{equation}
The total number of axions produced per unit volume and time can now be written as
\begin{eqnarray}
    \frac{d\dot{N}_a}{dVdk}&=&\frac{k^2 n_p}{(2\pi)^3 2k}\int d\Omega_\bk \frac{d^3\bp_1}{(2\pi)^3 2p_1} \frac{d^3\bp_2}{(2\pi)^3 2p_2} \frac{d^3\bq}{(2\pi)^3} \,(2\pi)^4 \delta^{(4)}(P_1-P_2-K-Q)
    \nonumber\\[1.5ex]
    &&\kern5em{}\times f(\bp_1)\left[1-f(\bp_2)\right] |\mathcal{M}|^2,
\end{eqnarray}
where $f(\bp)$ is the electron Fermi-Dirac distribution and $\Omega_\bk$ the solid angle associated with the axion direction.

To proceed further, we need to understand the different tendencies associated with the factors in the integral. In the strongly degenerate limit $\mu\gg T$, relevant for SNe, the Fermi-Dirac distribution functions are strongly peaked in the region in which $p_1, p_2\sim \mu\gg k\sim T$. The denominator $(P_2-P_1)^4$ leads to a strongly peaked distribution with $\bp_2$ nearly aligned with $\bp_1$; in fact, if we call $x$ the cosine of the angle between the two, after imposing energy conservation $E_2=E_1-k$, where $E_1$ and $E_2$ are the energies of the electron before and after emission (we need to keep here the electron mass correction), we find to lowest order in the effective electron mass $m_e$
\begin{eqnarray}
    -(P_2-P_1)^2+m_\gamma^2&\simeq&2p_1(p_1-k)\left[1-x+\frac{m_e^2 k^2}{2p_1^2 (p_1-k)^2}+\frac{m_\gamma^2}{2p_1(p_1-k)}\right]
    \nonumber\\[1ex]
    &\simeq&2p_1^2\left[1-x+\frac{m_\gamma^2}{2p_1^2}\right].
\end{eqnarray}
Notice that the effect of the electron mass is completely negligible compared with that of the photon mass, because of the suppression factor $k^2/p_1^2$ appearing in front of $m_e^2$. Thus, the emission is dominated by strongly collinear scattering, with the electron being deflected by $\delta\theta\sim m_\gamma/\mu\sim \sqrt{\alpha} $, where we have used $m_\gamma\sim \sqrt{\alpha}\mu$. Because this factor appears quadratically in the denominator, this is by far the strongest tendency in the integral. We can therefore use in all the other terms of the integral an expansion for $x\simeq 1$. 

The additional factor in the denominator is $q^2(q^2+\kS^2)$. If we neglect $m_e$ and the axion momentum $k\ll p_1$ altogether, we find $q^2\simeq 2p_1^2(1-x)$. However, we need to be more careful in the expansion and keep terms of higher order. To do so, let us introduce a parameterization for all the momenta; specifically, we write $\bp_1=p_1(1,0,0)$, so that the $x$-axis is aligned with the initial electron momentum. The axion momentum is $\bk=k(y,\sqrt{1-y^2}\cos\phi,\sqrt{1-y^2}\sin\phi)$. Finally, for the final momentum of the electron, we use energy conservation and expand to lowest order in the electron mass to obtain
\begin{equation}
    \bp_2=\left[p_1-k-\frac{k m_e^2}{2p_1 (p_1-k)}\right](x,\sqrt{1-x^2},0).
\end{equation}
For $x\sim 1$ we may expand, and in the coefficients of the expansion we can keep $k\ll p_1$, so that
\begin{equation}
    q^2=2p_1^2(1-x)+2p_1 k \sqrt{1-x}\sqrt{2(1-y^2)}\cos\phi+2k^2(1-y)+\frac{m_e^4 k^2}{4(k-p_1)^2 p_1^2}.
\end{equation}
The typical values of $q^2\sim\kS^2$; on the other hand, the terms containing the small difference $1-x$ here are of the order of $p_1^2(1-x)\sim m_e^2 k^2/p_1^2\ll \kS^2$. This means that we can neglect here all terms including the small difference $1-x$, as well as the terms depending on $m_e$ which are also very small, and use the simplified expression
\begin{equation}
    q^2=2k^2(1-y).
\end{equation}
Overall we find
\begin{equation}
    |\mathcal{M}|^2=16\pi^2\alpha^2 g_{a\gamma}^2\frac{4p_1^4 k^2(1-x)(1-y^2)\sin^2\phi}{2k^2(1-y)[2k^2(1-y)+\kS^2]4p_1^4(1-x+\delta^2)^2},
\end{equation}
where $\delta=m_\gamma/\sqrt{2}p_1$. 

We can now write the full expression as 
\begin{equation}
    \frac{d\dot{N}_a}{dVd|\bk|}=\frac{n_p k}{(2\pi)^8 8}\int p_1 dp_1 p_2 dp_2 f(p_1)\left[1-f(p_2)\right]\delta(p_1-p_2-k)\mathcal{J},
\end{equation}
where
\begin{equation}
    \mathcal{J}=8\pi^2\int dx dy d\phi |\mathcal{M}|^2=\frac{32g_{a\gamma}^2 \pi^5 \alpha^2}{k^2} I_x I_y
\end{equation}
and
\begin{equation}
    I_y=\int \frac{1+y}{1-y+\frac{\kS^2}{2k^2}}dy=\left(2+\frac{\kS^2}{2k^2}\right)\log\left[\frac{\kS^2+4k^2}{\kS^2}\right]-2.
\end{equation}
Instead, the integral over $x$
\begin{equation}
    I_x=\int \frac{1-x}{(1-x+\delta^2)^2}dx\simeq \int \frac{dx}{1-x+\delta^2}
\end{equation}
should not be done up to $x=-1$, because it is bound from below from the condition of validity of our approximations that $p_1^2(1-x)\ll\kS^2$. Thus, we may cut off the integral from below at such a typical value with logarithmic precision, so that
\begin{equation}
    I_x=2\log\left[\frac{\sqrt{2} \kS}{m_\gamma}\right].
\end{equation}
We can now perform the integral over the energy $p_1$ and $p_2$ easily, since we note that $p_1$ can be replaced everywhere with $\pF=\mu$, the Fermi momentum of the electrons, except in the distribution functions. Using the result
\begin{equation}
    \int dp_1 dp_2 f(p_1) [1-f(p_2)]\delta(p_1-p_2-k)=\frac{T}{e^{k/T}-1}\log\left[\frac{e^{k/T}+1}{e^{-k/T}+1}\right]=\frac{k}{e^{k/T}-1},
\end{equation}
we are finally led to
\begin{eqnarray}
     \frac{d\dot{N}_a}{dVdk}&=&\frac{n_p \alpha^2 g_{a\gamma}^2 \mu^2}{32\pi^3}\frac{1}{e^{k/T}-1}
     \nonumber\\[1ex]
     &&\kern5em{}\times\log\left[\frac{\sqrt{2}\kS}{m_\gamma}\right]\left[\left(2+\frac{\kS^2}{2k^2}\right)\log\left[\frac{\kS^2+4k^2}{\kS^2}\right]-2\right].
\end{eqnarray}
As estimated in Eq.~\eqref{eq:plasma-screening-ratio}, approximately $\kS^2/m_\gamma^2\simeq \kS^2/\wP^2\simeq \mu/T$ so that $\log(\sqrt2 \kS/m_\gamma)\simeq\frac{1}{2}\log(2\mu/T)$.
By replacing $k\to E$ and comparing with Eq.~\eqref{eq:Primakoff-emission-rate}, we see that the ratio of electro-Primakoff and Primakoff emissivity is
\begin{equation}
    \frac{\left(\frac{d\dot{N}_a}{dVdE}\right)_{\rm e-Prim}}{\left(\frac{d\dot{N}_a}{dVdE}\right)_{\rm Prim}}=\frac{\alpha \mu^2}{4\pi E^2}\log\left(\frac{2\mu}{T}\right).
\end{equation}
While the $(\mu/E)^2$ factor, for typical energies $E\sim 3T$, is larger than 1 (usually in SN we have $\mu\sim 6$--$7\,T$) and tends to enhance the electro-Primakoff in comparison with Primakoff due to the vast amount of electrons in comparison to photons, the additional factor $\alpha$ suppresses it. In the end, for $E\sim 3T$, the pre-logarithmic factor is of the order of $10^{-2}$, making electro-Primakoff at most a few percent effect.

\section{Nucleon-nucleon bremsstrahlung}
\label{app:Bremsstrahlung}

We here motivate the adopted parametric form Eq.~\eqref{eq:emission_rate_bremsstrahlung} of the bremsstrahlung emission rate by protons interacting with the nuclear background medium. In the long-wavelength approximation, defined by ignoring the axion momentum transfer to the nucleons, the axion number emissivity per unit mass can be expressed in the general form, according to Eq.~(4.21) of Ref.~\cite{Raffelt:1996wa},
\begin{equation}\label{eq:bremsstrahlung-form}
    \frac{d\dot{N}_a}{dE dM}=\left(\frac{C_{ap}}{2f_a}\right)^2\, \frac{Y_p}{m_u}\,\frac{E^3}{4\pi^2}\,S_\sigma(-E),
\end{equation}
where $Y_p$ is the proton fraction per nucleon and $m_u=1.661\times10^{-24}~{\rm g}$ the atomic mass unit. The dynamical proton spin structure function $S_\sigma(\omega)$ is the Fourier transform of the spin-spin autocorrelation function, where a negative energy transfer $\omega$ means energy lost by the medium. Absorption of axions with positive energy $E$ is accounted for by  $S_\sigma(+E)$ and detailed balance implies $S_\sigma(\omega)/S_\sigma(-\omega)=e^{\omega/T}$. 

All nuclear-physics issues are contained in the function $S_\sigma(\omega)$ that can be written in the general parametric form
\begin{equation}
    S_\sigma(\omega)=\frac{\Gamma_\sigma}{\omega^2+\Gamma_\sigma^2/4}\,
    \frac{2}{e^{-\omega/T}+1}\,s(\omega/T),
\end{equation}
where $s(x)$ is an even function of its argument and normalized to $s(0)=1$. The second factor is also unity at $\omega=0$ and represents detailed balance. If $s(x)=1$ everywhere, one finds the normalization $\int_{-\infty}^{+\infty} d\omega\,S_\sigma(\omega)=2\pi$, actually the same that follows from the first factor alone. This normalization would have to hold even for the true $s(x)$ if there were no spin-spin correlations between different protons, i.e., this sum rule constrains $s(x)$.

In the soft limit ($\omega\to 0$) and ignoring multiple scattering, one finds $S_\sigma(\omega)=\Gamma_\sigma/\omega^2$, and then $\Gamma_\sigma$ has the interpretation of the proton spin fluctuation rate. In a perturbative calculation, the $\omega^{-2}$ divergence derives from the nucleon propagator, but also appears in classical bremsstrahlung calculations. In the soft limit, the elastic collision rate and the radiation process factorize and $\Gamma_\sigma$ is uniquely related to the proton elastic scattering rate and can be extracted from nuclear scattering data, as was done in Ref.~\cite{Hanhart:2000ae} for the case of $nn$ scattering. Unfortunately, the authors report their results only in the form of a reduction factor relative to OPE rather than an absolute value, but in any event, the OPE result in the $nn$ system is roughly a factor of 5 too large. On the other hand, this finding only applies in the soft limit, i.e., the function $s(x)$ remains unknown in this approach.

Instead of our chosen detailed-balance factor, sometimes $e^{\omega/T}$ for negative $\omega$ and $1$ for positive $\omega$ was used, and then also fulfills detailed balance and has the same value at $\omega=0$, but is not differentiable and provides a different result for $\int_{-\infty}^{+\infty} d\omega\,S_\sigma(\omega)$ for $s(x)=1$. Another possible form follows from interpreting the dynamical structure function as the imaginary part of the spin susceptibility, suggesting $-x/(e^{-x}-1)$ with $x=\omega/T$, which has the same value and derivative at $x=0$ as our form and also fulfills detailed balance. One disadvantage is that $\int_{-\infty}^{+\infty} d\omega\,S_\sigma(\omega)$ now diverges, although, of course, the true behavior depends on the unknown function $s(x)$. Beyond linear expansion at $\omega=0$, the behavior follows from the nuclear interaction potential, modified by the nuclear medium, and depends also on the modified nucleon dispersion relation and degeneracy.

In the long-wavelength approximation, our representation of Eq.~\eqref{eq:bremsstrahlung-form} is general, but purely parametric in that we need to make assumptions about the function $s(x)$ and the parameter $\Gamma_\sigma$. By definition, $s(x)$ is even and $s(0)=1$. One would assume that the f-sum, the integral $\int_{-\infty}^{+\infty} d\omega\,\omega\, S_\sigma(\omega)$, should not diverge \cite{Sigl:1995ac}, suggesting a decreasing function of $x$, whereas the OPE result provides an increasing function. Without more information, we take $s(x)=1$ everywhere as presumably it will not strongly vary in the relevant range of energy transfers, and so we finally adopt Eq.~\eqref{eq:emission_rate_bremsstrahlung}. In this form, the emissivity is determined entirely by a single parameter, the spin fluctuation rate $\Gamma_\sigma$. Our simple estimate for $\Gamma_\sigma$ was given in Eq.~\eqref{eq:Gamma_sigma} in the main text.

While $\Gamma_\sigma$ can be derived, in principle, from nucleon scattering data to describe soft bremsstrahlung, it is also the soft limit where multiple-scattering modifications become important because the emission process is interrupted by further collisions and one cannot model the emission process as a sequence of uncorrelated events. This effect is accounted for by $\Gamma_\sigma^2/4$ in the denominator of Eq.~\eqref{eq:emission_rate_bremsstrahlung}, although one cannot expect this prescription to be suitable for a precision calculation when $\Gamma_\sigma$ itself is not small. In the OPE approximation, where $\Gamma_\sigma\gg T$, large multiple-scattering effects would have suppressed the bremsstrahlung rate so that the OPE estimate, after including this suppression, did not vastly overestimate axion emission \cite{Raffelt:2006cw}. The insight that more realistically $\Gamma_\sigma\sim T$ implies that multiple scattering is only a small correction, less important than nucleon degeneracy. Based on Eq.~\eqref{eq:emission_rate_bremsstrahlung}, the total number of emitted axions is reduced by some 15\% if $\Gamma_\sigma=T$, in agreement with the findings of Ref.~\cite{Carenza:2019pxu}. In other words, keeping the multiple-scattering denominator is mostly a vestige of the OPE epoch of axion emission calculations.

In the absence of a direct comparison with actual nuclear scattering data, we do not believe that sophisticated calculations are more reliable than our approach, especially when relying on the Born approximation and semi-phenomenological nuclear potentials dealing only with a part of the nuclear interactions (e.g.~the long-range component in the OPE case). Of course, even comparisons with nuclear scattering data do not directly reveal the modifications introduced by the dense nuclear environment or the semi-relativistic motion of nucleons. In this sense, the axion emission rate from a supernova core, as derived by any method in the literature, is ultimately not more reliable than an ``educated dimensional analysis,'' as previously conceded in Ref.~\cite{Raffelt:2006cw}. A parametric implementation, informed by experimental data, offers the advantage of simplicity, reproducibility, and a flexible modeling of the uncertainties, by encoding them in a few intuitive parameters.

\section{Pion conversion}
\label{app:PionConversion}

In our study, we have omitted the pionic axion production process, despite its recent popularity within the axion community, due to the significant uncertainties and even contradictions present in the existing literature. Below, we outline our primary concerns, for which we have not identified straightforward solutions.

\subsection{Pions in nuclear matter}

The role of pions in SNe remains poorly understood. The possibility of pion condensation in dense nuclear matter dates back to the early work by Migdal~\cite{Migdal:1972zz, Migdal:1971cu}. In nuclear matter, this effect is often identified with the formation of a condensate of negatively charged pions, as originally proposed by Sawyer~\cite{Sawyer:1972cq} and Scalapino~\cite{Scalapino:1972fu}, but the situation is considerably more complex. As discussed in detail in Ref.~\cite{Migdal:1990vm}, $\pi^-$ condensation is disfavored by the large energy shift due to s-wave pion-nucleon refraction. This energy shift, which applies to zero-momentum pions, is quantified, e.g., in Ref.~\cite{Baym:1974vzp}, and is of the order of 40--$50\,\mathrm{MeV}$ at a density of $3\times 10^{14}\,\mathrm{g/cm}^3$. A recent estimate from chiral perturbation theory~\cite{Fore:2023gwv} finds even larger energy shifts, disfavoring $\pi^-$ condensation even further.

Meanwhile, however, as discussed in Ref.~\cite{Migdal:1990vm}, the nuclear medium may be most unstable to excitations of the positive pionic field, usually denoted as $\pi^+_s$ . These excitations do not correspond to in-vacuum pion fields, but rather to zero-sound waves with the same quantum numbers. A $\pi^+_s$ condensate would have the same effect as, e.g., the Peierls instability in a lattice, leading to a spatially periodic restructuring of the nucleons. A similarly unstable field is associated with the $\pi^0$ quantum numbers and is denoted as $\pi^0_s$. 

At even larger densities, an instability involving the negatively charged pions might appear, not through the direct $\pi^-$ production via $n\to p \pi^-$---the numerical calculations reviewed in Ref.~\cite{Migdal:1990vm} all agree that this is impossible primarily due to the s-wave energy shift and nucleon-nucleon repulsion---but rather through $\pi^+_s \pi^-$ pair production. However, to firmly establish whether this instability truly appears, one should look at the instability conditions for the medium already including the $\pi^+_s$, that forms at lower densities inducing $\pi\pi$ self-interactions, as emphasized in Ref.~\cite{Migdal:1990vm}. 

One may think that it is consistent to neglect the possibility of pion condensation, and include purely the effects of thermal pions. However, the neglected condensate comes back to haunt us even in this case, as the abundance of thermal pions depends on their dispersion relation. The latter is strongly renormalized by the nuclear medium; in fact, the presence of condensates shows up precisely in the alteration of the dispersion relation; for example, for the $\pi^-$, at sufficiently large densities the dispersion relation might even acquire an effective Landau mass which turns negative, signaling the $\pi^+_s \pi^-$ instability that we have discussed above (see Fig.~10.6 in Ref.~\cite{Migdal:1990vm}). Such peculiar dispersion relations are of course unphysical because, first, they would be altered by the neglected $\pi\pi$ interaction due to the preceding formation of the $\pi^+_s$ condensate, and second, they signal an instability and therefore are inconsistent. 

Notice that according to Ref.~\cite{Migdal:1990vm}, the $\pi^+_s$ condensation instability occurs around nuclear saturation density, below that of a SN core. Therefore, determining the dispersion relation without accounting for the $\pi^+_s$ (and in principle also $\pi^0_s$) condensate formation may not be consistent. The most recent studies of the pionic dispersion relation in Refs.~\cite{Fore:2019wib, Fore:2023gwv} do not consider $\pi^+_s$ condensation, and reach somewhat opposite conclusions. In Ref.~\cite{Fore:2019wib}, the dispersion relation was adjusted to match the thermal abundance obtained by the virial expansion, but the very small energy shift assumed at zero momentum seems to be in conflict with early estimates. In Ref.~\cite{Fore:2023gwv}, the zero-momentum energy shift (``pion mass'') was determined by heavy-baryon chiral perturbation theory and found a large energy shift of even 100~MeV at nuclear saturation density for a neutron-rich medium, more along the lines of generic estimates~\cite{Ericson:1988gk}.

\subsection{Parametric estimate of the pion process}

Of course, one could take a pragmatic approach and use a plausible parametric form $E_\pi(p_\pi)$ for the $\pi^-$ dispersion relation. This is essentially what was done in the literature, which has focused on the $\pi^-p\to n a$ process. Of course, depending on the $\pi^0$ dispersion relation, a comparable contribution could come from $\pi^0 n\to n a$. While naively, the $\pi^-$ population is enhanced in comparison to $\pi^0$ by the chemical potential, the exact amount depends on the unknown dispersion relations, and the neutron population is also enhanced relative to protons, so the two  tendencies might counter. Without knowing the pion dispersion relations, one can not quantify such comparisons. In any case, purely parametrically, we can estimate the potential importance of the $\pi^- p\to n a$ process.

To this end, we ignore all of these issues and first calculate the emitted axion spectrum under simplifying assumptions. In terms of the kinetic energy $K_\pi(p_\pi)$ and the pion energy at vanishing momentum $E_{\pi,0}$, and using the pion chemical potential~$\mu_\pi$ (the subscript $\pi$ now always referring to $\pi^-$), the thermal Bose-Einstein pion occupation number is
\begin{equation}
    f_\pi(p_\pi)=\frac{z_\pi}{e^{K_\pi(p_\pi)/T}-z_\pi},
\end{equation}
where the fugacity is defined as
\begin{equation}
    z_\pi=e^{-(E_{\pi,0}-\mu_\pi)/T}.
\end{equation}
The vacuum dispersion relation would imply $E_{\pi,0}=m_\pi$, but according to a recent calculation based on heavy-baryon chiral perturbation theory, $E_{\pi,0}$ could be much larger, 200--260~MeV at nuclear density, and even larger at larger density \cite{Fore:2023gwv}.

The largest possible fugacity is $z_\pi=1$, at which point condensation of pions with $p_\pi=0$ obtains, forcing the overall system to adjust its equilibrium properties such this limiting value is not exceeded. On the other hand, so long as the system is not too close to condensation, $z_\pi$ can be taken to be small and neglected in the denominator, corresponding to Maxwell-Boltzmann statistics, whereas $z_\pi$ in the numerator is a global factor for the pion distribution. In this case, the pion number distribution can be written in the form
\begin{equation}
    \frac{dn_\pi}{dp_\pi}=Y_\pi n_B\,\frac{p_\pi^2}{\hat n_\pi}\,e^{-K_\pi/T},
\end{equation}
where $Y_\pi$ is the assumed number fraction of $\pi^-$ relative to baryons, which have the density $n_B$, $K_\pi$ is the kinetic energy, and the normalization factor is $\hat n_\pi = \int dp_\pi\,p_\pi^2 e^{-K_\pi/T}$. If the kinetic energy is taken to have the nonrelativistic form $K_\pi=p_\pi^2/2m_\pi$, the explicit form is $\hat n_\pi=\sqrt{\pi/2}\,(m_\pi T)^{3/2}$, but in general depends on the pion dispersion relation that determines $K_\pi(p_\pi)$. The pion energy spectrum is
\begin{equation}
    \frac{dn_\pi}{dE_\pi}=\frac{dn_\pi}{dK_\pi}
    =Y_\pi n_B\,\frac{p_\pi^2}{\hat n_\pi K'_\pi(p_\pi)}\,e^{-K_\pi/T},
\end{equation}
where $K'_\pi=dK_\pi/dp_\pi$ is the Jacobian and everywhere $p_\pi(K_\pi)$ must be inserted. We assume that $K_\pi$ is a monotonically increasing function of $p_\pi$. The nonrelativistic vacuum dispersion relation $K_\pi=p_\pi^2/2m_\pi$ implies $p_\pi=\sqrt{2 m_\pi K_\pi}$ and $K'=\sqrt{2K_\pi/m_\pi}$. 

Assuming the rate for ``pion to axion conversion'' $\pi^-\to a$ in the presence of nucleons is $\Gamma_{\pi a}$, accompanied by the recoil-free conversion of $p$ to $n$, the emitted axion spectrum per unit mass (which requires to divide by $n_B m_u$ with $m_u$ the atomi mass unit) is
\begin{equation}
    \frac{d \dot N_a}{dE_a dM}
    =\frac{Y_\pi}{m_u}\,\frac{[p_\pi(E_a)]^2}{\hat n_\pi K'_\pi(E_a)}\,e^{-K_a/T}\,\Gamma_{\pi a}.
\end{equation}
Here $K_a=E_a-E_{\pi,0}$ is the axion energy shifted by the zero-point of the pion energy, which is also the lowest emitted axion energy, or conversely, $E_a=E_{\pi,0}+K_a$. Of course,  $K_a$ is not the kinetic energy of the essentially massless axions, it is simply the shifted axion energy. For the nonrelativistic dispersion relation, this expression simplifies to
\begin{equation}
    \frac{d \dot N_a}{dE_a dM}
    =\frac{Y_\pi}{m_u}\,\frac{2\pi\,K_a^{1/2}}{(\pi T)^{3/2}}\,e^{-K_a/T}\,\Gamma_{\pi a},
\end{equation}
where the pion mass has completely dropped out. The only property of the nonrelativistic dispersion relation that has here entered is its quadratic dependence on $p_\pi$, not the coefficient for $p_\pi^2$. As a cross check we observe that $\int dK_a\,2\pi K_a^{1/2} e^{-K_a/T}/(\pi T)^{3/2}=1$.

We next need $\Gamma_{\pi{\to}a}=v_\pi\sigma_{\pi{\to}a}\,n_p$, where $n_p=Y_p n_B$ is the proton density. The cross section for $\pi^- p\to n a$ is related to the one for $\pi^- p\to n \pi^0$, the charge-exchange cross section that we will call $\sigma_{\rm cex}$. Notice that $\pi^0$ can be interpreted as an ALP with the mass $m_a=m_{\pi^0}$ and the couplings to nucleons
\begin{equation}\label{eq:pioncouplings}
    \frac{C_{a p}}{2f_a}\to\frac{g_A}{2f_\pi}
    \quad\mathrm{and}\quad
    \frac{C_{a n}}{2f_a}\to-\frac{g_A}{2f_\pi},
\end{equation}
where in vacuum the axial coupling is $g_A=1.27$ and the pion decay constant is $f_\pi=92.4$~MeV. In other words, $\pi^0$ is an ALP with purely isovector interaction, meaning that the translation to axions works directly only for isovector axions with $C_{ap}=-C_{an}$. We define the axion and $\pi^0$ iso-couplings as 
\begin{equation}\label{eq:isocouplings}
    C_{a\pm}=\frac{C_{ap}\pm C_{an}}{2}
    \quad\mathrm{and}\quad
    C_{\pi^0\pm}=\frac{C_{\pi^0 p}\pm C_{\pi^0 n}}{2}.
\end{equation}
For $\pi^0$, the structure is purely isovector, i.e., $C_{\pi^0+}=0$ and $C_{\pi^0-}=g_A$. Moreover, we also use the dimensionless axion couplings $g_{a\pm}=C_{a\pm}m_N/f_a$. After these substitutions, the iso-vector cross section is
\begin{equation}\label{eq:cex-1}
    \sigma_{\pi a}=\frac{g_{a-}^2f_\pi^2}{g_A^2m_N^2}\,\sigma_{\rm cex}.
\end{equation}
We anticipate that the charge exchange process is dominated by the $\Delta(1232)$ resonance as well as the pion-nucleon contact interaction. In other words, the isoscalar cross section would be much smaller and here we simply neglect it. In this case, for axions coupling purely to protons, the cross section is
\begin{equation}
    \sigma_{\pi a}=g_{ap}^2\left(\frac{f_\pi}{2m_N g_A}\right)^2\,\sigma_{\rm cex}.
\end{equation}
After these substitutions, the axion emission rate is
\begin{equation}
    \frac{d \dot N_a}{dE_a dM}
    =\frac{g_{ap}^2}{4m_N^2}\frac{Y_\pi}{m_u}\,\frac{2\pi\,K_a^{1/2}}{(\pi T)^{3/2}}\,e^{-K_a/T}\,\frac{f_\pi^2}{g_A^2}\,\Gamma_{\pi^-\pi^0},
\end{equation}
where $\Gamma_{\pi^-\pi^0}$ is the rate for $\pi^-$ to convert to $\pi^0$ in the nuclear background.

The charge-exchange process $\pi^- p\to n \pi^0$ can also be interpreted as $p\to n$ conversion under pionic attack, and the rate for a proton to convert in the background of the other particles must be $\Gamma_{pn}=(Y_\pi/Y_p)\,\Gamma_{\pi^-\pi^0}$. In this way, we can write the pionic emission rate in strong analogy to the bremsstrahlung one of Eq.~\eqref{eq:emission_rate_bremsstrahlung} in the form
\begin{equation}
    \frac{d \dot N_a}{dE_a dM}
    =\frac{g_{ap}^2}{4m_N^2}\frac{Y_p}{m_u}\,\frac{f_\pi^2\,2\pi\,K_a^{1/2}}{g_A^2(\pi T)^{3/2}}\,e^{-K_a/T}\,\Gamma_{np}.
\end{equation}
Integrating over energies and assuming an energy-independent $\Gamma_{np}$, the axion number emission rate per unit mass is
\begin{equation}
    \frac{d \dot N_a}{dM}\Big|_{\pi^- p\to n a}
    =\frac{g_{ap}^2}{4m_N^2}\frac{Y_p}{m_u}\,\Gamma_{np}\,\frac{f_\pi^2}{g_A^2}.
\end{equation}
We can compare this directly with the bremsstrahlung rate under the assumption that $\Gamma_\sigma$ so so small that we can ignore it in the denominator of 
Eq.~\eqref{eq:emission_rate_bremsstrahlung} and find
\begin{equation}
    \frac{d \dot N_a}{dM}\Big|_{\rm brems}
    =\frac{g_{ap}^2}{4m_N^2}\frac{Y_p}{m_u}\,\Gamma_{\sigma}\,\frac{T^2}{24}.
\end{equation}
Therefore, the ratio of number emission rates is
\begin{equation}
    \frac{{\rm pionic}}{{\rm brems}}=\frac{24 f_\pi^2}{g_A^2 T^2}\,
    \frac{\Gamma_{np}}{\Gamma_\sigma}=
    1.4\times10^2\left(\frac{30~{\rm MeV}}{T}\right)^2
    \frac{\Gamma_{np}}{\Gamma_\sigma}.
\end{equation}
Of course, the pionic emission produces axions with larger energies.

How large is the pion conversion rate? Typical pion momenta are $\sqrt{3 m_\pi T}\simeq 100$~MeV for which $\sigma_{\rm cex}\simeq6$~mb \cite{Breitschopf:2006gn}. Taking schematically $v_\pi\simeq 2/3$ and including a neutron degeneracy suppression schematically by a factor of 1/2 leads to $\Gamma_{\pi^-\pi^0}\simeq 2~{\rm mb}\,Y_p n_B$ as a rough estimate, which numerically is
\begin{equation}
    \Gamma_{\pi^-\pi^0}\simeq 7~{\rm MeV}\,
\frac{Y_p\,\rho}{3\times10^{14}~{\rm g}~{\rm cm}^{-3}}.
\end{equation}
Recalling that $\Gamma_{pn}=(Y_\pi/Y_p)\,\Gamma_{\pi^-\pi^0}$ as discussed earlier and with $\Gamma_\sigma$ of Eq.~\eqref{eq:Gamma_sigma}, assuming $T=30$~MeV, the ratio is $33\,Y_\pi$. If the $\pi^-$ abundance is around $0.03$ as found by Fore and Reddy \cite{Fore:2019wib}, the ratio is around one. 

\subsection{Pion dispersion relation}

This ratio is indeed roughly the result found in the pion-axion literature. Of course, the energy loss rate of the pionic process would be larger because every emitted axion carries both the rest energy $E_{\pi,0}$ as well as its kinetic energy $K_\pi$. As we have seen, the pion mass has dropped out, which happened because it was assumed to be quadratic in the momentum, i.e., of the form $K_\pi=p_\pi^2/2m_\pi^*$, where $m_\pi^*$ could be any large value. It is well known that the average kinetic energy of a nonrelativistic particle in thermal equilibrium is $\langle K_\pi\rangle=\frac{3}{2}\,T$, irrespective of how large is the mass. Relativistically, one could assume the dispersion relation to be
\begin{equation}
    E_\pi=E_{\pi,0}+\sqrt{(m_\pi^*)^2+p_\pi^2},
\end{equation}
where $m_\pi^*$ is called the Landau mass. In vacuum, of course, $E_{\pi,0}=m_\pi^*=m_\pi$. So for any assumed Landau mass, the kinetic energy is thermally distributed, whereas the total energy is shifted to larger energies by $E_{\pi,0}$. The expected axion spectrum then has the often-shown double-bump structure, one quasi-thermal bump from bremsstrahlung, the other a shifted one from pions.

The momentum distribution depends more sensitively on the dispersion relation than the kinetic energy distribution. For a larger $m_\pi^*$, the function $p_\pi^2/2m_\pi^*$ is ``flatter'', meaning that for the same kinetic energy, the corresponding momentum is larger. The pion-nucleon scattering cross section, being dominated by p-wave amplitudes, thus depends on what are typical pion momenta. Moreover, the dispersion relation need not be quadratic in $p_\pi$ and need not even be monotonically increasing. In neutron-rich matter and for $p_\pi\to 0$, one expects the energy to be increased by tens of MeV through a repulsive s-wave interaction as explained, for instance, around Eq.~(5.107) in Ref.~\cite{Ericson:1988gk}. Very recently, a calculation based on heavy-baryon chiral perturbation theory found $E_{\pi,0}$ of at least 200~MeV at nuclear density with a large upward uncertainty \cite{Fore:2023gwv}, but no information on the $p_\pi$ dependence was provided. Generally one expects an attractive p-wave forward scattering amplitude, dominated by the intermediate $\Delta$ resonance, implying that the energy shift $\Sigma_\pi(p_\pi)$ relative to the vacuum dispersion relation should be negative and soften the dispersion relation. Different forms for the dispersion relation were used in the literature on thermal pions in SNe \cite{Mayle:1993uj, Vijayan:2023qrt, Pajkos:2024iry, Fischer:2021jfm}, none of which is consistent with the other or based on a true calculation. We have concluded that we cannot resolve this question in a quantitatively meaningful way based on information available in the literature.

\section{Solar Maximum Mission Data}
\label{app:SMM}

We here derive our SMM fluence limit stated in Eq.~\eqref{eq:SMM-fluence-limit}, applicable to the 25--100~MeV channel. When considering photons from ALP decays, depending on the mass, the signal may have much longer duration than the actual emission period of a few seconds, which motivated the authors of Ref.~\cite{Hoof:2022xbe} to digitize the published plots that show the registered GRS event counts on the SMM satellite, and it is helpful to use this digital information, which is available on their github page. In a first paper \cite{Chupp:1989kx} (GRS-89), the counts of the GRS instrument were shown in 10.24~s bins for 9.3~min (557.7~s) around the SN~1987A neutrino burst, which is tagged by the arrival time $t_\nu$ of the first IMB neutrino that was detected  on 23 February 1987 at Universal Time (UT) 07:35:41.374 (27341.374~s). IMB was the only one of the three data-reporting detectors that had a reliable clock. It had a timing uncertainy of only $\pm50$~ms, allowing for a precise correlation with other experiments (for a recent review of the SN~1987A neutrino detection see Ref.~\cite{Fiorillo:2023frv}). A few years later, another paper \cite{Oberauer:1993yr} (GRS-93) showed counts for 152.6~s before and 223.232~s after $t_\nu$ in 2.048~s bins, implying 74 bins before and 109 bins after $t_\nu$. The better time resolution makes this data set more attractive to study transient features.

Besides their binning, the data sets differ in other ways. In the 25--100~MeV channel, GRS-89 showed only the counts from the CsI detectors (as indicated in the annotation of their Fig.~4 without explicit discussion), whereas GRS-93 showed the sum of the counts of the CsI and NaI detectors in this and the other channels, as stressed by the authors. Correspondingly, these papers state different effective areas, namely 63 and $90~{\rm cm}^2$, a ratio of 7/10, implying that in GRS-93, 70\% of the counts come from CsI and 30\% from NaI. The stated average count rates of 6.3 and 8.8~Hz reflect the same ratio. Therefore, there is no inconsistency between the two papers, they report different data.

Unfortunately, this point was not realized by the authors of Ref.~\cite{Hoof:2022xbe}, who instead speculated that somehow the data of GRS-93 were overstating the counts and that these should be corrected by a factor 7/10. To obtain the time structure of the signal, they digitized the figures with a procedure explained in their Appendix~A. For the GRS-93 data, they included the correction factor of 0.70, only in the 25--100~MeV channel, and always forced integer counts in each bin, so that finally they obtained a list of 183 integer counts. The timestamps are in the center of the bins and are such that there are 74 bins exactly before $t_\nu$ and 109 bins afterward. Of course, the SMM satellite did not know about the neutrino detection, so the temporal bins cannot be exactly aligned with $t_\nu$. Still, the exact alignment, such that the first ``signal'' bin begins exactly at $t_\nu$, looks like a reasonable approximation.  The digitized data for both reported data sets in all channels are available on their github page and were also used by other authors~\cite{Manzari:2024jns}.

We can apply a few simple tests on the consistency of the data. In the 10--25~MeV channel, both papers report the same data, the sum of the CsI and NaI counters. Therefore, re-binning the GRS-93 data into 10.24~s bins should provide exactly the same counts as reported in GRS-89. Based on the digitized data, we find that this is indeed the case, with three minor caveats. (i)~The GRS-93 data seem to have slipped by one bin width of 2.048~s relative to $t_\nu$, otherwise the rebinned data do not correspond to the GRS-89 ones. (ii)~There is one outlier in the data. (iii)~There is a nearly linear trend of a small difference developing with time, plausibly caused by a distortion in the printed plots or the digitazation procedure. Otherwise, the data show very clear correspondence and there is no doubt that the two papers show the same data in this channel.

We can repeat the same exercise in the 25--100~MeV channel. Here, the rebinned data of GRS-93 clearly do not correspond to those of GRS-89. By construction, the reported digitized data provide nearly the same averages, but no bin by bin correspondence. A more subtle test concerns the statistical fluctuation of the data. If we reconstruct the true GRS-93 data by multiplying the digitized ones by 10/7, which is correct up to rounding errors caused by the digitized data being integers, these original GRS-93 data visually show a nice Gaussian distribution. Quantitatively, if we have a data set of $n_i$ counts with an average $\overline n$, the expected variance is also $\overline n$, or the rms variation is $\sqrt{\overline n}$. Therefore, if the distribution is indeed Gaussian, one expects that the variance of the data is $\langle (n_i-\overline{n})^2\rangle\simeq\overline{n}$, which the reconstructed GRS-93 data fulfill within a few percent. Conversely, the reported digitized data, which include the 7/10 downscaling, do not fulfill this relation because the variance and $\overline n$ scale differently with the scaling factor. These tests prove the inconsistency of what was done in Ref.~\cite{Hoof:2022xbe}. In any event, there is no doubt about the true origin of the seeming discrepancy of the GRS-89 and GRS-93 data.

To finally determine our adopted constraints, we need the observed counts in the relevant bins. Beginning with GRS-89 with its coarse 10.24~s bins, we only need the counts in the one bin that follows $t_\nu$. According to the digitized data~\cite{Hoof:2022xbe}, these are $N_1=61$ counts, where the subscript denotes the number of the bin after $t_\nu$. The next one is $N_2=73$, and by visual inspection of Fig.~4 of GRS-89, one can easily identify this upward fluctuating bin to the right of the arrow that marks $t_\nu$. On the other hand, the arrow seems to point toward the tail of the bin before $t_\nu$, not directly on the boundary. However, $N_{-1}=60$ is very similar to $N_{1}=61$, so the result is not affected by strong fluctuations between these bins adjacent to $t_\nu$ or a slight misalignment between bins and neutrino signal. Anyway, axion emission would only commence around 1~s after the first neutrinos because the SN core first has to heat up. Therefore, we adopt $N=61$ for the count in the bin that includes the period of axion emission, to be compared with a background expectation of $B=65$.

For the GRS-93 data, we extract from the digitized data the counts
$(13,15,14,8,12)$ for the 2.048~s bins Nos.~$(-1,1,2,3,4)$, and visual inspection of Fig.~1 of GRS-93 allows one to easily identify the third bin to the right of the dotted line with a downward fluctuation. However, the alignment of the dotted line with these bins is visually not entirely unambiguous, and earlier we found a misalignment in the digitized data by 1~bin in the lower-energy channel. For an axion signal of duration of less than 5~s, we need the sum of three bins. So we have resolved to use $N_{-1}+N_{1}+N_2=42$ and avoid the downward fluctuation $N_3$ that might artifically improve the bound. After multiplying with 10/7 to reconstruct the true GRS-93 counts, our signal is here finally $N=\frac{10}{7}\,42=60$, to be compared with a background expectation of $B=54$ for the sum of three bins. Visual inspection of the original GRS-93 figure clearly reveals these three bins around the vertical dashed line and the sum of the counts in these bins is indeed around 60, with uncertainty of at most a few, without attempting a formal digitization.

We next determine the $3\sigma$ constraint on a putative signal $S>0$ that may contribute to the observed counts $N$. The likelihood for the number of counts is assumed to be Gaussian 
\begin{equation}
    P(N,S)\,dN=\frac{1}{\sqrt{2\pi\,(S+B)}}\,\exp\left[-\frac{(N-S-B)^2}{2\,(S+B)}\right]dN.
\end{equation}
We use the asymptotic likelihood test of Ref.~\cite{Cowan:2010js} and define a test statistic (TS) for upper bounds that is equal to $\Lambda=-2\bigl[\log[P(N,S)]-\log[P(N, \hat{S}(N)]\bigr]$ for $S>\hat{S}(N)$, and $\Lambda=0$ for $S<\hat{S}(N)$, where $\hat{S}(N)=(N-B)\,\Theta(N-B)$ is the best-fit value of $S$ for a given observed number of events, and $\Theta$ is the Heaviside function. Therefore, overall the TS is
\begin{equation}
    \Lambda=\frac{(N-S-B)^2}{(S+B)}\,\Theta(S+B-N)-\frac{(N-B)^2}{B}\,\Theta(B-N).
\end{equation}
We here neglect the $\sqrt{S+B}$ factors that appear outside of the exponential in the likelihood, and focus on the exponential factors only. Then the TS thus defined, under the assumption of a number $S$ of signal events, is expected to be distributed as a half-chi-squared distribution
\begin{equation}
    P_{\rm TS}(\Lambda)=\frac{1}{2}\delta(\Lambda)+\frac{1}{2}\frac{e^{-\Lambda/2}}{\sqrt{2\pi \Lambda}},
\end{equation}
where $\delta(\Lambda)$ is the Dirac delta. The $3\sigma$ constraint can be obtained by requiring that the one-sided interval $\Lambda>\overline{\Lambda}$ carries a probability smaller than $0.0027$, i.e., $\int_{\overline{\Lambda}}^{+\infty} P_{\rm TS}(\Lambda)d\Lambda=0.0027$, which gives $\overline{\Lambda}=7.7$. This in turn implies
\begin{equation}
    \frac{B+S-N}{\sqrt{2(B+S)}}>\chi=\sqrt{\frac{7.7}{2}}=1.96.
\end{equation}
For $N<B$, we instead have
\begin{equation}
    \frac{(N-S-B)^2}{S+B}-\frac{(N-B)^2}{B}>2\chi^2=7.7.
\end{equation}
Therefore, we finally find
\begin{subequations}\label{eq:limiting_signal}
\begin{eqnarray}
\kern-2em S&<&N-B+\chi^2+\sqrt{2N \chi^2+\chi^4}
\kern10.3em
\quad\mathrm{for}\quad N\geq B
\\[2ex]
\kern-2emS&<&\frac{N^2-B^2+2B \chi^2+\sqrt{8 B^3 \chi^2+(B^2-N^2-2B \chi^2)^2}}{2B}
\quad\mathrm{for}\quad 0\leq N\leq B
\end{eqnarray}
\end{subequations}
with $\chi=1.96$. For $N=B$ the two formulas are the same. The best constraints are obtained if $N\ll B$ and $B\gg 1$, in which case the expansion gives $S<2\chi^2$, whereas for $N\gg B$ we effectively obtain a measurement of $S$, but a formal limit is $S<N$. 

For $N=B+\Delta N$, with the positive or negative number of excess counts $\Delta N$ much smaller than the background as in our case, and assuming $\chi\ll B$ as well, a good approximation is
\begin{equation}
    S<\chi\sqrt{2B}+\Delta N +\frac{(\Delta N)^2}{2\chi\sqrt{2B}},
\end{equation}
showing the characteristic square-root behavior of constraints based on Poisson statistics in a background-dominated measurement. If $\Delta N=0$, the bound is $S<\chi\sqrt{2B}=2.8 \sqrt{B}$, very close to a naive $3\sigma$ fluctuation of $3\sqrt{B}$.

With this result, we can now obtain the $3\sigma$ constraint on the photon fluence. For GRS-89, using $B=65$ and $N=61$, we find that the upper bound on the number of signal events is $S<22.3$; with the effective area of $63\,\mathrm{cm}^2$, this gives an upper bound on the fluence of $\Phi_\gamma<0.35\,\mathrm{cm}^{-2}$. For GRS-93, using $B=54$ and $N=60$, we find $S<31.7$; with the effective area of $90\,\mathrm{cm}^2$, the upper limit $\Phi_\gamma<0.35\,\mathrm{cm}^{-2}$ is essentially the same. 

We note that at the time of SN~1987A, SMM was looking at the Sun, meaning that $\gamma$-rays from the direction of the SN where hitting the satellite roughly at $90^\circ$ sideways and had to penetrate $11.45~{\rm g}~{\rm cm}^2$ of CsI shield and $2.5~{\rm g}~{\rm cm}^2$ of inert spacecraft aluminum \cite{Chupp:1989kx}. The effective areas quoted in this paper take this effect into account, however assuming a power-law photon spectrum with the shape $E_\gamma^{-2}$ in the two higher-energy channels. In a much later paper deriving ALP constraints, Payez et al. \cite{Payez:2014xsa} comment, in their Sect.~4.2.4, that they could not reproduce these effective areas, but it is somewhat unclear what exactly they mean with these remarks. The effective areas of Ref.~\cite{Chupp:1989kx} were provided precisely for the $90^\circ$ off-axis situation---we cannot do better than the original SMM scientists.

We can also estimate the modification of the bound if one were to use the scaled data of Ref.~\cite{Hoof:2022xbe} who assumed the true background rate was a factor $\xi=0.7$ smaller than the reported one and that the effective area was also smaller by the same factor, i.e., both the expected signal and the background are smaller by the same factor $\xi$. From Eqs.~\eqref{eq:limiting_signal}, using both $N$ and $B$ to be a factor $\xi$ smaller, the limiting number of signal events is $S=26.4$; since the fluence itself is however larger by a factor $0.7$, the limit on the fluence should be weaker by about $26.4/0.7\times 31.7\simeq 1.19$.

Finally, we can test the impact of using $2\sigma$ rather than $3\sigma$ bounds, as in Ref.~\cite{Manzari:2024jns}. This corresponds to using $\chi=1.16$, so the bounds on the number of signal events is strengthened to $S=20.1$, i.e., by a factor 0.63.

\bibliographystyle{JHEP}
\bibliography{Biblio_cleaned.bib}

\end{document}